\documentclass{sig-alternate-10pt} % Aptara syntax

\usepackage{url}
\usepackage{mathrsfs}
\usepackage{enumitem}
\usepackage{MnSymbol}
\usepackage{amsbsy}
\usepackage{clrscode}
\usepackage{flushend}

\newcounter{descriptcount}

\newtheorem{definition}{Definition}
\newtheorem{theorem}{Theorem}
\newtheorem{lemma}{Lemma}
\newtheorem{proposition}{Proposition}

\newtheorem{corollary}{Corollary}
\newtheorem{example}{Example}

\def\core{\mathop{\mathit core}}
\def\dom{\mathop{\mathit dom}}
\def\chase{\mathop{\mathit Chase}}

\def\arity{\mathop{\mathit arity}}

\def\vars{\mathop{\mit vars}}

\newcommand{\SigmaTheta}{\ensuremath{\Sigma_{\scriptscriptstyle \Theta}}}

\newcommand{\SigmaLR}{\ensuremath{\Sigma_{\scriptscriptstyle L\!R}}}

\newcommand{\SigmaR}{\ensuremath{\Sigma_{\scriptscriptstyle \mathcal{R}}}}

\newcommand{\Rarrow}{\ensuremath{\rightarrow_{\scriptscriptstyle \mathcal{R}}}}

\newcommand{\sctaa}{\ensuremath{{{\mathsf{CT}}^{\mathsf{std}}_{\forall\forall}}}}
\newcommand{\sctae}{\ensuremath{{{\mathsf{CT}}^{\mathsf{std}}_{\forall\exists}}}}
\newcommand{\sctia}[1]{\ensuremath{{{\mathsf {CT}}^{\mathsf{std}}_{{#1}\forall}}}}
\newcommand{\sctie}[1]{\ensuremath{{{\mathsf {CT}}^{\mathsf{std}}_{{#1}\exists}}}}

\newcommand{\octaa}{\ensuremath{{{\mathsf{CT}}^{\mathsf{obl}}_{\forall\forall}}}}
\newcommand{\octae}{\ensuremath{{{\mathsf{CT}}^{\mathsf{obl}}_{\forall\exists}}}}
\newcommand{\octia}[1]{\ensuremath{{{\mathsf {CT}}^{\mathsf{obl}}_{{#1}\forall}}}}
\newcommand{\octie}[1]{\ensuremath{{{\mathsf {CT}}^{\mathsf{obl}}_{{#1}\exists}}}}

\newcommand{\soctaa}{\ensuremath{{{\mathsf {CT}}^{\mathsf{sobl}}_{\forall\forall}}}}
\newcommand{\soctae}{\ensuremath{{{\mathsf {CT}}^{\mathsf{sobl}}_{\forall\exists}}}}
\newcommand{\soctia}[1]{\ensuremath{{{\mathsf {CT}}^{\mathsf{sobl}}_{{#1}\forall}}}}
\newcommand{\soctie}[1]{\ensuremath{{{\mathsf {CT}}^{\mathsf{sobl}}_{{#1}\exists}}}}

\newcommand{\cctaa}{\ensuremath{{{\mathsf {CT}}^{\mathsf{core}}_{\forall\forall}}}}
\newcommand{\cctae}{\ensuremath{{{\mathsf {CT}}^{\mathsf{core}}_{\forall\exists}}}}
\newcommand{\cctia}[1]{\ensuremath{{{\mathsf {CT}}^{\mathsf{core}}_{{#1}\forall}}}}
\newcommand{\cctie}[1]{\ensuremath{{{\mathsf {CT}}^{\mathsf{core}}_{{#1}\exists}}}}

\newcommand{\actaa}{\ensuremath{{{\mathsf{CT}}^{\star}_{\forall\forall}}}}
\newcommand{\actae}{\ensuremath{{{\mathsf{CT}}^{\star}_{\forall\exists}}}}
\newcommand{\actia}[1]{\ensuremath{{{\mathsf {CT}}^{\star}_{{#1}\forall}}}}
\newcommand{\actie}[1]{\ensuremath{{{\mathsf {CT}}^{\star}_{{#1}\exists}}}}

\newcommand{\depenrich}[1]{\ensuremath{\hat{#1}}}
\newcommand{\enrich}[1]{\ensuremath{\widehat{#1}}}
\newcounter{foo}
\newcounter{qcounter}

\begin{document}

\title{The data-exchange chase under the microscope}

\numberofauthors{2} 
\author{
\alignauthor Gosta Grahne\\
\affaddr{Concordia University}\\
\affaddr{Montreal, Canada, H3G 1M8}\\
\email{grahne@cs.concordia.ca}
\alignauthor Adrian Onet
\titlenote{Contact author.}\\
\affaddr{Concordia University}\\
\affaddr{Montreal, Canada, H3G 1M8}\\
\email{a\_onet@cs.concordia.ca}
}

\maketitle

\begin{abstract}
A lot of research activity has recently taken place around the chase procedure, 
due to its usefulness in data integration, data exchange, query optimization, 
peer data exchange and data correspondence, to mention a few. As the chase 
has been investigated and further developed by a number of research groups 
and authors, many variants of the chase have emerged and associated results 
obtained. Due to the heterogeneous nature of the area it is frequently 
difficult to verify the scope of each result. In this paper we take closer 
look at recent developments, and provide additional results.
Our analysis allows us create 
a taxonomy of the chase variations and the properties they satisfy. 

Two of the most central problems regarding the chase is termination, 
and discovery of restricted classes of sets of dependencies that guarantee 
termination of the chase. 
The search for the restricted classes has been motivated by 
a fairly recent result that shows that it is undecidable to 
determine whether the chase with a given dependency set will 
terminate on a given instance. 
There is a small dissonance here, since the quest has been 
for classes of sets of dependencies guaranteeing termination of the chase 
on {\em all} instances, even though the latter problem was not 
known to be undecidable. 
We resolve the dissonance in this paper by showing that determining 
whether the chase with a given set of dependencies terminates on all 
instances 
is {\sf coRE}-complete. 
For the hardness proof we use a reduction from 
word rewriting systems, thereby  
also showing the close connection between the chase and word rewriting.
The same reduction also gives us the aforementioned 
instance-dependent {\sf RE}-completeness result as a byproduct. 
For one of the restricted classes
guaranteeing termination on all instances, 
the {\em stratified} sets dependencies,
we provide new complexity results for
the problem of testing whether a given set
of dependencies belongs to it.
These results rectify some previous claims that have occurred
in the literature.
\end{abstract}



\keywords{Chase, Date Exchange, Data Repair, Incomplete databases, Undecidability Complexity}

%-- INTRODUCTION 
\bigskip 
\section{Introduction}\label{introduction}

\bigskip
The chase procedure was initially developed 
by \cite{DBLP:conf/icalp/BeeriV81} for testing
logical implication between sets of dependencies, 
by \cite{DBLP:conf/xp/Mendelzon81,DBLP:journals/jacm/ImielinskiL84}
for determining 
equivalence of database instances known to satisfy 
a given set of dependencies,
and by \cite{320091} to determine query equivalence under database 
constraints.
Recently the chase has experienced a revival due to
its application in data integration, data exchange, data 
repair, query optimization, ontologies and data correspondence.
In this paper we will focus on 
constraints in the form of
embedded dependencies
specified by sets of tuple and equality generating dependencies
as specified by \cite{DBLP:journals/jacm/Fagin82}.
A tuple generating dependency (tgd) 
is a first order formula of the form
$$
\forall\bar{x} \forall\bar{y}\; \big( \alpha(\bar{x},\bar{y}) 
\rightarrow \exists \bar{z}\; \beta(\bar{x},\bar{z}) \big),
$$
where $\alpha$ and $\beta$ are conjunctions of relational 
atoms, and $\bar{x}$, $\bar{y}$, and $\bar{z}$ are sequences of variables.
We refer to $\alpha$ as the body and $\beta$ as the head of the dependency.
Sometimes, for simplicity, the tgd is written as  $\alpha\rightarrow \beta$.
An equality generating dependency (egd) 
is a first order formula of the form
$$
\forall \bar{x}\; \big( \alpha(\bar{x}) 
\rightarrow x=y \big),
$$
where $\alpha$ is a conjunction of relational 
atoms, $\bar{x}$ is a sequence of variables that contains both variables 
$x$ and $y$.
Sometimes, for simplicity, the egd is written as  $\alpha\rightarrow x=y$.
Intuitively the chase procedure repeatedly applies
chase steps to database instances that violate
some dependency. 
In case a tgd is not satisfied by the instance a chase step takes 
a set of tuples that witness the violation,
and adds new tuples
to the database instance so that the resulting instance 
does satisfy the tgd with respect to those witnessing tuples.
In case an egd is not satisfied by the instance a chase step takes
a set of tuples that  witness the violation and try to
unify in the instance
 the elements corresponding to the variables equated by the egd.
In case such a unification is not possible, because 
both matching elements are constants, the chase process will fail.
As an example consider a database instance \;\;
$I=\{$ {\sl{Std(S01, john), \;\; Aff(S01,MIT),\; 
Aff(S02,MIT),\; ParkResv(S02, mike, R03)}} $\}$
over a student database containing student information {\sl Std(sid,name)},
student affiliation {\sl Aff(sid, univ)}, and
student parking reservation {\sl ParkResv(sid, name, spot)} .
Consider also the foreign key constraint that states 
that each {\sl sid} value in the {\sl Aff} relation needs to be
a valid student id, that is. 
it needs to also occur in the 
{\sl Student} relation. 
This foreign key constraint can be expressed using the following 
tgd: 
$$
\forall \mbox{\sl sid } \forall \mbox{\sl univ\,}\;\big(\mbox{\sl  Aff(sid,univ)} 
\rightarrow 
\exists \mbox{\sl name\, Std(sid,name)}\big).
$$ 
In our example tuple {\sl Aff(S02,MIT)} violates
this constraint because student id {\sl S02} 
is not part of the {\sl Student}
relation. In this case the chase step will simply add 
to 
instance $I$ a new tuple
{\sl Std(S02, $x$)}, where $x$ represents an unknown value.
Consider also the following constraint that say that
each student id is associated with a unique student name:
\begin{eqnarray*}
& &\hspace{-0.5cm}\forall \mbox{\sl sid}\;
\forall \mbox{\sl name1}\;
\forall \mbox{\sl name2}\;
\forall \mbox{\sl spot}\; \\ 
& &\hspace{-0.5cm}\big(\mbox{\sl  Std(sid,name1), ParkResv(sid, name2, spot)} \\
& &\hspace{5cm} \rightarrow  \mbox{\sl name1=name2} \big).
\end{eqnarray*}
In our example the newly added tuple {\sl Student(S02, $x$)}
together with the tuple {\sl ParkResv(S02, mike, R03)}
violate the constraint because student id 
{\sl S02} is associated with unknown value $x$ and value {\sl mike}.
In this case the chase step will change value $x$ in the first
tuple to {\sl mike}.

Given an instance $I$ and a set of tgds $\Sigma$, 
a {\em model} of $I$ and $\Sigma$ is a database instance
$J$ such that there is a homomorphism from $I$ to $J$, 
and $J$ satisfies $\Sigma$.
A {\em universal model} of $I$ and $\Sigma$ is
a {\em finite} model of $I$ and $\Sigma$
that has a homomorphism into {\em every}
model of $I$ and`$\Sigma$.
It was shown by \cite{DBLP:conf/icdt/FaginKMP03,DBLP:conf/pods/DeutschNR08} 
that the chase computes a universal model
of $I$ and $\Sigma$, whenever $I$ and $\Sigma$
has one. In case
$I$ and $\Sigma$ does not have a universal model
the chase doesn't terminate (in this case 
it actually converges at a countably infinite model).

%\begin{table*}
%\centering
%\caption{The chase step complexity.}
%\begin{tabular}{lllll} 
% & \multicolumn{2}{c}{Data Complexity} & \multicolumn{2}{c}{Combined %Complexity}  \\
%\hline 
% Chase & Existence & Check & Existence & Check  \\
%\hline
%\hline
%standard & $O(n^{|\alpha|+|\beta|})$ & $O(n^{|\beta|})$ & $\Sigma^P_2$-complete %& {\sf coNP}-complete \\
%oblivious & $O(n^{|\alpha|})$ & $O(n)$ & {\sf NP}-complete & $O(n^2)$ \\
%\hline     
%\end{tabular}
%\end{table*}

As the research on the chase has progressed,
several variations of the chase have evolved.
As a consequence it has become difficult
to determine the scope of the results obtained.
We scrutinize the four most important chase variations,
deterministic and non-deterministic,
namely the {\em standard},
{\em oblivious},
{\em semi-oblivious}, and
{\em core} chase.
We will analyze, 
for each of these chase variations,
the data and combined complexity of testing
if the chase step is applicable 
for a given instance and tgd.
The data complexity measures the 
required computation time as a function of the 
number of tuples in the instance,
while the combined complexity also takes
the size of the dependency into account. 
It didn't come as a surprise
that the 
oblivious and semi-oblivious chase
variations share the same complexity,
and that the standard chase has a slightly 
higher complexity.
The table below shows the data and combined complexity
for the following problem:
given an instance with $n$ tuples
and a tgd $\alpha\!\rightarrow\!\beta$, 
is the core/standard/oblivious/semi-oblivious chase step applicable? 
By $|\alpha|$ we mean the number of atoms in $\alpha$,
and similarly for $|\beta|$.
Note that in the case of egds the chase applicability complexity 
is the same as the applicability of the oblivious tgd chase step.

\medskip 
\begin{center}
\begin{tabular}{l||ll}
%\hline 
\;\;\;\;\;\;\;\;\;\;\;\;Chase  & {Data} & {Combined}  \\
\;\;\;\;\;\;\;\;\;\;\;\;Step & {Complexity} & {Complexity}  \\
\hline
\hline
& & \\
standard/core & $O(n^{|\alpha|+|\beta|})$ & $\mathsf{\Sigma^P_2}$-complete  \\
& & \\
oblivious/semi-oblivious & $O(n^{|\alpha|})$ & {\sf NP}-complete  \\
%\hline     
\end{tabular}
\end{center}

\medskip 
Thus, at a first look the oblivious and semi-oblivious 
chase procedures will be a more appropriate 
choice when it comes to a practical implementation.
Still, as we will show,
 the lower complexity comes with 
a price, that is  
the higher the complexity
for a chase variation
the more "likely" it is that the chase process 
terminates for a given instance and set of dependencies.
%Now let us consider the problem: 
%{\em Given instances $I$,$J$ and trigger $(\xi,h)$ is it
%that $J$ is obtained in one core/standard/oblivious/semi-oblivious chase step
%from $I$ with trigger $(\xi,h)$ }?
%It is easy to see that
%for the /standard/oblivious/semi-oblivious chase
%the problem is polynomial and
%from \cite{DBLP:conf/pods/FaginKP03}
%follows that for the core chase the problem is {\sf DP}-complete. 
On the other hand, 
\cite{DBLP:conf/pods/DeutschNR08}
showed that 
the core chase is complete in finding 
universal models, 
meaning that
if any of the chase variations
terminates for some input,
then the core chase terminates as well.
We next compare the semi-oblivious and
standard chase when it comes to the termination problem.
With this we show 
\begin{itemize}
%that the 
\item
The standard and semi-oblivious chases 
are not distinguishable
by most classes of dependencies developed
to ensure the standard chase termination.
%Furthermore,
%we show that
\item 
The number of semi-oblivious chase steps
needed to terminate  remains 
the same as for the standard chase,
namely polynomial.
\end{itemize}
This raises the following question: 
\begin{itemize}
\item
What makes a class of dependency sets
that ensures termination for {\em all} input instances under 
the standard chase, 
also ensure termination for the
semi-oblivious chase as well?
\end{itemize}
We answer this question by giving a 
sufficient syntactical condition for a set of dependencies 
to also guarantee the semi-oblivious chase termination.
As we will see, 
most of the  known
classes of dependencies built to ensure standard chase 
termination on all instances actually
guarantee termination for the computationally 
less expensive semi-oblivious chase variation.

It has been known for some time, from
\cite{DBLP:conf/pods/DeutschNR08,DBLP:conf/kr/CaliGK08,DBLP:conf/pods/Marnette09},
that it is undecidable to determine if the chase with a given set
of tgds terminates on a given instance. 
This has spurred a quest for restricted classes of
tgds guaranteeing termination.
Interestingly, these classes all guarantee {\em uniform} termination,
that is,
termination on {\em all} instances. 
This, even though it was 
only known that the problem is undecidable for
a given instance. We remediate the situation
by proving that 
(perhaps not too surprisingly)
the uniform version of the termination problem is
undecidable as well, and show that it is not
recursively enumerable. 
\begin{itemize}
\item
We show that determining whether
the core chase with a given set of dependencies
terminates on all instances 
is a {\sf coRE}-complete problem.
\end{itemize}
We achieve this using a reduction
from the uniform termination problem
for word-rewriting systems (semi-Thue systems).
As a byproduct we obtain the result from 
\cite{DBLP:conf/pods/DeutschNR08} showing that testing
if the core chase terminates for a given instance and 
a given set of dependencies is {\sf RE}-complete.
We will show also that the same complexity result
holds for testing whether the standard chase
with a set of dependencies
is terminating on at least one
execution branch.
Next we will show that by using a single denial constraint
(a ``headless'' tgd)
in our reduction, 
the same complexity result holds also for 
the standard chase termination 
on all instances on all execution branches. 
It remains an open problem if this holds without
denial constraints.

Many of the restricted classes guaranteeing termination
rely on the notion of a set $\Sigma$ of dependencies
being {\em stratified}. Stratification involves
two conditions, one 
determining a partial order between tgds in
in $\Sigma$,
and the other on $\Sigma$ as a whole. 
It has been claimed by \cite{DBLP:conf/pods/DeutschNR08} 
that testing 
the partial order between tgds 
is in {\sf NP}.
\begin{itemize}
\item
We show that testing the partial order between tgds
cannot be in {\sf NP} 
%this cannot be the case
(unless {\sf NP}={\sf coNP}), by proving
that the problem is at least {\sf coNP}-hard.
\end{itemize}
We also prove a $\mathsf{\Delta^p_2}$
upper bound
for the problem.
Finding matching upper and lower bounds
remains an open problem.

This papers only focuses on constraints represented by finite
sets of tgds and egds, which covers most of the constraints used
in practice today.
Not covered here
is a whole body of work
involving more general classes of constraints, 
where tgds are extended
by allowing negated atoms in the body
and disjunctions of atoms in the head 
(e.g.\ the work of \cite{DBLP:conf/pods/HernichKLG13,DBLP:conf/ijcai/MagkaKH13})
and also by allowing disjunctions between the atoms in the body 
(e.g.\ the work of \cite{DBLP:conf/pods/DeutschNR08}).

\subsubsection*{Paper outline}

The next section contains the preliminaries
and describes the chase procedure and its variations.
Section 3 considers the
complexity of testing if for an instance and a dependency
there exists an ``applicable'' chase step.
Section 4 deals with problems related to the chase termination. 
We define termination classes for each of
the chase variations and then determine the
relationship between these classes.
Section 5 contains our main technical result, namely,
that it is {\sf coRE}-complete to test if the 
core or standard chase 
variations with a given 
set of dependencies 
terminate on all instances. 
This result is obtained via a 
reduction from the uniform termination problem for
word-rewriting systems.
In Section 6 we review the main
restricted classes that ensure termination
on all instances,
and relate them to different chase variations
and their termination classes.
Finally, in Section 7
we provide complexity results related to 
the membership problem for the 
stratification based classes of dependencies 
that ensure the standard chase termination.
Conclusions and directions for further work 
appear in the last section.

%-- PRELIMINARIES 
\bigskip
\section{Preliminaries}\label{prelim}

\bigskip
For basic definitions and concepts we
refer to \cite{DBLP:books/aw/AbiteboulHV95}. We will consider
the complexity classes
{\sf PTIME},
{\sf NP},
{\sf coNP},
{\sf DP},
$\mathsf{\Delta^P_2}$,
{\sf RE},
{\sf coRE},
and the first few levels
of the polynomial hierarchy.
For the definitions of these classes
we refer to \cite{DBLP:books/daglib/0072413}.
 
We start with some preliminary notions.
We will use the symbol $\subseteq$ for 
the subset relation, and $\subset$
for proper subset.
A function $f$ with a finite set
$\{x_1,\ldots,x_n\}$ as domain,
and where $f(x_i)=a_i$,
will be described as
$\{x_1/a_1,\ldots,x_n/a_n\}$. 
The reader is cautioned that the symbol
$\rightarrow$ will be overloaded;
the meaning should however be clear
from the context. 

\bigskip
\noindent
{\bf Relational schemas and instances}.
A {\em relational schema} is a finite set
$\mathbf{R} = \{R_1,\ldots,R_n\}$ of relational symbols
$R_i$, each with an associated positive integer $\arity(R_i)$.
Let {\sf Cons} be a countably infinite set of
constants, usually denoted $a,b,c,\ldots$,
possibly subscripted, and let {\sf Nulls} be
a countably infinite set of nulls denoted
$x,y_1,y_2,\ldots$. 
A {\em relational instance}
over a schema $\mathbf{R}$ is a function that
associates for each relational symbol $R\in\mathbf{R}$
a finite subset $R^{I}$ of 
$({\sf Cons} \; \cup \; {\sf Nulls})^{\arity(R)}$. 

A {\em relational atom} is an expression of the
form $R(\bar{x})$, where $R\in\mathbf{R}$,
and $\bar{x}$ is a sequence of nulls
and constants, and the sequence is of length $\arity(R)$.
If the sequence $\bar{x}$ contains only constants,
we denote it $\bar{a}$, and call $R(\bar{a})$
a {\em ground atom}.

We shall 
frequently identify 
an instance $I$ with the
set $\{R(\bar{x})  : (\bar{x})\in R^I, R\in\mathbf{R}\}$
of atoms, assuming appropriate lengths 
of the sequence $\bar{x}$ for each
$R~\in~\mathbf{R}$.
By the same convenience,
the atoms $R(x_1\ldots,x_k)$ 
will sometimes be called {\em tuples}
of relation $R^I$ and denoted $t,t_1,t_2,\ldots$.
By $\dom(I)$ we mean the set of
all constants and nulls occurring in the
instance $I$, and by $|I|$ we mean the number
of tuples in $I$. 

\bigskip
\noindent
{\bf Homomorphisms}.
Let $I$ and $J$ be instances,
and $h : \dom(I)\rightarrow\dom(J)$
a mapping
that is the identity on the constants.
We extend $h$ to tuples $(\bar{x})=(x_1,\ldots,x_k)$ by
$h(x_1,\ldots,x_k) = (h(x_1),\ldots,h(x_k))$.
By our notational convenience we can thus
write $h(\bar{x})$ as $h(R(\bar x))$,
when $(\bar{x})\in R^I$.
We extend homomorphism $h$ to instances by
$h(I) = \{h(t) : t\in I\}$.
If $h(I)\subseteq J$ we say that 
$h$ is a {\em homomorphism} from $I$ to $J$.
If $h(I)\subseteq I$,
we say that $h$
is an {\em endomorphism}.
%, and if $h$ also 
%is idempotent, $h$ is called a {\em retraction}.
If $h(I)\subseteq J$,
and the mapping $h$ is a bijection,
and if also $h^{-1}(J)~=~I$,
the two instances
are {\em isomorphic},
which we denote $I\cong J$.
If both $h(I)\subseteq J$, and
$g(J)\subseteq I$, for some homomorphisms
$h$ and $g$, we say that $I$ and $J$
are {\em homomorphically equivalent}.
Note that isomorphic instances are
homomorphically equivalent, but not
vice versa.

A subset $I'$ of $I$ is said to be a
{\em core} of $I$, if there is a
endomorphism $h$, such that $h(I)\subseteq I'$,
and there is no
endomorphism $g$ such that
$g(I')\subset I'$.
It is well known that all cores
of an instance $I$ are isomorphic,
so for our purposes we can consider 
the core unique,
and denote it $\core(I)$.
The core of an instance $I$ 
plays an important role as it is the 
smallest instance
that is homomorphically equivalent to $I$.

\bigskip
\noindent
{\bf Tuple generating dependencies}.
A {\em tuple generating dependency  (tgd)} is
a first order formula of the form
$$
\forall \bar{x},\bar{y}\; 
\big( \alpha(\bar{x},\bar{y})\rightarrow 
\exists \bar{z}\;
\beta(\bar{x},\bar{z}) \big),
$$
where $\alpha$ and $\beta$ are 
conjunctions of relational atoms,
and $\bar{x}$,$\bar{y}$ and $\bar{z}$ 
are sequences of variables.
Occasionally we will abuse notation and  write
 $x\in\bar{x}$ to mean that $x$ 
occurs in the sequence.
We assume that
the variables occurring in tgds come from a countably
infinite set {\sf Vars} disjoint from {\sf Nulls}.
We also allow constants from {\sf Cons} in the tgds.
In the formula we call $\alpha$ the {\em body} of the tgd.
Similarly we refer to $\beta$ as the {\em head}
of the~tgd. If there are no existentially
quantified variables the dependency is said
to be {\em full}.

When $\alpha$ is the body of a tgd and 
$h : {\sf Vars}\hspace{0.03cm}\cup\hspace{0.03cm}{\sf Const}
\rightarrow
{\sf Nulls}\hspace{0.03cm}\cup\hspace{0.03cm}{\sf Const}$
is a (partial) mapping 
that is identity on constants, 
we shall conveniently regard 
the set of atoms in $\alpha$ as an instance $I_{\alpha}$,
and write $h(\alpha)$ for 
the set $h(I_{\alpha})$. 
Then $h$ is a homomorphism
from $\alpha$ to an instance $I$,
if $h(\alpha)\subseteq I$.

%otherwise it is {\em embedded}.
%In case a body of a tgd contains only one atom, it is called
%a {\em Local-As-View tgd (LAV-tgd)}. If a LAV-tgd does not 
%contain repeated variables in the body
%it is called a {\em true LAV-tgd}.

Frequently, we omit the universal quantifiers
in tgd formulas.
Also, when the variables and constants 
are not relevant
in the context, we 
denote a tuple generating dependency 
$
\alpha(\bar{x},\bar{y})\rightarrow 
\exists \bar{z}\;
\beta(\bar{x},\bar{z})
$
simply as $\alpha\!\rightarrow\!\beta$. 
Tgds will be denoted by the letter $\xi$,
possibly subscripted.

Let $\xi = \alpha\!\rightarrow\!\beta$  
be a tuple generating dependency,
and $I$ be an instance. Then we say that
$I$ {\em satisfies} $\xi$, if 
$I\models\xi$
in the standard model theoretic sense
(see e.g.\ \cite{DBLP:books/daglib/0076838}), 
or equivalently, 
if for every homomorphism 
$h: {\sf Vars}\hspace{0.03cm}\cup\hspace{0.03cm}{\sf Const}
\rightarrow
{\sf Nulls}\hspace{0.03cm}\cup\hspace{0.03cm}{\sf Const}$, 
such that
$h(\alpha)\subseteq I$, there is an extension
$h'$ of $h$, such that $h'(\beta)\subseteq I$.

\bigskip
\noindent
{\bf Equality generating dependencies}.
An {\em equality generating dependency  (egd)} is
a first order formula of the form
$$
\forall \bar{x} 
\big( \alpha(\bar{x})\rightarrow x=y \big),
$$
where $\alpha$ is a 
conjunctions of relational atoms,
$\bar{x}$ is a sequence of variables 
and $x, y$ are variables from $\bar{x}$. 
We assume that
the variables occurring in egds come from a countably
infinite set {\sf Vars} disjoint from {\sf Nulls}.
In the formula we call $\alpha$ the {\em body} of the egd.

Similarly to the tgd case, frequently, 
we omit the universal quantifiers
in egd formulas.
Also, when the variables and constants 
are not relevant
in the context, we 
denote an equality generating dependency 
$
\alpha(\bar{x})\rightarrow x=y
$
simply as $\alpha\!\rightarrow\! x=y$. 

Let $\xi = \alpha\!\rightarrow\!x=y$  
be an equality generating dependency,
and $I$ be an instance. Then we say that
$I$ {\em satisfies} $\xi$, if 
$I\models\xi$
in the standard model theoretic sense
(see e.g.\ \cite{DBLP:books/daglib/0076838}), 
or equivalently, 
if for every homomorphism 
$h: {\sf Vars}\hspace{0.03cm}\cup\hspace{0.03cm}{\sf Const}
\rightarrow
{\sf Nulls}\hspace{0.03cm}\cup\hspace{0.03cm}{\sf Const}$, 
such that
$h(\alpha)\subseteq I$, it holds that $h(x)=h(y)$.

%\medskip
%\section{The Chase}\label{theChase}

\bigskip
\noindent
{\bf The Chase}.
The chase is a procedure that takes as input and instance $I$
and a set of constraints $\Sigma$, which in the present
paper will be a finite set of tgds and egds. 
The chase might not
terminate or it might {\em fail}, 
but if it does terminate successfully  (without failing)
it produces a finite instance $J$,
such that
\begin{enumerate}
\item
$J\models\Sigma$.
\item
There is a homomorphism
from $I$ to $J$.
\item
For every (finite or infinite) instance $K$,
if there is a homomorphism from $I$ to $K$
and $K\models\Sigma$, then there is a homomorphism
from $J$ to $K$.
\end{enumerate}
The instance $J$ produced by the chase on $I$
with $\Sigma$ is called an 
{\em universal model} of $I$ and~$\Sigma$ \cite{DBLP:conf/icdt/FaginKMP03}.
Universal models are crucial in {\em data exchange},
where they are used to materialize the data
imported into a target database
from a source database. In this context
the constraints $\Sigma$ describe the relationship
between the source schema and the target schema,

For a formal definition of the chase,
let $\Sigma$ be a (finite) 
set of tgds and egds, and $I$ an instance. 
A~{\em trigger} 
for $\Sigma$ on $I$ 
is a pair $(\xi,h)$, where 
$\xi$ is either a tgd 
%$\alpha\!\rightarrow\!\beta\in\Sigma$
or an egd, 
%$\alpha\!\rightarrow\!x=y \in\Sigma$,
and $h$ is a homomorphism from
$\alpha$ to $I$, i.e.\ 
$h(\alpha)\subseteq I$.
If $\xi$ is a tgd
$\alpha\!\rightarrow\!\beta$, the trigger
$(\xi,h)$ is said to be {\em active} if
there is no extension $h'$ of $h$,
such that $h(\beta)\subseteq I$, 
and 
if $\xi$ is an edg
$\alpha\!\rightarrow\!x=y$, 
the trigger $(\xi,h)$
is said to be active on $I$ if
$h(x)\neq h(y)$,

Let $(\xi,h)$ be a trigger for $\Sigma$ on $I$. 
To {\em fire} the trigger means:

\begin{itemize}
\item in case $\xi$ is a tgd, transforming $I$ into
the instance 
$J=I\cup\{h'(\beta)\}$, where $h'$ is 
a {\em distinct extension} of $h$,
i.e.\ an extension
of $h$ that assigns new fresh nulls
to the existential variables in $\beta$.
By ``new fresh'' we mean the next unused element
in some fixed enumeration
of the nulls.

The transformation is denoted
$I\xrightarrow{(\xi,h)}J$,
or just 
$I\rightarrow_{\xi} J$,
if the particular homomorphism $h$ and its distinct
extension $h'$ are 
irrelevant or understood from the context.

\item in case $\xi$ is an egd:

\begin{itemize}
\item 
if both $h(x)$ and $h(y)$ are nulls, 
transforming $I$ into instance $J$ such that all occurrences
of $h(y)$ are replaced with $h(x)$. Here we assume there is 
an enumeration of the variables, and that $h(x)<h(y)$ is this enumeration.
\item 
if one of $h(x)$ and $h(y)$ is a null and the other 
a constant, transforming $I$ into instance $J$ such that all occurrences 
of the null is replaced with the constant.
\item 
if both $h(x)$ and $h(y)$ are constants (note that $h(x)\neq h(y)$),
then the firing of the trigger {\em fails} on~$I$.
\end{itemize}
In case the trigger fails we denote the transformation
$I\xrightarrow{(\xi,h)} \bot$.
Otherwise, the transformation is denoted 
$I\xrightarrow{(\xi,h)}J$,
or just 
$I\rightarrow_{\xi} J$.
\end{itemize}

A (finite or infinite) sequence $I_0,I_1,I_2\ldots$ of instances 
is said to be a {\em chase sequence with
$\Sigma$ originating from} $I_0$, if
for all $n=0,1,2,\ldots$ 
there is a trigger $(\xi,h)$ 
%that is
%active 
on $I_n$,
such that
$I_n\xrightarrow{(\xi,h)}I_{n+1}$.
If $I_n\xrightarrow{(\xi,h)}\bot$ for some $I_n$
in the sequence we say that the sequence {\em fails}.
Otherwise, the sequence is said to be {\em successful}.
If there is an $I_n$ in a successful sequence, such that
there are no more active triggers for $\Sigma$ on $I_n$,
we say that the sequence {\em terminates}.
Otherwise, the successful sequence does not terminate.
A chase sequence can be 
generated by the following algorithm:

\bigskip 
\bigskip
\bigskip
\bigskip
\bigskip
\begin{codebox}
\Procname{{\sc Algorithm} {\em Standard-}$\chase_{\Sigma}(I)$}
%\Procname{Algorithm $\chase_{\Sigma}(I)$}
\li $I_0 := I$; $i := 0$; 
\li \If exists an active trigger $(\xi,h)$ for $I_i$. 
\li \Then 
\li \If $I_i \xrightarrow{(\xi,h)} \bot$
\li \Then \Return FAIL  
\li \Else $I_i\xrightarrow{(\xi,h')} I_{i+1}$; $i := i+1$ \End
\li \Else \Return $I_i$ \End
\li \Goto 2
\end{codebox}

Note that there can be several chase sequences for a given
$\Sigma$ and $I$, 
as shown by \cite{DBLP:conf/icdt/FaginKMP03}. 
This is reflected by the 
non-deterministic choice of a trigger at line 2.
%We therefore construe the chase process as a 
%{\em computation tree} with root $I_0$ and each chase sequence
%as a branch. As there are only finitely many triggers
%for any finite instance, the chase tree is
%finitely branching.
A successful chase sequence can be infinite, as in the following example.
\begin{example}\label{finiteInfiniteChaseExample}
Consider instance $I=\{R(a,b)\}$ and tgd:
\begin{eqnarray*}
\xi &=& R(x,y) \rightarrow \exists z\;R(y,z). 
\end{eqnarray*}

One possible chase sequence is
\medskip

\begin{minipage}[b]{0.10\linewidth}
\centering
\begin{tabular}{rl} 
\multicolumn{2}{c}{$I_0=I$} \\
\multicolumn{2}{c}{$R$} \\
\hline 
$a$ & $b$  \\
\end{tabular}
\newline
\vspace{0.3cm}
\end{minipage}
\begin{minipage}[b]{0.05\linewidth}
$$\xrightarrow{(\xi,h_{1})}$$
\newline
\newline
\vspace{-0.1cm}
\end{minipage}
\begin{minipage}[b]{0.10\linewidth}
\hspace{-0.05cm}
\centering
\begin{tabular}{rl} 
\multicolumn{2}{c}{$I_1$} \\
\multicolumn{2}{c}{$R$} \\
\hline 
$a$ & $b$ \\
$b$ & $x_1$ 
\end{tabular}
\vspace{-0.132cm}
\end{minipage}
\begin{minipage}[b]{0.10\linewidth}
\hspace{-1.0cm}
$$\xrightarrow{(\xi,h_{2})}$$
\vspace{0.7cm}
\end{minipage}
\begin{minipage}[b]{0.05\linewidth}
\hspace{0.0cm}
$\ldots $
\vspace{1.4cm}
\end{minipage}
\begin{minipage}[b]{0.03\linewidth}
\hspace{-3.5cm}
$$\xrightarrow{(\xi,h_{n})}$$
\vspace{0.7cm}
\end{minipage}
\begin{minipage}[b]{0.10\linewidth}
\hspace{-0.5cm}
\begin{tabular}{ll} 
\multicolumn{2}{c}{$I_n$} \\
\multicolumn{2}{c}{$R$} \\
\hline 
$a$ & $b$ \\
$b$ & $x_1$ \\
$x_1$ & $x_2$ \\
\multicolumn{2}{c}{$\ldots$} \\
$x_{n-1}$ & $x_n$
\end{tabular}
\end{minipage}
\begin{minipage}[b]{0.10\linewidth}
\hspace{0.8cm}
$\xrightarrow{(\xi,h_{n+1})}$
\vspace{1.3cm}
\end{minipage}
\begin{minipage}[b]{0.10\linewidth}
\hspace{0.2cm}
$\ldots $
\vspace{1.4cm}
\end{minipage}
\end{example}

As the next example shows 
there are cases when some chase sequences fail,
while other sequences are successful and do not terminate.

\begin{example}\label{finiteegdInfiniteChaseExample}
Let $\Sigma = \{\xi_1,\xi_2,\xi_3\}$, where
\begin{eqnarray*}
\xi_1 &=&  R(x,y) \rightarrow T(y,x) \\
\xi_2 &=& T(x,y) \rightarrow x=y \\
\xi_3 &=& R(x,y) \rightarrow \exists z\;R(y,z). 
\end{eqnarray*}

\noindent 
Let $I=\{ R(a,b) \}$. One possible chase sequence
is 
$$
\{R(a,b)\}\xrightarrow{(\xi_1,\{ x/a,y/b \})}
\{R(a,b),T(a,b)\}\xrightarrow{(\xi_2, \{ x/a,y/b \})}\bot 
$$
Another possible chase sequence, namely
\begin{eqnarray*}
& &\hspace{-0.5cm}\{R(a,b)\}\xrightarrow{(\xi_3,\{ x/a,y/b \})}
\{R(a,b),T(a,x_1)\} \\ 
& &\;\;\;\;\xrightarrow{(\xi_3, \{ x/a,y/x_1 \})}
\{R(a,b),T(a,x_1),T(x_1,x_2)\xrightarrow{\tau}
\cdots
\end{eqnarray*}

will be infinite.
\end{example}

In order to avoid 
exhaustively choosing the same dependencies in the 
nondeterministic step 2 in the standard chase algorithm,
the following notion is introduced.

\begin{definition}\label{DEF:fairChase}
Let $I=I_0$ be an instance and $\Sigma$ a set of tgds and egds.
An infinite chase sequence $I_0, I_1, I_2\ldots$ is said to be {\em fair},
if for all $i$ and for all active triggers $(\xi,h)$ for
$I_i$, where $\xi\in \Sigma$,  
there exists
$j\geq i$ such that either $I_j \xrightarrow{(\xi,h)} I_{j+1}$,
or the trigger
$(\xi,h)$ is no longer active on $I_j$
\end{definition}

From an algorithmic point of view the choice
of the next trigger to fire
%from all "applicable" triggers at that step,
is essential.
Based on this, the following variations
of the chase process have been 
considered in the literature
(for a comprehensive review of different chase variation see \cite{Onet12}).

\medskip 
\begin{enumerate}
\item
{\em The standard chase} as presented by \cite{DBLP:conf/icdt/FaginKMP03}.
This is algorithm {\em Standard-}$\chase_{\Sigma}(I)$.
In other words,
the next trigger is chosen nondeterministically
from the subset of current triggers that are {\em active}. 
As shown by \cite{DBLP:conf/icdt/FaginKMP03}, 
if all possible chase sequences are successful and terminate,
then all the leaves in the tree of chase sequences are
homomorphically equivalent. 
Thus the nondeterminism is of type ``don't care,''
and we write $\chase_{\Sigma}(I)=J$, where $J$
is a representative of this homomorphism class.
It is also shown by \cite{DBLP:conf/icdt/FaginKMP03},
that if
all sequences are successful, but some sequences
are infinite, the results are nevertheless all homomorphically
equivalent.\footnote{An infinite chase sequence 
$I_0, I_1, I_2, \ldots, I_n, \ldots$ results in the
infinite instance $\bigcup_{n=0}^{\infty} I_n$.}

\medskip
\item
{\em The oblivious chase} as presented by \cite{DBLP:conf/kr/CaliGK08}.
The next trigger is chosen nondeterministically
from the set of all current triggers, active or not,
but each trigger is fired only once in a chase sequence.
The oblivious chase can be obtained from the
algorithm for the standard chase, by changing
line 2 to
\begin{quote}
2\;\; {\bf if} exists a trigger  $(\xi,h)$ for $I_i$, 
and $(\xi,h)$ has not been fired before 
\end{quote}
It is shown by \cite{DBLP:conf/kr/CaliGK08} that for 
$\Sigma$ containing tgds only, if one
oblivious chase sequence with $\Sigma$ on an instance $I$ terminates, 
so do all oblivious
chase sequences with $\Sigma$ on $I$, 
and furthermore, that all the leaves in
the tree of sequences are isomorphic. Fairness, in the
context of the oblivious chase means that all triggers
are eventually fired. This can be achieved e.g.\ by
placing the triggers (as they arise) in a First-In-First-Out queue,
or, as shown by \cite{DBLP:conf/kr/CaliGK08}, by enumerating the triggers
based on an enumeration of the constants, nulls, and dependencies. 
  
We note that the oblivious chase in 
\cite{DBLP:conf/kr/CaliGK08} 
considers Skolemized versions of the tgds,
where each existentially quantified variable
is replaced by a so called Skolem function.
For instance, the Skolemized version of
$\xi = R(x,y) \rightarrow \exists z, w \; S(x,z,w)$ is
$R(x,y) \rightarrow  S(x,f_{\xi,2}(x,y),f_{\xi,3}(x,y))$.
The subscripts in $f_{\xi,2}$ indicates that each
tgd is assigned a unique Skolem function for each
existentially quantified position in the head.
The Skolemization is however not necessary,
as long as the new nulls are generated 
by the distinct extension described in the
chase step for tgds.

\medskip
\item
{\em The semi-oblivious chase} as presented by \cite{DBLP:conf/pods/Marnette09}.
This is a slight variation of the oblivious chase.
Let $\xi$ be a tgd $\alpha(\bar{x},\bar{y})\rightarrow\beta(\bar{x},\bar{z})$.
Then triggers $(\xi,h)$ and $(\xi,g)$ are considered
equivalent if $h(\bar{x})=g(\bar{x})$.
The semi-oblivious chase works as the oblivious one,
except that exactly one trigger
from each equivalence class is fired in a sequence.
For this, we change line 2 of the standard chase algorithm to
\begin{quote}
2\;\; {\bf if} exists a trigger $(\xi,h)$ for $I_i$, and no
trigger $(\xi,g)$, where $g(\bar{x})=h(\bar{x})$\\ 
\hspace*{0.5cm}has been fired before 
\end{quote}
We note that the semi-oblivious chase by 
\cite{DBLP:conf/pods/Marnette09} also considers
the Skolemized versions of the tgds.
The Skolemization is slightly different from
the one used in the oblivious chase, reflecting
the slightly different choice of trigger to fire.
For instance, the semi-oblivious Skolemized version for the 
dependency 
$\xi = R(x,y) \rightarrow \exists z, w \; S(x,z,w)$ is
$R(x,y) \rightarrow  S(x,f_{\xi,1}(x),f_{\xi,2}(x))$.
As in the oblivious chase, the Skolemization is however not necessary
for the semi-oblivious chase,
as long as the new nulls are generated 
by the distinct extension described in the
chase step for tgds.

\medskip
\item
{\em The core chase}  as presented by \cite{DBLP:conf/pods/DeutschNR08}.
At each step, all currently active triggers are fired
in parallel, and then the core of the 
union of the resulting instances is
computed before the next step.
Note that this makes the chase process deterministic
and also fair.
The core chase algorithm is described below.

\end{enumerate}

\begin{codebox}
\Procname{{\em Core-}$Chase_{\Sigma}(I)$}
\li $I_0 := I$; $i := 0$;
\li  \If exists an active egd trigger $(\xi,h)$ for $I_i$ 
\li \Then 
\li \If ${I_i}\xrightarrow{(\xi,h)} \bot$
\li \Then \Return FAIL  
\li \Else $I_i \xrightarrow{(\xi,h)}{I_{i+1}}$; $i := i+1$ \Goto 2 \End \End
\li  \If exists active tgd triggers for $I_i$ 
\li \Then 
\li  \For all $n$ active tgd triggers $(\xi,h)$ for $I_i$ 
\li   \Do compute in parallel ${I_i}\xrightarrow{(\xi,h)}{J_{j}}$ \End
\li ${I_{i+1}} := core(J_1\cup\cdots\cup J_n)$;\;\; $i := i+1$
\li \Else 
\li \Return $I_i$ \End
\li \Goto 2
\end{codebox}

To illustrate the
difference between these chase variations, 
consider dependency set
 $\Sigma=\{\xi\}$, where
$\xi$ is the tgd 
$R(x,y) \rightarrow \exists z\; S(x,z)$
and instance $I_0$ below:

\begin{center}
\begin{tabular}{l} 
 \ \ \ \ $I_0$ \\
\hline 
$R(a,b)$ \\
$R(a,c)$ \\
$S(a,d)$
\end{tabular}
\end{center}

There are two triggers for the set $\Sigma$ on instance $I_0$,
namely $({\xi},{\{ x/a, y/b \}})$ and
$({\xi},{\{x/a, y/c \}})$.
Since $I_0\models\xi$ neither of the
triggers is active, so the standard chase
will terminate at $I_0$.
The core chase will
also terminate at $I_0$.
In contrast, both the oblivious and semi-oblivious chase
will fire the first trigger, resulting in instance
$I_1 = I_0\cup\{S(a,z_1)\}$.
The semi-oblivious chase will terminate at this point,
while the oblivious chase will fire the second trigger,
and then terminate in 
$I_2 = I_1\cup\{S(a,z_2)\}$.

Let $\Sigma$ be a set of tgds and egds,
and $I$ an instance, such that all chase
sequences  
generated by $\Sigma$ on $I$
(use any of the 4 chase variations presented) terminate. 
The all leaves of the tree of chase sequences 
are homomorphically equivalent.
Furthermore, if the chase algorithm is fair,
then all leaves are homomorphically equivalent
even if some sequences are infinite.
\cite{DBLP:conf/icdt/FaginKMP03}
also showed that if the chase fails on one sequence,
it will also fail for all fair sequences.

%-- THE CHASE 
\bigskip 
\section{Complexity of the chase step}\label{SEC:complexityChaseStep}

%\bigskip
%{\bf Chase complexity}.
In this subsection we'll review the complexity of the chase
step. We consider only the
chase steps for tgds. The complexity of a 
\raisebox{-0.2ex}{$\star\,$}-\,chase 
step for egds is the same as the 
complexity of the oblivious-chase tgd step,
where $\star\in\{{\sf std},{\sf obl},{\sf sobl},{\sf core}\}$.

Algorithmically, 
there are two problems to consider.
For knowing when to terminate the chase, 
we need to
determine whether for a given instance $I$ and tgd $\xi$
there exists a homomorphism $h$ such that
$(\xi,h)$ is a trigger on $I$.
This pertains to the
oblivious and semi-oblivious variations.
The second problem pertains to the standard
and core chase: given an instance $I$
and a tgd $\xi$, is there a homomorphism $h$,
such that $(\xi,h)$ is an {\em active} trigger  
on $I$. We call these problems the
{\em trigger existence} problem, 
and the {\em active trigger existence} problem,
respectively.
The {\em data complexity} of these problems
considers $\xi$ fixed, and in the {\em combined complexity}
both $I$ and $\xi$ are part of the input.
%Note that the trigger existence problem is equivalent 
%with testing if the chase step is applicable for a
%given instance with a given tgd.
The following theorem gives the 
combined and data complexities of the two problems.

\begin{theorem}
Let $\xi$ be a tgd and $I$ an instance.
Then
\begin{enumerate}
\item
For a fixed $\xi$,
testing whether there exists a trigger
or an active trigger on a given $I$
is polynomial.
\item 
Testing whether there exists a trigger
for a given $\xi$ on a given $I$
is {\sf NP}-complete.
\item 
Testing whether there exists an active trigger 
for a given $\xi$ and a given $I$
is $\mathsf{\Sigma^p_2}$-complete. 
\end{enumerate}
\end{theorem}

\noindent
{\em Proof}:
The polynomial cases can be verified by 
checking all homomorphisms from the body of the dependency 
into the instance. For the active trigger problem we also need to consider, 
for each such homomorphism, if it has an extension that maps
the head of the dependency into the instance.
These tasks can be carried out in 
%$O(n^{|\alpha|})$ and  $O(n^{|\alpha|+|\beta|})$ time, respectively.
$O(|I|^{|\vars(\alpha)|})$ and  $O(|I|^{|\vars(\alpha\cup\beta)|})$ time, respectively.

It is easy to see that the trigger existence problem
is {\sf NP}-complete in combined complexity,
as the problem is equivalent to testing 
whether there exists a homomorphism between two instances 
(in our case $\alpha$ and $I$); 
a problem shown by \cite{DBLP:conf/stoc/ChandraM77} to be {\sf NP}-complete.

For the combined complexity of the active trigger existence problem,
we observe that it is in $\mathsf{\Sigma^P_2}$,
since one may guess a homomorphism $h$
from $\alpha$ into $I$, 
and then use an {\sf NP} oracle to verify
that there is no extension $h'$ of $h$,
such that $h'(\beta)\subseteq I$.
For the lower bound we
will reduce the following problem 
to the active trigger existence problem.
Let $\phi(\bar{x},\bar{y})$ be a Boolean formula in 3CNF
over the variables in $\bar{x}$ and $\bar{y}$.
Is the formula
$$
\exists \bar{x}\; \neg \big(\exists \bar{y} \, \phi(\bar{x},\bar{y})\big)
$$
true? 
This problem by \cite{Rutenberg:1986:CGG:22416.22478}
is a variation 
of the standard $\exists\forall$-QBF problem
by \cite{DBLP:journals/tcs/Stockmeyer76}.

For the reduction,   
let $\phi$ be given.
We construct an instance $I_{\phi}$
and a tgd 
$\xi_{\phi}$. 
The instance $I_{\phi}$ is as below:

\medskip

\begin{minipage}[b]{0.50\linewidth}
\centering
\hspace{2cm}
\begin{tabular}{lll} 
\multicolumn{3}{c}{$F$}  \\
\hline 
1&0&0  \\ 
0&1&0  \\ 
0&0&1  \\ 
1&1&0  \\ 
1&0&1  \\ 
0&1&1  \\ 
1&1&1  

\end{tabular}
\end{minipage}
\begin{minipage}[b]{0.20\linewidth}
\begin{tabular}{ll} 
\multicolumn{2}{c}{$N$}  \\
\hline 
0&1 \\
1&0 \\
\\
\\
\\
\\
\\
\end{tabular}
\end{minipage}

\bigskip
The tgd $\xi_{\phi} = \alpha\!\rightarrow\!\beta$
is constructed as follows.
For each variable $x \in \bar{x}$ in 
$\phi(\bar{x},\bar{y})$,
the
body $\alpha$ will
contain the atom  $N(x,x')$ 
($x'$ is used to represent~$\neg x$).
The head $\beta$ is existentially quantified
over that set 
$\bigcup_{y\in{\bar{y}}}\{ y,y'\}$
of variables. 
For each conjunct $C$ of $\phi$, 
we place in $\beta$ an atom
$F(x,y,z)$, 
where $x,y$ and $z$ are the variables in $C$, 
with the convention that if variable $x$ is
negated in $C$,
then $x'$ is used in the atom.
Finally for each $y \in \bar{y}$, 
we place in $\beta$ the atom $N(y,y')$,
denoting that $y$ and $y'$ 
should not have the same truth assignment.

Suppose now that the formula
$\exists \bar{x}\; \neg \big(\exists \bar{y} \, \phi(\bar{x},\bar{y})\big)$
is true. This means that there is a $\{0,1\}$-valuation $h$
of $\bar{x}$ such that for any $\{0,1\}$-valuation $h'$ of $\bar{y}$,
the formula
$\phi(h(\bar{x}),h'(\bar{y}))$ is false.
It is easy to see that $h(\alpha) \subseteq I$.
Also, since $\phi(h(\bar{x}),h'(\bar{y}))$ is false
for any valuation $h'$, for each $h'$
there must be an atom $F(x,y,z)\in\beta$, such that  
$h'\circ h(F(x,y,z))$
is false,
that is, 
either $h'\circ h(F(x,y,z))= F(0,0,0)\notin I_{\phi}$,
or $h'$ assigns for some existentially quantified 
variables non-Boolean values. 
Consequently
the trigger $(\xi,h)$ is active on $I_{\phi}$.

For the other direction, 
suppose that there
exists a trigger $(\xi,h)$ which is 
active on $I_{\phi}$,
i.e.,  $h(\alpha) \subseteq I_{\phi}$
and $h'(\beta) \not \subseteq I$,  
for any extension $h'$ of $h$.
This means that 
for any such extension $h'$, either
$h'$ is not $\{0,1\}$-valuation,
or that the atom $F(0,0,0)$ is in $h'(\beta)$.
Thus the formula
$\exists \bar{x}\; \neg \big(\exists \bar{y} \, \phi(\bar{x},\bar{y})\big)$
is true.
$_{\blacksquare}$

\bigskip
\medskip
Note that the trigger existence relates to the oblivious and semi-oblivious 
chase variations,
whereas the active trigger existence relates to the standard and the core chase.
This means that the oblivious and semi-oblivious chase variations 
have the same complexity. 
This is not the case for the standard and the core chase.
because the core chase step applies all active triggers in parallel and 
also involves the core computation for the resulted instance. 
\cite{DBLP:conf/pods/FaginKP03} 
have shown that computing the core involves a
{\sf DP}-complete decision problem.

%\medskip
%{\bf Termination classes}.
\bigskip
\bigskip
\section{Chase termination questions}

From the previous section we know that we may determine if 
the chase can continue at any given step by checking the 
existence of trigger/active trigger. On the other hand,
testing if the chase process terminates, even when considering only tgds, 
as we will see, is undecidable. 
In the following subsection we show that for any set $\Sigma$ of tgds and egds
there is a set $\Sigma^{tgd}$ of tgds only, 
such that for any instance $I$ and
$\star\in\{{\sf std},{\sf obl},{\sf sobl},{\sf core}\}$, 
if the $\star$-chase for $I$ and $\Sigma^{tgd}$ terminates, 
then so does the $\star$-chase for $I$ and $\Sigma$.

The different chase variations have
different termination behaviors,
so in the second subsection we introduce some notions 
that will help to distinguish them.

\bigskip
\subsection{A sufficient condition for egds and tgds}

In this subsection we  present a rewriting technique that 
transforms a set of tgds and egds into a new set of tgds only, 
such that the chase termination on the new set will
guarantee the chase termination on the initial set.

Let $E$ be a new binary relational symbol 
outside the schema $\mathbf{R}$.
Intuitively $E(x,y)$ will mean that 
elements $x$ and $y$ are equal. 
Given a set $\Sigma$ of egds and tgds, 
by $\Sigma^{tgd}$
we denote the set of tgds constructed as follows, 

\begin{enumerate}
\item 
If $\xi \in \Sigma$ and $\xi$ is a tgd, then add $\xi$ to $\Sigma^{tgd}$.
\item 
For each egd  $\alpha \rightarrow x=y$ from $\Sigma$, add 
the tgd $\alpha \rightarrow E(x,y),E(y,x)$ to $\Sigma^{tgd}$.
\item 
For each predicate symbol $R\in\mathbf{R}$ used in $\Sigma$ 
and for each integer
$i$ such that $1 \leq i \leq arity(R)$, 
add the following tgd to $\Sigma^{tgd}$: 
\begin{eqnarray*}\label{egd2tgd}
& &\hspace{-1.0cm}E(x,y),R(x_1, x_2, \ldots,, x_{i-1},x, x_{i+1},\ldots, x_{arity(R)}) \\ 
& &\hspace{0.6cm} \rightarrow R(x_1, x_2, \ldots,, x_{i-1},y, x_{i+1},\ldots, x_{arity(R)})
\end{eqnarray*}
\end{enumerate}
Note that for any $\Sigma$ the transformation into $\Sigma^{tgd}$
is polynomial in the size of $\Sigma$.

\begin{example}
As an example of the transformation,
consider $\Sigma$ containing 
the following two dependencies:

\begin{eqnarray*}
R(x,x) &\rightarrow& \exists y,z\; S(x),R(y,z) \\
R(x,y) &\rightarrow& x=y
\end{eqnarray*}

In this case $\Sigma^{tgd}$ will contain the following set of tgds:
\begin{eqnarray*}
R(x,x) &\rightarrow& \exists y,z\; S(x),R(y,z) \\
R(x,y) &\rightarrow& E(x,y),E(y,x) \\
E(x,y),R(z,y) &\rightarrow& R(z,x) \\
E(x,y),R(y,z) &\rightarrow& R(x,z) \\
E(x,y),S(x) &\rightarrow& S(y)
\end{eqnarray*}
\end{example}

The following theorem shows that the chase termination on the
rewritten set $\Sigma^{tgd}$ of tgds ensures chase termination for the initial
set $\Sigma$ of both tgds and egds. 

\begin{theorem}
Let $\Sigma$ be a set of tgds and egds, let $I$ be an instance,
and let
$\star\in\{{\sf std},{\sf obl},{\sf sobl},{\sf core}\}$.
If the 
$\star$-chase
on $I$ with $\Sigma^{tgd}$ terminates,
then so does the
$\star$-chase  
on $I$ with $\Sigma$. 
\end{theorem}

\noindent
{\em Proof}:  
We will prove the theorem only for the standard-chase case, for the other
variations the proof is similar.
Let $I$ be an instance, $\Sigma$ a set of tgds and egds,
and $I_0=I, I_1, I_2, \ldots$ 
a fair successful standard chase sequence on $I$ with $\Sigma$.
We will show that there 
exists a fair standard chase sequence 
$J_0=I, J_1, J_2, \ldots$
on $I$ with $\Sigma^{tgd}$, such that
for each $I_k$ in the first sequence there exists
a $J_{\ell}$ in the second sequence with $I_k\subseteq J_{\ell}$.
We will construct the $J$-sequence inductively on the length of the
$I$-sequence.

For the base step of the induction we have $J_0=I=I_0$.
For the inductive step, suppose that 
there exists a sequence $J_0, \ldots, J_{\ell}$, 
and for all $I_i$, where $i\leq k$, in the $I$-sequence,
there exists an instance $J_j$ in the $J$-sequence, 
such that $I_i \subseteq J_j$.
We need to show that for $k+1$ there exists a positive integer $\ell'$
such that $J_0=I, J_1, J_2, \ldots, J_{\ell}, \ldots, J_{\ell'}$
is (a prefix of) a fair chase sequence
with $\Sigma^{tgd}$, and that $I_{k+1}\subseteq J_{\ell'}$. 

From the definition of a chase sequence
we have that 
$I_k\xrightarrow{(\xi,h)}I_{k+1}$,
where $h(body(\xi)) \subseteq I_{k}$,
and $(\xi,h)$ is a trigger active on $I_k$.
If $\xi$ is a tgd, then $\xi \in \Sigma^{tgd}$,
and since $h(body(\xi)) \subseteq I_{k} \subseteq J_{\ell}$, 
it follows that either
$I_{k+1} \subseteq J_{\ell}$, 
or the trigger $(\xi,h)$ is active on $J_{\ell}$.
Thus $J_{\ell}\xrightarrow{(\xi,h)}J_{\ell+1}$,
and $I_{k+1} \subseteq J_{\ell+1}$.
In other words, $\ell' = \ell+1$.

\smallskip

Suppose then that $\xi$ is an egd 
$\alpha \rightarrow x=y$.
This means that $\Sigma^{tgd}$ will contain the tgd 
$\alpha \rightarrow E(x,y), E(y,x)$, 
and for each relational symbol $R$ and $i$, where $1\leq i \leq arity(R)$,
the set $\Sigma^{tgd}$ will contain a tgd of the form~\eqref{egd2tgd}.
Now
$I_k\xrightarrow{(\alpha\!\rightarrow\!x=y,\,h)}I_{k+1}$,
and all occurrences of $h(y)$ in $I_k$ have been
replaced by $h(x)$ in $I_{k+1}$ 
(assuming $h(x)<h(y)$ in the enumeration of variables).
Clearly $(\alpha\rightarrow{E}(x,y),E(y,x),\, h)$,
is a trigger for $\Sigma^{tgd}$ on $J_{\ell}$. 
Thus $J_{\ell}\xrightarrow{(\alpha\rightarrow{E}(x,y),E(y,x),\, h)}J_{\ell+1}$, 
where the instance $J_{\ell+1} = J_{\ell}\cup\{E(h(x),h(y)),E(h(y),h(x))\}$. 
For each tuple $R(\ldots,h(y),\ldots)$ in $I_{k}\subseteq J_{\ell}$
there will be a tgd
$\xi = E(x,y),R(\ldots,y,\ldots)\rightarrow R(\ldots,x,\ldots)$
in $\Sigma^{tgd}$. 
We then take the chase step
$J_{\ell+1}\xrightarrow{(\xi,h)}J_{\ell+2}$
which adds tuple $R(\ldots,h(x),\ldots)$ to $J_{\ell+2}$
If there are $m$ occurrences of $h(y)$ in $J_{\ell}$,
we repeat similar chase steps $m-1$ times, and arrive at instance
$J_{\ell+m}$ which clearly satisfies $I_{k+1}\subseteq J_{\ell+m}$.
This concludes the inductive proof and the claim of
the theorem.
$_{\blacksquare}$

\bigskip
With this theorem we have showed that it suffices to check if 
the standard chase with $\Sigma^{tgd}$ on $I$ terminates,
in order to infer that the standard 
chase with $\Sigma$ (containing both tgd's and egds) on  $I$
will terminate.
Note also that in case $\Sigma$ contains only egds and full tgds, 
then $\Sigma^{tgd}$ will contain only full tgds ensuring that 
the chase terminates.
The theorem can easily be extended to all instances as follows:

\begin{corollary}
Let $\Sigma$ be a set if tgds and egds. 
If all $\star$-chase sequences with $\Sigma^{tgd}$ 
terminates on all instances, 
then they also terminate for $\Sigma$ on $I$,
where $\star\in\{{\sf std},{\sf obl},{\sf sobl},{\sf core}\}$.
\end{corollary}

The previous results gives us a sufficient way to test termination
for sets of egds and tgds. Thus, one may use to test if
$\Sigma^{tgd}$ belongs to one of the known termination classes,
as presented in the following subsection, 
in order to infer that $\Sigma$ terminates. Thus,
from now one will consider only classes of tgds.

%Note that only testing the termination for the set of tgds from $\Sigma$
%does not suffice, as it is shown by Example \ref{egdConverse}.

\bigskip
\subsection{Termination classes}\label{Sec:terminationClasses}

Let $\star\in\{{\sf std},{\sf obl},{\sf sobl},{\sf core}\}$,
corresponding to the chase variations introduced in 
Section~\ref{prelim}, 
and let $\Sigma$ be a set of tgds.
If there exists a terminating $\star$-chase sequence 
%$I\!=\!I_0, I_1,I_2,\ldots$ 
with $\Sigma$ on $I$, 
we say that the
$\star$-chase terminates for {\em some} branch
on instance~$I$,
and denote this as $\Sigma\in \actie{I}$.
Here $\actie{I}$ is thus to be understood
as the class
of all sets of tgds for which the $\star$-chase terminates
on some branch on instance $I$.
Likewise, $\actia{I}$ denotes
the class of all sets of tgds for which the $\star$-chase 
with $\Sigma$ on $I$ terminates on {\em all} branches.
From the definition of the chase variations
it is easy to observe that any trigger
applicable by the standard chase step on an instance $I$
is also applicable by the semi-oblivious and oblivious 
chase steps on the same instance. 
Similarly, 
all the triggers applicable by the
semi-oblivious chase step on an instance $I$
are also applicable by the oblivious chase step on
instance $I$. 
%Note that the converse is not always true.
Thus
$ \octie{I} \; \subseteq \;\soctie{I}\; \subseteq \;\sctie{I}.$
It is also easy to verify that
$ \sctia{I} \; \subseteq \; \sctie{I}$, 
and that
$\soctia{I} \; \subseteq \; \sctia{I}$.
We also observe that for the  oblivious and 
semi-oblivious chase 
each sequence will fire
(eventually, in case of infinite sequences)
the same set of triggers.
From this, and the fairness property, 
it directly follows that
$ 
\actie{I}  
= 
\actia{I}
$
for $\star\in\{{\sf obl},{\sf sobl}\}$.

The following proposition shows that these 
results can be strengthened to strict inclusions:

\begin{proposition} 
For any instance $I$ we have:
\begin{eqnarray*}
& & \octie{I}
= 
\octia{I}
\; \subset \;
\soctia{I}
= 
\soctie{I}
\; \subset \;
\sctia{I}
\; \subset \; 
\sctie{I}.
\end{eqnarray*}
\end{proposition}

\medskip

\noindent
{\em Proof:} (Sketch)
For the strict inclusion
$\octia{I} \; \subset \; \soctia{I}$
consider
$$
I=\{R(a,b)\} \mbox{ and } \Sigma = \{R(x,y) \rightarrow \exists z\; R(x,z)\}.
$$

%instance $I=\{ R(a) \}$
%and
%$\Sigma$ 
%containing a single (tautological) dependency 
%$ \xi = R(x) \rightarrow \exists z\; R(z)$.
Then the oblivious chase sequence will look like

{\small
\begin{eqnarray*}
& &\hspace{-0.5cm}\{R(a,b)\}\xrightarrow{(\xi,\{x/a,y/b\})}
\{R(a,b),R(a,z_1)\} \\ 
& &\xrightarrow{(\xi,\{x/a,y/z_1\})}
\{R(a,b),R(a,z_1),R(a,z_2)\}\xrightarrow{(\xi,\{x/a,y/z_2\})}
\ldots
\end{eqnarray*}
}\\
which will converge only at the infinite instance
$$
\bigcup_{n\geq 1}\{R(z_n)\}\cup\{R(a)\}.
$$
The semi-oblivious chase, on the other hand, 
will terminate at the instance
$\{R(a,b),R(a,z_1)\}$.
We conclude that $\Sigma \in \soctia{I}$
but $\Sigma \notin \octie{I}$.

For the second strict inclusion
$\soctie{I} \; \subset \; \sctia{I}$,
consider 
$$
I=\{S(a,a)\} \mbox{ and } \Sigma=\{ S(x,y) \rightarrow \exists z\; S(y,z)\}.
$$
%instance $I=\{ S(a,a) \}$
%and $\Sigma=\{ S(x,y) \rightarrow \exists z\; S(y,z) \}$.
Because $I \models \Sigma$
it follows that  $\Sigma \in \sctia{I}$.
On the other hand, 
the semi-oblivious 
chase will converge only at the infinite instance
$$
\bigcup_{n\geq 1}\{S(z_n,z_{n+1})\} \cup \{S(a,a),S(a,z_1)\},
$$
and thus $\Sigma \notin \soctie{I}$.

For the final strict inclusion
$\sctia{I}
\; \subset \; 
\sctie{I}$, 
let instance 
$
I=\{S(a,b),R(a)\}$  and set of dependencies 
$\Sigma=\{ S(x,y)\rightarrow \exists z\; S(y,z); \;\; R(x) \rightarrow S(x,x)\}.
$
%$I=\{ S(a,b), R(a) \}$
%and 
%$\Sigma=\{ S(x,y)\rightarrow \exists z\; S(y,z); \;\; R(x) \rightarrow S(x,x) \%}$. 
It is easy to see that any standard chase sequence that
starts by firing the trigger based on the first tgd will not terminate
as it will generate new tuple $S(b,z_1)$ and this will fire an infinite 
chase sequence on any branch.
On the other hand, if we first fire the trigger based on the second tgd
the standard chase will terminate after one step.
$_{\blacksquare}$

\bigskip
The next questions are whether all or some $\star$-chase
sequences terminate on {\em all} instances.
The corresponding classes of sets of tgds are
denoted $\actaa$ and
$\actae$, respectively.
Obviously
$ \sctaa \; \subset \; \sctae $.
Similarly to the instance dependent 
termination classes, 
$ \octaa \; \subset \; \soctaa $
and
$ \actaa \; = \; \actae $,
for $\star\in\{{\sf obl},{\sf sobl}\}$.
We can relate the oblivious, semi-oblivious and standard 
chase termination classes as follows:

\begin{theorem}\label{PROP:semiObliviousChaseTerm}
$
\octaa
\; = \;
\octae 
\; \subset \; 
\soctaa=\soctae 
\; \subset \; 
\sctaa 
\; \subset \; 
\sctae.
$

%\begin{eqnarray*}
%\octaa
%\; = \;
%\octae 
%\; \subset \; 
%\soctaa=\soctae 
%\; \subset \; 
%\sctaa 
%\; \subset \; 
%\sctae.
%\end{eqnarray*}
\end{theorem}

\medskip
\noindent
{\em Proof}:
We will only show the strict inclusion parts of the theorem.
For the first inclusion, let $\mathbf{R}=\{R\}$, and
$\Sigma=\{\xi\}$, where  
$$
\xi =\{ R(x,y) \rightarrow \exists z\; R(x,z) \}.
$$
Let $I$ be an arbitrary non-empty instance.
Then $I$ contains at least one tuple, say, $R(a,b)$,
and we can write $I$ as 
$$
\{R(a,b)\}\cup J,
$$
for some (possible empty) instance $J$. 
The oblivious chase will generate the sequence

{\small
$$
\{R(a,b)\}\cup{J}\xrightarrow{(\xi,\{x/a,y/b\})}
\{R(a,b),R(a,z_1)\}\cup{J}\xrightarrow{(\xi,\{x/a,y/z_1\})}
%\{R(a,b),R(a,z_1),R(a,z_2)\}\cup{J}\xrightarrow{(\xi,\{x/a,y/z_2\})}
\ldots
$$
}

and thus converge in (possibly a superset of) 
the infinite instance

$$
\bigcup_{n\geq 1}\{R(z_n,z_{n+1}\} \cup \{R(a,b),R(a,z_1)\}\cup{J}.
$$

The semi-oblivious chase 
will consider the triggers
$(\xi,\{x/a,y/b\})$,
$(\xi,\{x/a,y/z_1\})$,
$(\xi,\{x/a,y/z_2\}),\ldots\;$
that are all equivalent, since they all map the
only common variable $x$, from the body and the head of the tgd,
 to $a$.
Therefore the semi-oblivious chase will
terminate in instance
$\{R(a,b),R(a,z_1)\}\cup{J}.$
We conclude that
$\Sigma\in \soctaa\setminus\octae$.

The second strict inclusion 
$\soctaa \; \subset \; \sctaa$ 
is more intricate, 
as most sets of  dependencies in $\sctaa$ are also in $\soctaa$.
To distinguish the two classes,
let $\Sigma = \{\xi_1,\xi_2\}$, where

\vspace{-0.1cm}
\begin{eqnarray*}\label{EQ:oblivNonStandardTermination}
\xi_1 & = & {R}(x) \rightarrow \exists z\; {S}(z),{T}(z,x),\mbox{ and}\\
\xi_2 & = & {S}(x) \rightarrow \exists z'\; {R}(z'),{T}(x,z').
\end{eqnarray*}

\vspace{0.1cm}
\noindent
Let $I$ be an arbitrary non-empty instance,
and suppose that 
$$
I = \{R(a_1),\ldots,R(a_n),S(b_1),\ldots,S(b_m)\}.
$$
There is no loss of generality,
since if the standard chase with $\Sigma$ on $I$ terminates, 
then it will also terminate even if the initial instance
contains atoms over the relational symbol $T$.
It is easy to see that all standard chase sequences 
with $\Sigma$ on $I$
will terminate in the instance 
$A\cup B$, where

\vspace{-0.1cm}
\begin{eqnarray*}
A & = & \bigcup_{i=1}^n \{R(a_i),S(z_i),T(z_i,a_i)\},\mbox{ and}   \\
B & = & \bigcup_{i=1}^m \{R(z'_i),S(b_i),T(b_i,z'_i)\}.
\end{eqnarray*}

\vspace{0.1cm}
On the other hand, all semi-oblivious chase
sequences will converge only at the infinite instance
$$
\bigcup_{k\geq 1} \big(C_k \cup D_k\big) \cup A \cup B,$$
where

\vspace{-0.1cm}
{\small{
\begin{equation*}
\begin{split}
C_k =&\bigcup_{i=1}^{m} \{ S(z_{kn+(k-1)m+i}), 
T(z_{kn+(k-1)m+i},z'_{(k-1)m+(k-1)n+i}) \} \cup                 \\
   &   \bigcup_{j=1}^{n} \{S(z_{kn+km+j}), T(z_{kn+km+j},z'_{km+(k-1)n+j}) \},                             \\
D_k=& \bigcup_{i=1}^{n} \{ R(z'_{km+(k-1)n+i}),T(z_{(k-1)n+(k-1)m+i},z'_{km+(k-1)n+i}) \} \cup                 \\
   &\bigcup_{j=1}^{m} \{R(z'_{km+kn+j}),T(z_{kn+(k-1)m+j},z'_{km+kn+j}) \}.
\end{split}
\end{equation*}
}}

For the last inclusion $\sctaa \subset \sctae$,  consider 
$\Sigma=\{ R(x,y)\rightarrow R(y,y);\;\; R(x,y)\rightarrow \exists z\; R(y,z) \}$. 
Clearly $\Sigma \in \sctae$, 
because for any instance, 
all chase sequences that start by firing the first tgd will terminate.
On the other hand  $\Sigma \notin \sctaa$, 
as the standard chase with $\Sigma$
does not terminate on
$I=\{ R(a,b) \}$
whenever
the second tgd is applied first.
$_{\blacksquare}$

\bigskip

%\vspace{0.1cm}
Note that for any 
$\star\in\{{\sf std},{\sf obl},{\sf sobl},{\sf core}\}$,
and for any non-empty instance $I$, we have that 
$\actaa \subset \actia{I}$
and 
$\actae \subset \actie{I}$.

\bigskip
We now turn our attention to the core chase.
Note the core chase is deterministic
since all active triggers are fired
in parallel, 
before taking the
core of the result. Thus we have: 

\begin{proposition}
$\cctia{I} = \cctie{I}$
and 
$\cctaa = \cctae$.
\end{proposition}

\medskip
It is well known that all here considered chase
variations
compute a finite {\em universal model}
of $I$ and $\Sigma$
when they terminate,
as shown by \cite{DBLP:conf/icdt/FaginKMP03,DBLP:conf/pods/DeutschNR08,DBLP:conf/kr/CaliGK08,DBLP:conf/pods/Marnette09}.
In particular,
\cite{DBLP:conf/pods/DeutschNR08}, 
showed that if $I\cup\Sigma$ has a finite universal model,
the core chase will terminate in an instance that
is the core of all universal models (which are homomorphically equivalent). 
The standard chase, on the other hand,
might have to choose the correct sequence
in order to terminate. The example above, 
with
$I=\{S(a,b),R(a)\}$ and 
$\Sigma = \{S(x,y)\rightarrow\exists z S(y,z);\;
R(x)\rightarrow S(x,x)\}$,
serves as an illustration of the importance
of ``the correct sequence.''
We have:

\begin{proposition}\label{THEO:coreChaseTerminVSStandTermin}
\begin{enumerate}
\item
$\sctie{I} \, \subset \, \cctia{I}$,
for any $I$.
\item
$\sctae \; \subset \; \cctaa$.
\end{enumerate}
\end{proposition}

\medskip

\noindent
{\em Proof:} (Sketch)
To see that the inclusion in part $(2)$ of the proposition
is strict,
let 
$$
{I}=\{{R}(a)\} \mbox{ and }  
\Sigma = \{ {R}(x) \rightarrow \exists z\; {R}(z),{S}(x)\}.
$$
%$\Sigma = \{ {R}(x) \rightarrow \exists z\; {R}(z),{S}(x)\}$, 
%and 
%${I_0}=\{ {R}(a) \}$. 
In this setting 
there will be exactly one 
active trigger at each step, 
and the algorithm will converge 
only at the infinite instance 
$$
\bigcup_{i\geq 2} \{ R(z_i), S(z_{i-1}) \} \cup \{ R(a),S(a), R(z_1) \}.
$$
From this, it follows that $\Sigma \notin \sctaa$ and $\Sigma \notin \sctae$.
Note that for any $i>0$, the core of ${I}_i$
is $\{ {R}(a),{S}(a) \}$. 
Thus the core chase will 
terminate at instance 
$I_1 = \{ {R}(a),{S}(a) \}$.
$_{\blacksquare}$

\bigskip
The following Corollary  highlights the 
relationship between the termination classes.

\begin{corollary}\label{COR:concl}
$\octaa \subset \soctaa \subset \sctaa \subset \sctae \subset \cctaa\mbox{.}$
\end{corollary}

\bigskip
\begin{figure}[!htbp]
  \begin{center} 
\includegraphics[scale=0.22]{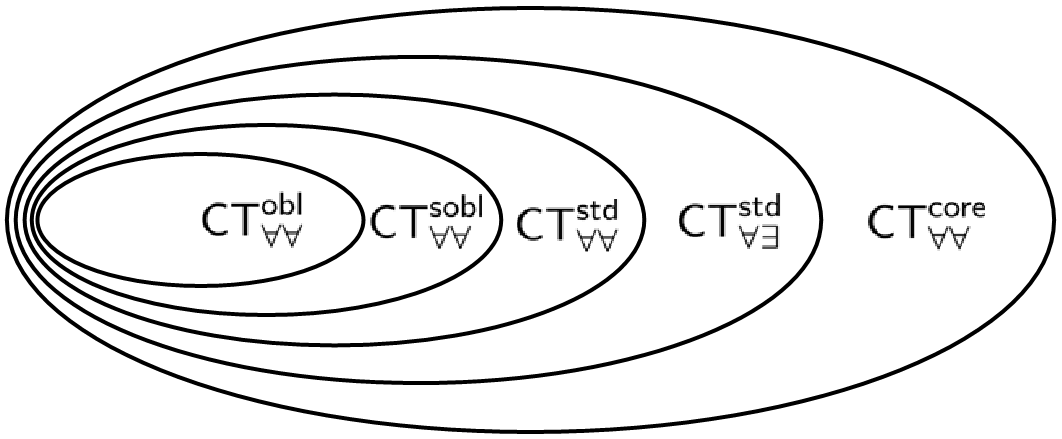}
 \caption{Termination classes for chase variations.}
  \end{center}
\end{figure}

%-- UNDECIDABILITY 
\bigskip 
\section{Undecidability of termination}

\bigskip
\setcounter{theorem}{4}
\begin{theorem}\label{A:hitherto}
\begin{enumerate}
\item
$\sctia{I}$ and 
$\sctie{I}$
are {\sf RE}-complete
{\em \cite{DBLP:conf/pods/DeutschNR08}}.
\vspace*{2ex}
\item
$\cctia{I}=\cctie{I}$,
and both sets are {\sf RE}-complete
{\em \cite{DBLP:conf/pods/DeutschNR08}}. 
\vspace*{2ex}
\item
$\soctia{I}=\soctie{I}$,
and both sets are {\sf RE}-complete
{\em \cite{DBLP:conf/pods/Marnette09}}.
\vspace*{2ex}
\item
$\soctaa=\soctae$,
and both sets are {\sf RE}-complete
{\em \cite{Gogaz2014}}.
\vspace*{2ex}
\item
$\octaa=\octae$,
and both sets are {\sf RE}-complete by {\em \cite{Gogaz2014}}.
\vspace*{2ex}
\item
Let $\Sigma$ be a set of guarded\footnote{A tgd 
is guarded if its body contains an atom called guard that covers
all variables occurring in the body.}
tgds as introduced by
{\em \cite{DBLP:conf/kr/CaliGK08}}.
Then the question $\cctia{I}$
is decidable 
{\em \cite{DBLP:conf/icdt/Hernich12}}.
\end{enumerate}
\end{theorem}

\bigskip

The proof in \cite{DBLP:conf/pods/DeutschNR08} 
of part 1 of Theorem \ref{A:hitherto} encodes
a Turing machine $M$ with
an initial blank tape as a set $\Sigma_M$ of tgds,
such that a chase sequence using $\Sigma_M$
mimics the computation history of $M$.
Then 
\cite{DBLP:conf/pods/DeutschNR08} 
show 
that $M$ halts if and only if
the standard chase with $\Sigma_M$ on the empty instance
terminates. The dependencies in $\Sigma_M$ are constructed
so that some chase sequence terminates 
if and only if all chase sequences terminate.
\cite{DBLP:conf/pods/DeutschNR08} also show that if the standard
chase terminates, then so does the core chase, thus
yielding them part 2 of Theorem \ref{A:hitherto}.

However, the encoding used by \cite{DBLP:conf/pods/DeutschNR08},
using the empty tape as input to $M$, 
does not allow for input instances to the chase
other than the empty one. Thus their proof cannot be used
to determine whether the chase terminates on 
{\em all} input instances, i.e.\ to determine
the complexity of sets 
$\actaa$ and~$\actae$, 
for $\star\in\{{\sf std}, {\sf core}\}$.

The proof of part 3 of Theorem \ref{A:hitherto} by
\cite{DBLP:conf/pods/Marnette09} 
first shows with a proof similar to
\cite{DBLP:conf/pods/DeutschNR08} that the termination of the semi-oblivious
chase on the empty instance is an {\sf RE}-complete problem.
\cite{DBLP:conf/pods/Marnette09} then shows that if the semi-oblivious chase with
a set $\Sigma$ of tgds 
terminates on a special ``critical instance,''
it terminates
on all instances, thus obtaining the {\sf RE}-completeness
result for $\soctaa$.

\cite{Gogaz2014} use
a reduction from the halting problem for
three-counter automata for showing that
$\octae$ is 
{\sf RE}-complete. They then apply the 
``critical instance'' technique to obtain
the {\sf RE}-completeness result for
$\octaa$.

\medskip

This leaves the complexity of the sets
$\actaa$ and~$\actae$ open,
for $\star\in\{{\sf std}, {\sf core}\}$.
We show that
%, perhaps surprisingly,
all of these sets
are {\sf coRE}-complete. 
Our proofs use reductions from
word rewriting systems,
which we feel are symbolically closer to the chase
than computation histories of Turing machines.
Rather interestingly, as a corollary
of our results, it follows that the
``critical instance'' technique is not
applicable to the core and standard chases.

Parts 1 and 2 of Theorem \ref{A:hitherto}
are also directly obtainable from our
reduction. We note that our reduction is
technically more involved than the one used
by \cite{DBLP:conf/pods/DeutschNR08},
where the  empty database instance was used in the reduction. 
This is because  here we also need to show termination on all
input instances, 
not just on a given input instance.  

%As mentioned, 
%our reductions will be from word rewriting systems,
%which we describe next.

\medskip
\noindent
{\bf Word rewriting systems}.
Let $\Delta$ be a finite set of symbols,
denoted $a,b,\ldots$, possibly
subscripted.
A {\em word} over $\Delta$ 
is a finite sequence $a_1a_2\ldots a_n$,
where each $a_i$ is a symbol from~$\Delta$.
The {\em empty} word ($n=0$) 
is denoted~$\epsilon$.
The {\em length} of a word 
$w=a_1a_2\ldots a_n$,
denoted $|w|$, is $n$.
The {\em concatenation} of two words
$u=a_1a_2\ldots a_n$ and $v=b_1b_2\ldots b_m$,
is the word $a_1a_2\ldots a_nb_1b_2\ldots b_m$,
denoted $uv$. We have $|uv|=n+m$,
and $\epsilon{u} = u\epsilon = u$,
for all words $u$. 
We say that a word $u$ is a {\em factor} of a word $v$,
if $v=xuy$, for some words $x$ and $y$.
If $x=\epsilon$ we say that $u$ is a {\em suffix}
of $v$, and if $y=\epsilon$ 
we say that $u$ is a {\em prefix} of $v$. 
Note that both $x$ and $y$ might 
be the empty word $\epsilon$,
so the factor relation is reflexive.
We let $\Delta^*$ denote the set of all
finite words over~$\Delta$.
%and from $\Delta^*$ will be denoted
%$u,v,w, \ldots$, possibly subscripted.

A pair $\mathcal{R}=(\Delta^*,\Theta)$,
where $\Theta$ is a finite subset of
$\Delta^*\!\times\Delta^*$,
is called a {\em word rewriting system}.
Treating each pair in $\Theta$ as a {\em rule},
the relation $\Theta$ gives rise to
a {\em rewriting relation}
$\Rarrow
\;\; \subseteq \;\; \Delta^*\!\times\Delta^*$
defined~as
$$
\{(u,v)
%\in\Delta^*\times\Delta^* 
\; : \; u = x\ell y, v = xry,\;
(\ell,r)\in\Theta,\;x,y\in\Delta^*\}. 
$$
We use the notation $u\Rarrow v$ 
instead of $\Rarrow\!(u,v)$.
If $\mathcal{R}$ is understood from the context we will simply
write $u\rightarrow v$.
If we want to emphasize
which rule $\rho\in\Theta$ was used we write
$u\rightarrow_{\rho} v$.
When $u\rightarrow_{\rho} v$ we say that
$v$ is obtained from $u$ by  a {\em rewriting step
(based on $\rho$)}.  
%By $u \rightarrow_n v$ we mean that
%$v$ can be obtained from $u$ in 
%at most $n$ rewriting steps. 

A sequence $w_0,w_1,w_2,\ldots$ 
of words from
$\Delta^*$ is said to be a 
$\mathcal{R}$-{\em derivation sequence}
(or simply a derivation sequence),
if $w_i\Rarrow w_{i+1}$ for 
all $i=0,1,2,\ldots$.
A derivation sequence might be finite or infinite. 
A derivation sequence $w_0,w_1,w_2,\ldots$ 
will sometimes also be written
$w_0\Rarrow
w_1\Rarrow
w_2\Rarrow
\cdots$, 
or simply
$w_0\rightarrow
w_1\rightarrow 
w_2\rightarrow \cdots$, 
when $\mathcal{R}$ is understood from the context.
A word $w\in\Delta^*$ is said to be in {\em normal form}
if there is no word $v\in\Delta^*$, $v\neq w$,
such that $w\Rarrow v$.  Note that normal forms are not
necessarily unique.

\begin{definition}\label{def:revterm}
Let $\mathcal{R}=(\Delta^*,\Theta)$
be a word rewriting systems.
\begin{enumerate}
\item
The {\em termination problem 
for $\mathcal{R}$ and a word} 
$w_0\in\Delta^*$,
is to determine whether
all derivation sequences
$w_0\Rarrow w_1\Rarrow w_2\Rarrow\cdots$ 
originating from $w_0$
are finite. 
\item
The {\em uniform termination problem for}
$\mathcal{R}$ is to determine
whether for {\em all} words $w_0\in\Delta^*$,
it holds that all derivation sequences
$w_0\Rarrow w_1\Rarrow w_2\Rarrow\cdots$ 
originating from $w_0$
are finite. 
\item
The {\em uniform normal form existence problem for}
$\mathcal{R}$ is to determine whether all
words in $\Delta^*$ have a (not necessarily unique)
normal form, i.e. if all words have some terminating
derivation sequence.
\end{enumerate}
\end{definition}
It is known that 
the termination problem is {\sf RE}-complete,
that the uniform termination problem
is {\sf coRE}-complete, and that the
normal form existence problem is complete
for level $\mathsf{\Pi^0_2}$ in the  
arithmetic hierarchy \cite{DBLP:books/daglib/0069623}.
Interestingly, none of our chase decision problems
go beyond $\mathsf{\Pi^0_1}$, 
that is, {\sf coRE}.

%It has long been known that
%the termination problem is {\sf RE}-complete
%\cite{david58},
%and that the uniform termination problem
%is {\sf coRE}-complete \cite{DBLP:books/daglib/0069623}.
%Since a word $w_0$ has a normal form if and only if
%there is (at least) one finite derivation sequence
%starting from $w_0$, it is easy to see that Rice's Theorem
%\cite{DBLP:journals/Rice53} gives us the {\sf RE}-completeness of
%the normalization problem for $\mathcal{R}$ and $w_0$
%from the {\sf RE}-completeness of the termination problem 1.
%{\color{red} ++++++++++++ still need to determine problem 4.
%Most likely it is {\sf RE}-complete ++++++++}

\bigskip
\noindent
{\bf The reduction.}
Let $\mathcal{R}=(\Delta^*,\Theta)$
we a word rewriting system.
We now describe our reduction 
$\mathcal{R}\mapsto\SigmaR$.
We assume without loss of generality that $\Delta=\{0,1\}$.
The tgd set $\SigmaR$
is over schema
$\mathbf{S}_{\scriptscriptstyle \mathcal{R}} = 
(E,L,R)$, 
and is defined as
$
\SigmaR  =  \SigmaTheta\cup\SigmaLR.
$

\medskip
We let $\SigmaTheta = \{\xi_{\rho} : \rho\in\Theta\}$,
% \cup TC \cup AD
%\cup S$, 
where $\xi_{\rho}$ is

\begin{eqnarray*}
\lefteqn{E(x_0,a_1,x_1),E(x_1,a_2,x_2),\ldots,E(x_{n-1},a_n,x_n) \; \rightarrow} \\ 
& & \exists \; y_0\ldots  
\exists \; y_m \; L(x_0,y_0),
E(y_0,b_1,y_1),E(y_1,b_2,y_2),\ldots,\\
& &\hspace{3.2cm}\ldots,E(y_{m-1},b_m,y_m), R(x_n,y_m),
\end{eqnarray*}

\noindent
when $\rho = (a_1\ldots a_n, b_1\ldots b_m)$.
Intuitively, we model a word 
$w=a_1a_2\ldots a_n$ as the line-graph
$
E(x_0,a_1,x_1)$, $E(x_2,a_2,x_3)$, $\ldots$ , 
$E(x_{n-1},a_n,x_n)$.

The effect of tgd 
$\xi_{(a_{1}a_{2}\ldots a_{n},b_{1}b_{2}\ldots b_{m})}$ 
is to transform the
line graph for $a_1a_2\ldots a_n$ to a grid:

\vspace{0.0cm}

  \begin{center} 
\includegraphics[scale=0.16]{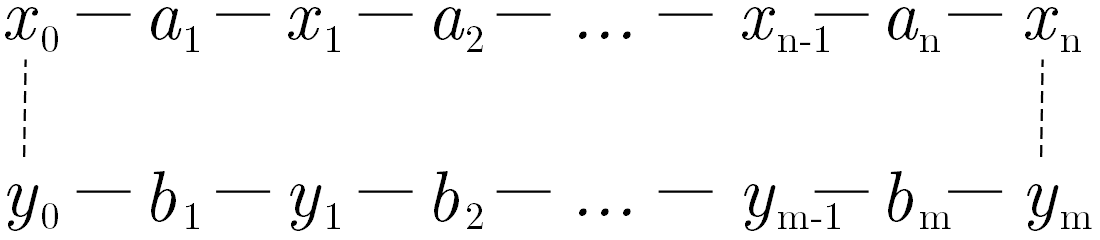}
  \end{center}
\vspace{0.1cm}

Since however a rewriting rule 
$\xi_{(a_{1}a_{2}\ldots a_{n},b_{1}b_{2}\ldots b_{m})}$  is applicable
to longer words, such as e.g.\
$$c_1\ldots c_ka_1a_2\ldots a_nd_{1}\ldots d_{p},$$
resulting in word
$c_1\ldots c_kb_1b_2\ldots b_md_{1}\ldots d_{p}$,
we need rules that copy the $c_1\ldots c_k$ and
$d_{1}\ldots d_{p}$ parts to the new line of the grid.
This is achieved by the following ``grid creation'' rules, 
using ``left'' $L$ and ``right'' $R$ predicates.

{\small{
\begin{eqnarray*}
\xi_{L_{0}} & = & E(x_0,0,x_1),L(x_1,y_1)  \; \rightarrow \;\;\; \exists \; y_0\; L(x_0,y_0), E(y_0,0,y_1) \\
\xi_{L_{1}} & = & E(x_0,1,x_1),L(x_1,y_1)  \; \rightarrow \;\;\; \exists \; y_0\; L(x_0,y_0), E(x_0,1,y_1) \\
\xi_{R_{0}} & = & R(x_0,z_0), E(x_0,0,x_1) \; \rightarrow \;\;\; \exists \; z_1\; E(z_0,0,z_1),R(x_1,z_1)  \\
\xi_{R_{1}} & = & R(x_0,z_0), E(x_0,1,x_1) \; \rightarrow \;\;\; \exists \; z_1\; E(z_0,1,z_1),R(x_1,z_1) 
\end{eqnarray*}
}}

\noindent
We now define 
$\SigmaLR=\{ \xi_{L_{0}}, \xi_{L_{1}}, \xi_{R_{0}}, \xi_{R_{1}} \}$.

\medskip

As an illustration, consider the rewrite rule
$\rho=(0,1)\in\Theta$ applied to word $w=1101$.
This gives us a rewrite step
$1101\rightarrow_{\rho}1111$.
However, in the corresponding core chase sequence
we will need three chase-steps as illustrated
below.

\bigskip
%\begin{verbatim}
%I_0            x--1--x--1--x--0--x--1--x
%
%
%I_1            x--1--x--1--x--0--x--1--x
%                           |     |
%                           x--1--x
%
%
%I_2            x--1--x--1--x--0--x--1--x
%                     |     |     |     |
%                     x--1--x--1--x--1--x
%
%
%I_3            x--1--x--1--x--0--x--1--x
%               |     |     |     |     |
%               x--1--x--1--x--1--x--1--x
%
%\end{verbatim}

  \begin{center} 
\includegraphics[scale=0.25]{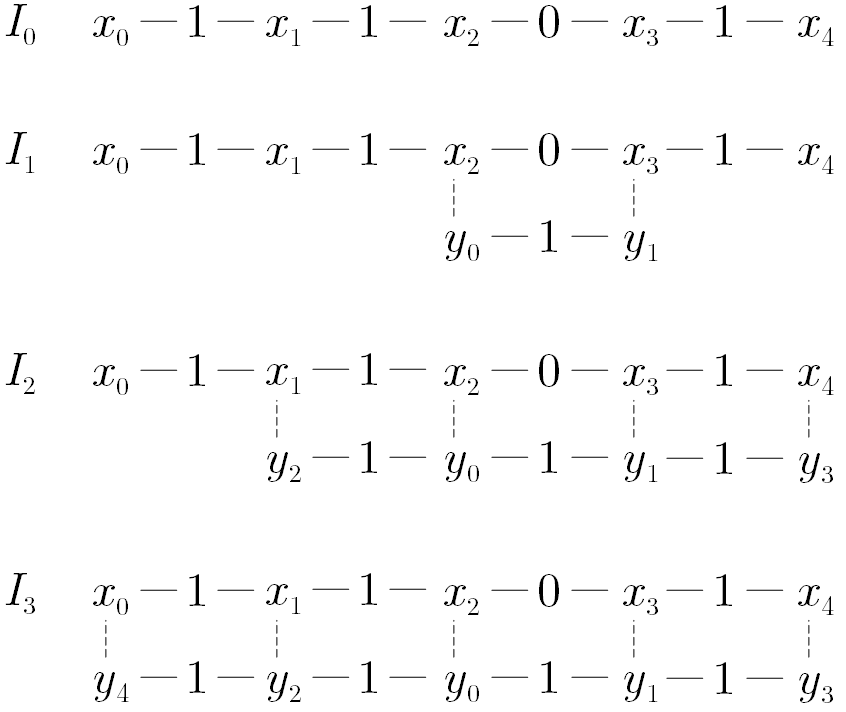}
  \end{center} 

\bigskip

We still need a few more notions.
A {\em path} of an instance~$I$ over 
schema $\mathbf{S}_{\scriptscriptstyle \mathcal{R}}$
is a set 
\begin{equation*}\label{equ:path-def}
\pi = \{E(x_0,a_1,x_1),E(x_1,a_2,x_2),\ldots,E(x_{n-1},a_n,x_n)\}
\end{equation*}
of atoms of~$I$, 
such that $\{a_1, a_2, \ldots, a_n\} \subseteq  \Delta$ 
(recall that $\Delta$ is the alphabet of 
the rewriting system $\Theta$),
and the $x_i$'s are pairwise distinct
nulls from {\sf Nulls}.
The {\em word spelled by the path} $\pi$ is
\begin{equation*}
word(\pi) = a_1a_2\ldots a_n.
\end{equation*}
A~{\em max-path} $\pi$ in an instance 
$I$ is a path, such that no other path in $I$
is a strict superset of~$\pi$.
We can now relate words and instances as follows:
Let $I$ be an 
%acyclic 
instance.  
We define
\begin{eqnarray*}
paths(I) & = & \{\pi : \pi \mbox{ is a max-path in } I\} \mbox{ and}, \\
words(I) & = & \{word(\pi) : \pi\in paths(I)\}.
\end{eqnarray*}

Clearly $paths(I)$ is finite, for any finite instance $I$.
Conversely, let $w=a_1a_2\ldots a_n \in \Delta^*$. We define
\begin{equation*}
I_w=\{ E(x_0, a_1, x_1), E(x_1, a_2, x_2), \ldots, E(x_{n-1}, a_n, x_n) \},
\end{equation*} 
where the $x_i$'s are pairwise distinct nulls
from {\sf Nulls}.
Clearly 
\begin{eqnarray*}
words(I_w) & = &\{w\}, \mbox{ and} \\
I_{word(\pi)} & \cong & \pi.
\end{eqnarray*}

In order to relate core chase sequences to word derivation
sequences, we shall look the core chase sequence in a
finer granularity. Let
$I_0, I_1, I_2, \ldots$  
be a core chase sequence, 
where $I_0=I_w$ for some $w\in\Delta^*$.
Since the core chase step fires all dependencies
in $\SigmaR$ in parallel, we will in the sequel
write the sequence $I_0, I_1, I_2, \ldots$ as
$I_0\rightarrow_{\SigmaR} I_1\rightarrow_{\SigmaR}  
I_2\rightarrow_{\SigmaR}  \cdots$.  
Consider a particular step
$I_{n}\rightarrow_{\SigmaR}  I_{n+1}$.
Suppose wlog the tgds that fired on $I_n$ were
$\xi_{1}, \xi_{2},\ldots,\xi_{k_{n}}$.
We can think of the core chase step
$I_{n}\rightarrow_{\SigmaR}  I_{n+1}$
as $k_n$ standard chase steps
$I_{n} \rightarrow_{\xi_{m}}\! I_{nm}$,
where
$m=1,2,\ldots,k_n$, performed in parallel, 
followed by the computation of
$I_{n+1}=core(I_{n1} \cup I_{n2} \cup \cdots \cup I_{nk_{n}})$.
This allows us to prove the following lemma.

\begin{lemma}\label{lemma:paths}
Let $w\in\Delta^*$, and
$I_w=I_0\rightarrow_{\SigmaR}\!
I_1\rightarrow_{\SigmaR}\!
I_2\rightarrow_{\SigmaR}\cdots$
be a core chase sequence. Then
$paths(I_{n1} \cup I_{n2} \cup \cdots \cup I_{nk_{n}}) =
paths(I_{n+1})$,
for all pairs of instances $I_n,I_{n+1}$ in the 
sequence.
\end{lemma}

{\em Proof:}
Denote the instance
$I_{n1} \cup I_{n2} \cup \cdots \cup I_{nk_{n}}$
with $J$.
If there were a path $\pi$,
such that 
$\pi\in paths(J)
\setminus paths(core(J))$,
there would have to be atoms of the form
$L(x,x)$ and $R(x,x)$ in $J$.
But such $L$ and $R$ atoms do not occur in $I_w$,
neither are they created by any of the rules
$\xi_{\rho},\xi_{L_{1}}, \xi_{L_{2}}, \xi_{R_{1}}$ or
$\xi_{R_{2}}._{\blacksquare}$

\bigskip

We are now ready for the following result.

\begin{theorem}\label{theorem:A}
For each derivation sequence
\begin{eqnarray*}\label{equ:der-seq}
w_0\Rarrow
w_1\Rarrow
\cdots
\Rarrow
w_n
\end{eqnarray*}
there is a sequence
$1\leq j_1< j_2 < \cdots < j_n$ of indices,
such that 
\begin{eqnarray*}\label{equ:core-seq}
I_{w_0}=I_0
\rightarrow_{\SigmaR}\cdots\rightarrow_{\SigmaR}
I_{j_1}
\rightarrow_{\SigmaR}\cdots\rightarrow_{\SigmaR}
I_{j_2}
\rightarrow_{\SigmaR}\cdots\rightarrow_{\SigmaR}
I_{j_n}
\end{eqnarray*}
is a core chase sequence,
and there is a path namely 
$\pi_{n}\in paths(I_{j_{n}}) \setminus paths(I_{j_{n-1}})$,
with $word(\pi_{n}) = w_n$.
\end{theorem} 

{\em Proof:}
We prove the claim
by an induction on $n$.

For the base case, we note that if
$w_0\Rarrow w_1$, then there must be 
a rule $(\ell,r)\in\Theta$, such that
$w_0 = x\ell y$ and $w_1=xry$. 
Then the dependency $\xi_{(\ell,r)}\in\Sigma_{\Theta}$
will fire on $I_{w_{0}}$ (perhaps along with other tgds
from $\SigmaR$), 
resulting in instance $I_1$,
such that, by Lemma  (\ref{lemma:paths}),
there is a path 
$\pi_1$ in $paths(I_1)\setminus paths(I_0)$
with $word(\pi_1)=r$.
If $|x|=|y|=0$, then $w_1=r$.
Otherwise, let $m=max\{|x|,|y|\}$.
The copy dependencies in $\SigmaLR$ will fire
at steps $2,\ldots,1+m$, and from Lemma (\ref{lemma:paths})
it follows
there will be a path $\pi_1$ in 
$paths(I_{1+m})\setminus paths(I_{1})$,
such that $word(\pi_1)=xry=w_1$.
Hence $j_1=1+m$.

As inductive hypothesis, suppose that there is
a path $\pi_n$ in $paths(I_{j_{n}})\setminus paths(I_{j_{n-1}})$,
such that $word(\pi_n)=w_n$.
Suppose that 
$w_n\Rarrow
w_{n+1}$,
for a word $w_{n+1}$, where $w_n=x\ell y$, 
$w_{n+1}=xry$, and $(l,r)\in\Theta$. 
Let $m=max\{|x|,|y|\}$.
Similarly to the base case there will be
a path $\pi_{n+1}$ in $I_{n+m}\setminus I_{n}$,
such that $word(\pi_{n+1})=w_{n+1}$.
Hence $j_{n+1}=n+m$.
$_{\blacksquare}$

\bigskip

We need a couple of more notions.
Let $w$ be a word in $\Delta^*$.
The {\em level $n$ derivation tree of} $w$,
is a $\Delta^*$-labeled tree $\mathcal{T}_{w,n}$,
such that $\mathcal{T}_{w,0}$ is a single
node labeled $w$, and
the tree $\mathcal{T}_{w,n+1}$
is obtained by adding a
node labeled $v$ as a child of a leaf node
in $T_{w,n}$ labeled $u$,
whenever $u\rightarrow_{\rho} v$,
for some $\rho\in\Theta$.
By $\mathcal{T}_{w}$,
the {\em derivation tree of} $w$, we mean the
tree $\bigcup_{n\geq 0}\mathcal{T}_{w,n}$.
%where the union is set union, not disjoint union.

%%%%%%%%%%%%%%%%%%%%%%%%%%%%%%%%%%%%%%%%%%%%%%%%%%%%%%%%%%%%%%%%%
%Furthermore, let 
%$I_{w_{0}}\!=I_0\rightarrow_{\SigmaR}
%I_1\rightarrow_{\SigmaR}I_2\rightarrow_{\SigmaR}\cdots$
%be a core chase sequence. Then clearly
%$paths(I_0)\subset paths(I_1)\subset paths(I_2)\subset\cdots$.
%For each path $\pi\in paths(\chase_{\SigmaR}(I_{w_{0}}))$,
%define $\level(\pi)=k$, where $I_k$
%is the first instance in the core chase sequence
%where the path appears.
%%%%%%%%%%%%%%%%%%%%%%%%%%%%%%%%%%%%%%%%%%%%%%%%%%%%%%%%%%%%%%%%%

\begin{theorem}\label{theorem:B}
For each $w_0\in\Delta^*$ and $n\geq 0$,
such that
$I_{w_{0}}\!=I_0\rightarrow_{\SigmaTheta}
I_1\rightarrow_{\SigmaTheta}\cdots\rightarrow_{\SigmaTheta}I_n$
is a core chase sequence,
there is a injection $\mu_n$ from $paths(I_n)$
to $\mathcal{T}_{w_0,n}$, such that
the label of 
$\mu_n(\pi)$ contains $word(\pi)$
as a factor,
for each $\pi\in paths(I_n)$.
\end{theorem}

{\em Proof:}
We do an induction on $n$.
For the base case 
we note that $paths(I_{w_{0}}) = \{\pi\}$,
where $word(\pi)=w_0$ is the label of the root
$\mathcal{T}_{w_{0},0}$.

For the inductive step,
consider the core chase step
$I_n\rightarrow_{\SigmaR}I_{n+1}$.
For each path $\pi\in paths(I_{n+1})\cap paths(I_{n})$,
we assign $\mu_{n+1}(\pi)=\mu_{n}(\pi)$.
Each path $\pi$ from the set $paths(I_{n+1})\setminus paths(I_{n})$
must have been created by a dependency.
For each $\xi\in\SigmaR$ 
that fired at step $n+1$
there are two possibilities.
\begin{enumerate}
\item
$\xi\in\SigmaTheta$.
Then there is a rewrite rule $(\ell,r)\in\Theta$,
and a path $\pi_n\in paths(I_{n})$,
such that $word(\pi_n)=x\ell y$,
for some $x,y\in\Delta^*$.
Furthermore, there is a path
$\pi_{n+1}\in paths(I_{n+1})\setminus paths(I_n)$,
with $word(\pi_{n+1})=r$.
By the inductive hypothesis,
the label of the node $\mu_n(\pi_n)$
in $\mathcal{T}_{w_{0},n}$ contains
$x\ell y$ as a factor,
i.e.\ the label of $\mu_n(\pi_n)$ is
$x'\ell y'$, where 
$x$ is a suffix of $x'$
and $y$ is a prefix of $y'$.
Now $\mathcal{T}_{w_{0},n+1}$ will
have a child of $\mu_n(\pi_n)$ labeled
$x'ry'$. We assign $\mu_{n+1}(\pi_{n+1})$
to be that child. Since $r$ is a factor
of $x'ry'$, the claim follows.

\item
$\xi\in\SigmaLR$.
Then there must be a path
$\pi_n\in paths(I_n)$,
(and possibly another dependency $\xi'\in\SigmaLR$),
such that
firing $\xi$
(together with $\xi'$),
creates a path $\pi_{n+1}\in paths(I_{n+1})\setminus paths(I_n)$,
such that $\pi_n\subset \pi_{n+1}$.
By the inductive hypothesis,
the label of the node $\mu_n(\pi_n)$
in $\mathcal{T}_{w_{0},n}$ contains
$word(\pi_n)$ as a factor.
From the construction of $\SigmaLR$ and $\mu_n$
it follows that $word(\pi_{n+1})$
also is a factor of the label of
$\mu_n(\pi_n)$. Since $\pi_n\subset \pi_{n+1}$,
it means that $\pi_n\notin paths(I_{n+1})$,
as $\pi_n$ no longer is a max-path in $I_{n+1}$.
Therefore we can assign 
$\mu_{n+1}(\pi_{n+1})=\mu_n(\pi_{n})$.
As $\mu_{n}(\pi_{n})$ also is a node
in $\mathcal{T}_{w_{0},n+1}$ the claim follows.$_{\blacksquare}$
\end{enumerate}

\begin{theorem}\label{THEO:reprove}
Let $\mathcal{R}=(\Delta^*\!,\Theta)$ 
be a word rewriting system,
and let $w_0\in\Delta^*$.
Then the core chase sequence 
$I_{w_{0}}\!=I_0\rightarrow_{\SigmaR}  
I_1\rightarrow_{\SigmaR} 
I_2\rightarrow_{\SigmaR} \cdots$
is infinite
if and only if 
there is an infinite derivation 
$w_0\Rarrow
w_1\Rarrow
w_2\Rarrow\cdots$. 
\end{theorem}

{\em Proof:}
Suppose there is an infinite 
derivation sequence 
$w_0\Rarrow
w_1\Rarrow
w_2\Rarrow\cdots$. 
It then follows from
Theorem \ref{theorem:A} that the core chase
sequence 
$I_{w_{0}}\!=I_0\rightarrow_{\SigmaR}  
I_1\rightarrow_{\SigmaR} 
I_2\rightarrow_{\SigmaR} \cdots$
is infinite as well.

Conversely, suppose that 
the core chase
sequence 
$I_{w_{0}}\!=I_0\rightarrow_{\SigmaR}  
I_1\rightarrow_{\SigmaR} 
I_2\rightarrow_{\SigmaR} \cdots$
is infinite.
At each core chase step $n$,
if at least one dependency from $\SigmaTheta$
fires, we have 
$|paths(I_{n})| > |paths(I_{n-1})|$,
and if only dependencies from $\SigmaLR$
fire, we have 
$|paths(I_{n})| = |paths(I_{n-1})|$.
It is however easy to see that
there can only be a finite number
of consecutive chase steps that only
fire dependencies from $\SigmaLR$.
Consequently 
%$\cchase_{\SigmaR}(I_{w_{0}})$
$\bigcup_{n\geq 0}I_n$
contains an infinite number of paths.
From Theorem (\ref{theorem:B}) it then follows that
$\mathcal{T}_{w_{0}}$ 
is infinite as well,
and since $\mathcal{T}_{w_{0}}$ is finitely branching,
K\"{o}nig's Lemma tells us that $\mathcal{T}_{w_{0}}$ 
has an infinite branch.
Let the labels on this path be, in order,
$w_0,w_1,w_2,\ldots$. 
This means that the derivation  
$w_0\Rarrow w_1\Rarrow w_2\Rarrow\ldots$
is infinite.
$_{\blacksquare}$

\bigskip
This theorem, 
together with the
{\sf RE}-completeness of
rewriting termination, 
yields the undecidability result of
\cite{DBLP:conf/pods/DeutschNR08} 
for core chase termination on a given instance.

\begin{corollary}\label{ATHEO:DeutschTheo}
The set $\cctia{I}$
is {\sf RE}-complete.
\end{corollary}

\bigskip
\noindent
{\bf The uniform case}.
Next we shall relate the uniform termination
problem with the set $\cctaa$. 
This means that we need to consider arbitrary instances,
not just instances of the form $I_w$, for $w\in\Delta^*$.
If the arbitrary instance $I$ is cyclic
the behavioral correspondence between
word derivations in $\mathcal{R}$ and
the core chase sequence with $\SigmaR$ on $I$ 
breaks down.
We therefore need to extend the schema
$\mathbf{S}_{\mathcal{R}}$ to include relational
symbols $D$ and $E^*$. Intuitively,
the unary  relation $D$ will hold the
active domain of an instance, and
the binary relation $E^*$ will hold
the transitive closure of the graph
obtained from the instance. 

The following tgd set $\Sigma_{\scriptscriptstyle AD}$ computes
$dom(I)\cup\Delta$
in relation $D$. 

%{\small{
\begin{eqnarray*}
            &\rightarrow& \; D(0), D(1)      \\
E(x,z,y) \; &\rightarrow& \; D(x),D(z),D(y)  \\
L(x,y)   \; &\rightarrow& \; D(x),D(y)       \\
R(x,y)   \; &\rightarrow& \; D(x),D(y)       \\
E^*(x,y) \; &\rightarrow& \; D(x),D(y) 
\end{eqnarray*}
%}}

\medskip

Given an instance $I$,
by {\em the graph of} $I$ we mean the graph
with edge set:

\begin{eqnarray*}
& & G_I = \{ (x,y)\;:\; E(x,z,y) \in I \mbox{ or } E^*(x,y) \in I \\
& &\hspace{3.5cm}\mbox{ or }  L(x,y)\in I \mbox{ or }  R(x,y)\in I \}.
\end{eqnarray*}

\medskip

The following set $\Sigma_{\scriptscriptstyle TC}$ 
computes in $E^*$ the transitive closure
of $G_I$: 

%{\small{
\begin{eqnarray*}
E(x,z,y) \;           &\rightarrow& \;  E^*(x,y) \\
L(x,y) \;             &\rightarrow& \;  E^*(x,y) \\
R(x,y) \;             &\rightarrow& \;  E^*(x,y) \\
E^*(x,y),E^*(y,z) \;  &\rightarrow& \;   E^*(x,z) 
\end{eqnarray*}
%}}

If $G_I$ has a cycle, 
the chase will eventually place
an atom of the form $E^*(v,v)$ in the instance.
Once an instance contains such a tuple,
the dependencies in the following 
``saturation'' set $\Sigma_{\scriptscriptstyle SAT}$
will be fired: 

\begin{eqnarray*}
E^*(v,v),D(x),D(z),D(y) \; &\rightarrow& \; E(x,z,y) \\
E^*(v,v),D(x),D(y) \;      &\rightarrow& \; L(x,y),R(x,y),E^*(x,y) 
\end{eqnarray*}

From here on, we assume that the schema
$\mathbf{S}_{\mathcal{R}}$ is $\{E,L,R,D,E^*\}$, and
that the reduction $\mathcal{R}\mapsto\SigmaR$ gives
$\SigmaR = \{\SigmaTheta,\SigmaLR,\Sigma_{\scriptscriptstyle AD},
\Sigma_{\scriptscriptstyle TC},\Sigma_{\scriptscriptstyle SAT}\}$.

\medskip

We denote by $H_I$ the Herbrand base of instance $I$,
i.e.\ the instance where, for each relation symbol $R\in\mathbf{S}_{\mathcal{R}}$,
the interpretation
$R^{H_{I}}$ contains all tuples (of appropriate arity) that can be formed
from the constants in 
$(dom(I)\cap{\sf Cons})\cup\Delta$.
The proof of the following lemma is straightforward:

\begin{lemma}\label{LEMMA:herbrandCore}
$H_I\models\SigmaR$, and
$core(I) = H_I$, whenever 
$H_I$ is a subinstance of $I$. 
\end{lemma}

\medskip
\begin{lemma}\label{LEMMA:cycleTermination}
Let $I_0$ be an arbitrary instance over schema $\mathbf{S}_{\mathcal{R}}$,
and let $I_0,I_1$,$I_2$,$\ldots$  be the core chase sequence 
with $\Sigma_{\mathcal{R}}$ on $I_0$. 
If there is an $n\geq 0$, and a
constant or variable $v$, such that $E^*(v,v)\in I_n$ 
(i.e.\ the graph $G_{I_{m}}$ is cyclic
for some $m\leq n$), then the core chase sequence is finite. 
\end{lemma}

\noindent 
{\em Proof:}  (Sketch)
First we note that $H_{I_{n}} = H_{I_{0}}$
for any instance $I_n$ in the core chase sequence,
since the chase does not add any new constants.
If the core chase does not terminate at
the instance $I_{n}$ from the chase sequence,
it follows that the dependencies in the set 
$\Sigma_{\scriptscriptstyle SAT}$
will fire at $I_n$ and generate $H_{I_0}$ as a subinstance.
It then follows from Lemma \ref{LEMMA:herbrandCore}
that $I_{n+1} = H_{I_{0}}$.
Since
$H_{I_{0}}\models\Sigma_{\mathcal{R}}$
the core chase will terminate at 
instance~$I_{n+1}$.
$_{\blacksquare}$

\bigskip

Intuitively, 
the previous lemma guarantees that 
whenever we have a cycle in the initial instance 
the core chase sequence will terminate. Thus,
in the following we will 
assume that the instances we consider 
are acyclic. Furthermore,
in order not to unnecessarily terminate
a core chase sequence, we need
the following lemma that ensures us 
the core chase with $\SigmaR$
on an acyclic instance will not create
any cycles.

\begin{lemma}\label{LEMMA:noCycle}
Let $I_0$ be an arbitrary instance over schema 
$\mathbf{S}_{\mathcal{R}}$,
such that $G_{I_{0}}$ is acyclic,
and let  $I_0,I_1,I_2,\ldots$ be the core chase
sequence with $\SigmaR$ on $I_0$. 
Then $G_{I_{n}}$ is acyclic,
for all instances $I_n$
in the sequence.
\end{lemma}

{\em Proof:}  (Sketch)
Suppose to the contrary that $G_{I_{n}}$ is cyclic,
for some $I_n$ in the sequence.
Wlog we assume that $I_n$ is the first such instance
in the sequence.
Clearly $n\geq\!{1}$.
This means that by applying all 
active triggers on $I_{n-1}$
will add a cycle 
(note that the taking the core cannot add a cycle).
Let $\Sigma'\subseteq\SigmaR$ 
%$\xi_1,\ldots,\xi_k$
be the dependencies that fired at $I_{n-1}$.
First, it is easy to see 
that
$\Sigma'\cap\SigmaLR 
%$\{\xi_1, \ldots, \xi_k\}\cap\SigmaLR = 
\emptyset$. 
This is because these dependencies 
do not introduce any new edges in $G_{I_{n}}$
between vertices in $G_{I_{n-1}}$,
they only add a new vertex into $G_{I_{n}}$ which
will have two incoming edges from vertices
already in $G_{I_{n-1}}$.
A similar reasoning shows 
that none of the tgds
in $\Sigma_{\scriptscriptstyle TC}$ or in $\Sigma_{\scriptscriptstyle AD}$
can be part of 
$\Sigma'$.
%the set $\{ \xi_1, \ldots, \xi_n \}$.
Finally, the dependencies in the set $\Sigma_{\scriptscriptstyle SAT}$ 
may introduce 
cycles and may thus be part of $\Sigma'$.
%the set 
%$\{ \xi_1, \ldots, \xi_k \}$.
But the dependencies in $\Sigma_{\scriptscriptstyle SAT}$ 
are fired only when
$E^*(x,x)\in I_{n-1}$,
which means that 
$G_{I_{n-1}}$ already contains a cycle,
namely the self-loop on $x$.
This contradicts our  assumption that
$I_{n}$ is the first instance 
in the chase sequence that contains a cycle.
$_{\blacksquare}$
%%%%%%%%%%%%%%%%%%%%%%%%%%%%%%%%%%%%%%%%%%%%%%%%%%%%%%%%%%%%%%%%%

\bigskip 

%We are going to transform an arbitrary
%instance $I$ into a finite set of instances
%$$
%I^*=\{I_w : w\inwords(I)\},
%$$ 
%such that $\chase^{core}_{\SigmaR}(I)$
%terminates if and only if
%$\chase^{core}_{\SigmaR}(J)$
%terminates for all $J\in I^*$.
We are going to transform an arbitrary
instance $I$ into an instance
$$
I^*=\bigcup_{\pi\in paths(I)}I_{word(\pi)},
$$ 
such that 
%$\chase^{core}_{\SigmaR}(I)$
the core chase sequence with $\SigmaR$ on $I$
terminates if and only if
it terminates on~$I^*$.
%$\chase^{core}_{\SigmaR}(I^*)$
%terminates.
In order for the construction to work,
we first need to ``chase out'' of $I$ all
the $\Sigma_{LR}$ dependencies.

\begin{lemma}\label{iterLemma}
Let $I$ be an arbitrary acyclic instance,
and $J$ the finite instance in which
the core chase with $\SigmaLR$ on $I$
terminates. Then the core chase sequence
$I=I_0\Rarrow I_1 \Rarrow I_2 \Rarrow \cdots$
is finite if and only if the core chase sequence
$J=J_0\Rarrow J_1 \Rarrow J_2 \Rarrow \cdots$
is finite.
%
%and $I=I_0\Rarrow I_1 \Rarrow I_2 \Rarrow \cdots$
%be a core chase 
%Then $\cchase_{\SigmaR}(I)$
%is finite if and only if
%$\cchase_{\SigmaR}(\cchase_{\SigmaLR}(I))$
%is finite.
%Then the core chase sequence
%$I_0\rightarrow_{\SigmaR} 
%I_1\rightarrow_{\SigmaR} 
%I_2\rightarrow_{\SigmaR}  
%\ldots$
%is finite
%if and only if
%the core chase sequence
%$I^\!*=J_0\rightarrow_{\SigmaR} 
%J_1\rightarrow_{\SigmaR}  
%J_2\rightarrow_{\SigmaR} 
%\ldots$ is finite.
\end{lemma}

{\em Proof:}
Since $I_0\subseteq J$,
the if-direction follows from the monotonicity of the chase
\cite{DBLP:conf/icdt/FaginKMP03}.
The only-if direction follows from
the observation that since
%$\cchase_{\SigmaLR}(I_0)$
$J$ is finite, and
$\SigmaLR\subseteq\SigmaR$, 
there must be
an $n\geq 0$, such that
$J \subseteq I_n$.
$_{\blacksquare}$

\medskip

The next lemma is key to showing that
$\mathcal{R}$ is uniformly terminating
is and only if the core chase with
$\SigmaR$ terminates on all instances.

\medskip
\begin{lemma}\label{istar}
Let $I$ be an arbitrary acyclic 
instance such that $I \models \Sigma_{LR}$,
and let $I=I_0,I_1,I_2,\ldots,I_n$ and
$I^*\!=J_0,J_1,J_2,\ldots,J_n$,
$n\geq 0$,
be core chase sequences with $\SigmaR$
on $I$ and $I^*$, respectively.
Then there is an bijection $\mu_n$ from
$paths(I_n)$ to $paths(J_n)$,
such that $word(\pi) = word(\mu_n(\pi))$,
for each $\pi\in paths(I_n)$.
%Then $|paths(\chase^{core}_{\SigmaR}(I))| =
%|paths(\chase^{core}_{\SigmaR}(I^*))|$.
%Then the core chase sequence
%$I_0\rightarrow_{\SigmaR} 
%I_1\rightarrow_{\SigmaR} 
%I_2\rightarrow_{\SigmaR}  
%\cdots$
%is finite
%if and only if,
%for all instances $J\in I^*_0$,
%the core chase sequence
%$J=J_0\rightarrow_{\SigmaR} 
%J_1\rightarrow_{\SigmaR} 
%J_2\rightarrow_{\SigmaR}
%\cdots$ 
%is finite. 
\end{lemma}

{\em Proof:}
We do an induction on $n$.
For the base case, 
$\mu_0$ maps each path $\pi\in paths(I_0)$
to the single element in
$paths(I_{word(\pi)})\subseteq paths(I^*_0)$.

For the inductive step,
let $\pi_{n+1}\in paths(I_{n+1})\setminus paths(I_{n})$.
Then $\pi_{n+1}$ must have been created by the firing
of a dependency from $\SigmaTheta$,
or by firing one or two dependencies from 
$\SigmaLR$.
In the first case there must be a path
$\pi_n\in paths(I_{n})$, such that
the dependency fired on $\pi_n$
creating the path $\pi_{n+1}$ in $I_{n+1}$.
By the inductive hypothesis,
there is a unique path $\mu_n(\pi_n)\in paths(J_n)$.
Since $\mu_n$ maps constants to themselves,
the same dependency will fire on $J_n$,
creating a unique path $\tau\in paths(I_{n+1})$
with $word(\tau)=word(\pi_{n+1})$.
We then assign $\mu_{n+1}(\pi_{n+1})=\tau$.

In the second case the path $\pi_{n+1}$
was created by firing one or two dependencies
from $\SigmaLR$. Then there must be a unique
path $\pi_n$ in $I_n$, such that the
dependency (or dependencies) fired
on $\pi_n$, extending it to a path $\pi_{n+1}$
in $I_{n+1}$. By the inductive hypothesis
$\mu_n(\pi_n)\in paths(J_n)$. Again,
the same dependency (or dependencies)
fire on $\mu_n(\pi_n)$ in $J_n$ and create a path 
$\tau$ in $J_{n+1}$, such that
$\mu_n(\pi_n)\subset\tau$.
Then $\mu_n(\pi_n)$ is no longer a path in $I_{n+1}$,
so we can assign $\mu_{n+1}(\pi_{n+1})=\tau$.
$_{\blacksquare}$

\medskip 
We are now ready for the main result of this section.
\begin{theorem}\label{ATHEO:terminationForAll}
A word rewriting system $\mathcal{R}$ is uniformly terminating 
if and only if
the core chase with $\SigmaR$
terminates on all instances over $\mathbf{S}_{\mathcal{R}}$
(i.e. iff $\SigmaR \in \cctaa$).
\end{theorem}

{\em Proof:}
First let us suppose that $\SigmaR \in \cctaa$.
Let $w \in \Delta^*$ be an arbitrary word.
Because $\SigmaR \in \cctaa$ it follows that 
the core chase will terminate also on instance $I_w$.
From this and Theorem \ref{THEO:reprove} it follows that 
the all derivation sequences in $\mathcal{R}$ originating
from $w$ are finite.

For the other direction
suppose that $\SigmaR \notin \cctaa$.
Then there exists an instance $I$,
such that the core chase sequence 
$I=I_0,I_1,I_2,\ldots$ 
with $\SigmaR$ 
is infinite.
From Lemma \ref{LEMMA:cycleTermination}
we know that $I$ is acyclic.
From 
Lemma \ref{istar}
it follows that the core chase of $I^*$
with $\SigmaR$
must be infinite as well.
Similarly to the second part of the proof of 
Theorem \ref{THEO:reprove},
there must be path $\pi\in I^{*}$
such that $\SigmaR$ admits an infinite derivation
starting from $word(\pi)$.
But this means that $\mathcal{R}$ is not uniformly 
terminating.
$_{\blacksquare}$

\medskip

Using the previous result, 
and the {\sf coRE}-completeness 
of the uniform termination problem
for word rewriting systems,
we now have the main theorem.

\setcounter{theorem}{5}
\begin{theorem}\label{ATHE:mainUndecidability}
The membership problem for $\cctaa$ is {\sf coRE}-complete.
$_{\blacksquare}$
\end{theorem}

\medskip
Up to here we proved the undecidability of the $\cctaa$ 
class (and the $\cctae$ class).
Next we will show that this result can be extended with some minor changes
to other termination classes as well.

\begin{theorem}\label{ATHEO:sctaeUndecidable}
The membership problem for $\sctae$ is {\sf coRE}-complete.
\end{theorem}

{\em Proof:}  (Sketch)
It is easy to see that the same $\SigmaR$
reduction works for the $\sctae$ case as well 
by choosing the branch that first applies 
all the  dependencies in $\Sigma_{\scriptscriptstyle AD}\cup 
\Sigma_{\scriptscriptstyle TC} \cup 
\Sigma_{\scriptscriptstyle SAT}$.
This is because in case the initial arbitrary instance contains a cycle
the full dependencies $AD\cup TC \cup S$ 
will saturate the instance and the standard chase will terminate.
If $I$ does not contain any cycles then, as we showed, during the 
chase process no cycles are added and the termination proof
is the same as for the core chase, mutatis mutandis.
$_{\blacksquare}$

\bigskip 
To show that the basic $\mathcal{R}\mapsto\SigmaR$ reduction cannot be used for 
the $\sctaa$ class. Consider the word rewriting system 
$\mathcal{R} = (\{0,1\}, (1,0))$
and $I=\{ E(x,0,x),L(x,y) \}$. It is easy to see that 
the branch that applies the $\xi_{L_{0}}$ dependency
first will not terminate as it will generate the following
infinite set of tuples:
\bigskip

\begin{minipage}[b]{0.50\linewidth}
\centering
\hspace{2.5cm}
\begin{tabular}{lll} 
\multicolumn{3}{c}{$E$}  \\
\hline 
a&0&a  \\ 
$x_1$&0&b  \\ 
$x_2$&0&$x_1$  \\ 
\multicolumn{3}{c}{$\ldots$}   \\ 
$x_n$&0&$x_{n-1}$  \\ 
\multicolumn{3}{c}{$\ldots$}   
\end{tabular}
\end{minipage}
\begin{minipage}[b]{0.30\linewidth}
\centering
\begin{tabular}{ll} 
\multicolumn{2}{c}{$L$}  \\
\hline 
a&b  \\ 
a&$x_1$  \\ 
a&$x_2$  \\ 
\multicolumn{2}{c}{$\ldots$}  \\
a&$x_n$  \\ 
\multicolumn{2}{c}{$\ldots$}  
\end{tabular}
\end{minipage}

\medskip
\noindent
On the other hand, it is clear that the 
word rewriting system
$\mathcal{R} = (\{0,1\}, (1,0)) \}$
is uniformly terminating.

\bigskip

The undecidability result can still be obtained for the $\sctaa$ class
if we allow 
{\em denial constraints}, i.e.\
tgds where the head is the constant {\em False}.
A denial constraint 
$\alpha\!\rightarrow \bot$ 
satisfied by an instance I if there is 
no homomorphism $h$, such that $h(\alpha)\subseteq I$. 
If a denial constraint is violated, the chase terminates (in failure). 
Then we simply define
$\Sigma_{\scriptscriptstyle F} =
\{E^*(x,x) \rightarrow \bot \} \cup
\SigmaR\setminus\Sigma{\scriptscriptstyle SAT}$.

\begin{theorem}\label{ATHEO:sctaaUndecidable}
Let $\Sigma$ be a set of tgds and one denial constraint.
The 
membership problem $\Sigma\in \sctaa$ is {\sf coRE}-complete.
\end{theorem}

{\em Proof:} (Sketch)
Similarly to the proof of Theorem \ref{ATHEO:sctaeUndecidable}
it is easy to see that if an arbitrary instance $I$ contains 
a cycle, then the standard chase on $I$ with $\Sigma_{F}$
will terminate on all branches.
This is because the fairness conditions
guarantees that the denial constraint will be fired,
and the chase will terminate.
$_{\blacksquare}$

\bigskip

We can now summarize the results of
this section together with the results
cited in Theorem \ref{A:hitherto}, yielding the
following table.

\bigskip

\begin{center}

\begin{tabular}{|l|cccc|}
\hline
\vspace*{-2ex}
& & & & \\
\large{$\star$} & $\mathsf{CT}^{\star}_{I \exists}$ & $\mathsf{CT}^{\star}_{I \forall}$ & $\mathsf{CT}^{\star}_{\forall\forall}$ &  $\mathsf{CT}^{\star}_{\forall\exists}$ \\  
\hline%\hline
                & &\multicolumn{1}{r|}{}  & & \\            
Core            & &\multicolumn{1}{r|}{}  & & \\                        
                & &\multicolumn{1}{r|}{}  & \multicolumn{2}{c|}{{\sf coRE}-complete}   \\
Standard        & &\multicolumn{1}{r|}{}  & & \\            
\cline{4-5}                         
                & & & & \\
Semi-Oblivious  & & & & \\            
                & \multicolumn{4}{c|}{{\sf RE}-complete}   \\
Oblivious       & & & & \\                 
                & & & & \\
\hline
\end{tabular}

\end{center}

%-- SUFFICIENT CLASSES
\bigskip
\section{Guaranteed termination}\label{sufficient}

%%%%%%%%%%%%%%%%%%%%%%%%enrichment%%%%%%%%%%%%%%%%%%%%%%%%%%%%%%
%%%%%%%%%%%%%%%%%%%%%%%%%%%%%%%%%%%%%%%%%%%%%%%%%%%%%%%%%%%%%%%%

To overcome the undecidability of chase termination,
a flurry of restricted classes of tgds have been 
proposed in the literature. These classes have
been put forth as subsets of
$\mathsf{CT}_{\forall\forall}$,
although at the time only $\mathsf{CT}_{I\forall}$
was known to be undecidable. 
In this section we review these 
restricted classes
with the purpose of determining 
their overall structure and
termination properties.

\bigskip

The termination of the oblivious chase can be related to the 
termination of the standard chase by using the
 {\em enrichment}
transformation introduced by \cite{DBLP:conf/amw/GrahneO11}.
The enrichment
takes a tgd 
$
\xi = 
\alpha(\bar{x},\bar{y})\rightarrow 
\exists \bar{z}\;
\beta(\bar{x},\bar{z})
$ over schema $\mathbf{R}$
and converts it into tuple generating dependency 
$
\depenrich{\xi} = 
\alpha(\bar{x},\bar{y})\rightarrow 
\exists \bar{z}\;
\beta(\bar{x},\bar{z}),{H}(\bar{x},\bar{y}),
$
where ${H}$ is a new relational symbol
that does not appear in~$\mathbf{R}$.
For a set $\Sigma$ of tgds 
the transformed set is 
$\enrich{\Sigma} = \{\depenrich{\xi} : \xi\in\Sigma\}.$
Using the enrichment notion the following
was shown.

\begin{theorem}\label{THEO:oblivious-and-hat}
{\em \hspace*{-1ex}\cite{DBLP:conf/amw/GrahneO11}}
$\;\;\Sigma\in \octaa$ if and only if 
$\enrich{\Sigma} \in \sctaa$. 
\end{theorem}

\noindent
{\em Proof:}
Let $\Sigma \in \octaa$ be a set of tgds.
Suppose now that there is an instance ${I}$ for which
the standard chase with $\enrich{\Sigma}$ does not terminate.
This means that there is an infinite standard-chase sequence
\begin{eqnarray*}\label{EQ:infiniteEnrichChase}
I = {I_0}\xrightarrow{(\depenrich{\xi}_0,h_0)}  
	 {I_1}\xrightarrow{(\depenrich{\xi}_1,h_1)}  
	 \ldots\ldots 
	 \xrightarrow{(\depenrich{\xi}_{n-1},h_{n-1})}  
	  {I_n}\xrightarrow{(\depenrich{\xi}_n,h_n)}  
    \ldots.  
\end{eqnarray*}

\noindent
Thus ${h}_i(body(\depenrich{\xi}_i))\subseteq I_i$,
for all $i\geq 0$. 
Since $body(\depenrich{\xi}_i)=body(\xi_i)$,
we have that
\begin{eqnarray*}\label{EQ:infiniteStdChase}
I = {J_0}\xrightarrow{({\xi}_0,h_0)}
	 {J_1}\xrightarrow{({\xi}_1,h_1)}
	 \ldots\ldots 
 	 \xrightarrow{({\xi}_{n-1},h_{n-1})}
	 {J_n}\xrightarrow{({\xi}_n,h_n)}
    \ldots.  
\end{eqnarray*}
where ${J}_i$ is the same as ${I}_i$ restricted to the atoms in initial
schema, is an infinite oblivious-chase sequence
with $\Sigma$ and ${I}$. From this it follows by contraposition that
$\Sigma\in \octaa$ implies $\enrich{\Sigma} \in \sctaa$.

For the second part, suppose that there is an instance 
 ${I}$ for which
the oblivious chase with $\Sigma$ does not terminate. 
Then there is an infinite oblivious-chase sequence
\begin{eqnarray*}\label{EQ:infiniteOblChase}
I = {I_0}\xrightarrow{({\xi}_0,h_0)} 
    {I_1}\xrightarrow{({\xi}_1,h_1)} 
	 \ldots\ldots 
    \xrightarrow{({\xi}_{n-1},h_{n-1})} 
    {I_n}\xrightarrow{({\xi}_n,h_n)} 
    \ldots.  
\end{eqnarray*}

Let ${J}_0 = {I}_0$, 
and for all $i\geq 0$, let
$${J}_{i+1} = {I}_{i+1}\cup\{
{H}(h_{i}(\bar{x}),h_{i}(\,\bar{y}))\},
$$ 
where $H$ is the enrichment atom used in $\depenrich{\xi}_i$.
We claim that
\begin{eqnarray*}\label{EQ:infiniteOblChaseAll}
I = {J_0}\xrightarrow{(\depenrich{\xi}_0,h_0)} 
	 {J_1}\xrightarrow{(\depenrich{\xi}_1,h_1)} 
	 \ldots\ldots 
	 \xrightarrow{(\depenrich{\xi}_{n-1},h_{n-1})} 
	 {J_n}\xrightarrow{(\depenrich{\xi}_n,h_n)} 
    \ldots.  
\end{eqnarray*}
\noindent
is an infinite standard-chase sequence with
$\enrich{\Sigma}$ and ${I}$.
Towards a contradiction, suppose it is not.
Then there must be an $i\geq 0$, 
such that 
the standard-chase step cannot be applied 
with ${h}_i$ and $\depenrich{\xi}_i$ on ${J}_i$.
Let 
$\xi_i = \alpha(\bar{x},\bar{y})\rightarrow \exists \bar{z}\;
\beta(\bar{x},\bar{z})$.
Then
$\depenrich{\xi}_i = \alpha(\bar{x},\bar{y})\rightarrow
\exists \bar{z}\; \beta(\bar{x},\bar{z}),{H}(\bar{x},\bar{y})$.
If  $({\depenrich{\xi}_i},{h_i})$
is not an active trigger for ${J}_i$,
there exists an extension $h_i'$ 
of $h_i$ such that
$h_i'(body({\depenrich{\xi}_i})) \subseteq {J}_i$.
Since $h_i'$ is an extension of ${h}_i$, 
it follows that
$h_i'(\bar{x})={h}_{i}(\bar{x})$ and 
$h_i'(\bar{y})={h}_{i}(\bar{y})$, meaning that
${H}({h}_{i}(\bar{x}),{h}_{i}(\,\bar{y}))) \in {J}_i$.
Because the facts over ${H}$ are only introduced by the standard chase,
it follows that the homomorphism ${h}_i$ has already been
applied with $\depenrich{\xi}_i$ earlier in the standard-chase
sequence. But then ${h}_i$ must also have been applied with
$\xi_i$ at the same earlier stage in the oblivious-chase sequence.  
This is a contradiction,
since it entails that trigger $({\xi_i},{h_i})$
would have been applied twice in the oblivious-chase sequence.$_{\blacksquare}$

\bigskip
To relate the termination of the semi-oblivious 
chase to the standard chase termination,
we use a transformation similar to the enrichment.
This transformation is called 
{\em semi-enrichment} and 
takes a tuple generating dependency
$
\xi = 
\alpha(\bar{x},\bar{y})\rightarrow 
\exists \bar{z}\;
\beta(\bar{x},\bar{z})
$ over a schema $\mathbf{R}$
and converts it into the tgd
$
\tilde{\xi} = 
\alpha(\bar{x},\bar{y})\rightarrow 
\exists \bar{z}\;
\beta(\bar{x},\bar{z}),{H}(\bar{x}),
$
where ${H}$ is a new relational symbol
which does not appear in~$\mathbf{R}$.
For a set $\Sigma$ of tgds defined on schema $\mathbf{R}$,
the transformed set is 
$\widetilde{\Sigma} = \{\tilde{\xi} : \xi\in\Sigma\}.$
Using the semi-enrichment notion, 
the standard and the semi-oblivious chase 
can be related as follows.

\begin{theorem}\label{THEO:semi-oblivious-and-hat}
$\Sigma\in \soctaa$ if and only if
$\widetilde{\Sigma} \in \sctaa$. 
\end{theorem}

{\em Proof}: 
Similar to the proof of Theorem
\ref{THEO:oblivious-and-hat}.
$_{\blacksquare}$

\bigskip
A  class of sets of tgds $\mathscr{C}$
is said to be {\em closed under enrichment}
if $\Sigma \in \mathscr{C}$
implies that $\enrich{\Sigma} \in \mathscr{C}$.
Using this notation together with Theorem \ref{THEO:oblivious-and-hat}
gives us a sufficient condition for a class of dependencies
to belong to $\octaa$:

\begin{proposition}\label{PROP:suffobl}
Let $\mathscr{C} \subseteq \sctaa$ such that $\mathscr{C}$
is closed under enrichment. 
Then
$\mathscr{C} \subseteq \octaa$.
\end{proposition}

\noindent
{\em Proof}: 
Follows directly from Theorem
\ref{THEO:oblivious-and-hat}.
$_{\blacksquare}$

\bigskip
Using this proposition we will reveal classes of 
dependencies that ensure termination for the 
oblivious chase.
Similarly we define the notion of {\em semi-enrichment closure} 
for classes of dependency sets.
The semi-enrichment closure property
gives a sufficient condition for the semi-oblivious chase
termination.

\begin{proposition}\label{PROP:suffsobl}
Let $\mathscr{C} \subseteq \sctaa$ such that $\mathscr{C}$
is closed under semi-enrichment. 
Then
$\mathscr{C} \subseteq \soctaa$.
\end{proposition}

\noindent
{\em Proof}: 
Follows directly from Theorem
\ref{THEO:semi-oblivious-and-hat}.
$_{\blacksquare}$

\bigskip 

As we will see next,
most of the known classes that ensure the 
standard chase termination are closed under semi-enrichment,
and thus those classes actually 
guarantee the semi-oblivious chase termination
as well. 
As we saw in Section \ref{SEC:complexityChaseStep}, 
the semi-oblivious chase has a lower complexity that
the standard chase.

\bigskip
\subsection*{Acyclicity based classes}

\bigskip
As full tgds do not generate any new nulls
during the chase, any sequence with a set of
full tgds will terminate
since there only is a finite number of tuples
that can be formed out of the elements of the
domain of the initial instance. The cause of
non-termination lies in the existentially
quantified variables in the head of the
dependencies. Most restricted classes thus
rely on restricting the tgds in a way that
prevents these existential variables
to participate in any recursion.
%We next briefly describe some of these
%restricted classes of tgds.

The class {\sf WA} of {\em weakly acyclic} sets
of tgds, introduced by 
\cite{DBLP:conf/icdt/FaginKMP03},
was one of the
first restricted classes to be proposed. 
Consider 
\begin{eqnarray*}
\Sigma_1 & = & \{R(x,y)\rightarrow\exists{z}S(z,x)\}.
\end{eqnarray*}

Let $(R,1)$ denote the first position in $R$,
and $(S,2)$ the second position in $S$, and so on.
In a chase step based on this dependency the values
from position $(R,1)$ get copied into
the position  $(S,2)$, whereas the value in position
$(R,1)$
``cause'' the generation of a new null value in $(S,1)$. 
This structure can been seen in the {\em dependency graph}
of $\Sigma_1$ that has a ``copy'' edge from vertex 
$(R,1)$ to vertex $(S,2)$,
and a ``generate'' edge from vertex $(R,1)$ to vertex $(S,1)$.
Note that the graph does not consider any
edges from $(R,2)$ because variable $y$ does not contribute
to the generated values.
The chase will terminate since there is no recursion
going through the $(S,2)$ position. By contrast, 
the dependency graph of 
\begin{eqnarray*}
\Sigma_2 & = & \{R(x,y)\rightarrow\exists{z}\;R(y,z)\}
\end{eqnarray*}
has a generating edge from
$(R,2)$ to $(R,2)$. It is the generating self-loop 
at $(R,2)$ which causes the chase on for example the instance
$\{R(a,b)\}$ to converge only at
the infinite instance 
$\{R(a,b),R(b,z_1)\}\cup\{R(z_i,z_{i+1}) :  i>0\}$.
The class of weakly acyclic tgds ({\sf WA}) is defined 
by \cite{DBLP:conf/icdt/FaginKMP03} to be
those sets of tgds whose dependency graph doesn't have
any cycles involving a generating edge.
It is easy to observe that 
the class {\sf WA} is closed under 
semi-enrichment
but it is not closed under enrichment.
This is because in the case of semi-enrichment 
the new relational symbol $H$ considered 
for each dependency contains only variables
that appears both in the body and the head of the 
dependency,
and the new $H$ atoms appear only in the heads of the 
semi-enriched dependency.
This means 
that the dependency graph for 
a semi-enriched set of {\sf WA}  tgds
will only add edges oriented into positions
associated with the new relational symbol.
The set $\Sigma=\{ R(x,y) \rightarrow \exists z\; R(x,z) \}$
shows that this is not the case for enrichment as
$\Sigma \in {\sf WA}$ but $\enrich{\Sigma} \notin {\sf WA}$.

The slightly smaller class of sets of tgds
with {\em stratified witness} ({\sf SW}) 
was introduced by \cite{DBLP:conf/icdt/DeutschT03} 
around the same time as {\sf WA}.
An intermediate class, the {\em richly acyclic} 
tgds ({\sf RA}) was introduced by 
\cite{DBLP:conf/pods/HernichS07} 
in a different context 
and it was later shown by
\cite{DBLP:conf/amw/GrahneO11}
that ${\sf RA}\in\mathsf{CT}^{\sf obl}_{\forall\forall}$.
It can be easily verified that both 
classes {\sf SW} and {\sf RA} are closed under 
enrichment.
The {\em safe dependencies} ({\sf SD}), introduced by 
\cite{DBLP:journals/pvldb/MeierSL09},
and the {\em super-weakly acyclic} ({\sf sWA}),
introduced by \cite{DBLP:conf/pods/Marnette09},
are both generalizations of the {\sf WA} class,
and both are close under semi-enrichment.

All of these classes have been proven to have {\sf PTIME}
membership tests, and have the following properties.

\begin{theorem}
{\em \cite{DBLP:conf/icdt/DeutschT03,DBLP:conf/icdt/FaginKMP03,DBLP:journals/pvldb/MeierSL09,DBLP:conf/pods/Marnette09,DBLP:conf/amw/GrahneO11}}
\begin{enumerate}
\item
${\sf SW} \; \subset \; {\sf RA} \; \subset \; {\sf WA}
\; \subset \; {\sf SD} \; \subset \; {\sf sWA}.$
\item
${\sf WA} \; \subset \; \mathsf{CT}^{\sf std}_{\forall\forall}$,
${\sf RA} \; \subset \; \mathsf{CT}^{\sf obl}_{\forall\forall}$, 
and
${\sf sWA} \; \subset \; \mathsf{CT}^{\sf sobl}_{\forall\forall}$. 
\end{enumerate}
\end{theorem}

\bigskip 
In order to 
complete the picture suggested by the previous
theorem we need a few more results.
Consider 
\begin{eqnarray*}
\Sigma_3 & =  & \{ {R}(x,y) \rightarrow \exists z\; {R}(x,z) \}.
\end{eqnarray*} 
Clearly $\Sigma_3 \in{\sf WA}$.
Let $I_0=\{ R(a,b) \}$,
and consider a semi-oblivious chase sequence
$I_0,I_1,I_2,\ldots$.
It is easy to see that for any $I_n$, 
where $n>0$, 
there 
exists a (non-active) trigger $(\xi,\{ x/a, y/z_{n} \})$,
meaning that the oblivious chase will not terminate.
Thus we have $\Sigma_3 \notin \octaa$.
On the other hand, for the set
\begin{eqnarray*} 
\Sigma_4 & = & \{ {S}(y),{R}(x,y)\rightarrow \exists z\; {R}(y,z)\},
\end{eqnarray*}
we have 
$\Sigma_4 \in \octaa$.
Furthermore,
$\Sigma_4 \notin{\sf WA}$,
since the dependency graph of $\Sigma_4$ will
have a generating self-loop on vertex $(R,2)$. 
This gives us

\begin{proposition}\label{PROP:WAvsSemiOblivious}
The classes ${\sf WA}$ and $\octaa$ are incomparable
wrt inclusion.
\end{proposition}

It was shown in \cite{DBLP:journals/pvldb/MeierSL09} 
that  ${\sf WA} \subset {\sf SD}$ and also that 
${\sf SD} \subset \sctaa$. We can now extend this result by showing
that, similarly to the {\sf WA} class,  the following holds:

\begin{proposition}\label{PROP:SDSemiOblivious}
The classes ${\sf SD}$ and $\octaa$ are incomparable
wrt inclusion.
\end{proposition}

\noindent
{\em Proof}: (Sketch)
The proof consists of showing that
$\Sigma_3\in{\sf SD}\setminus\octaa$,
and showing that $\Sigma_5 \in \octaa \setminus {\sf SD}$,
where 
\begin{eqnarray*}
\Sigma_5 & = & \{ R(x,x) \rightarrow \exists y\; R(x,y) \}.
\end{eqnarray*} 
Details are omitted. 
$_{\blacksquare}$

\medskip
From the semi-enrichment closure of the 
{\sf WA} and {\sf SD} 
classes and Proposition \ref{PROP:suffsobl}
we get the following result.

\begin{proposition}\label{PROP:soblTermClasses}
${\sf WA} \in \soctaa$
and ${\sf SD} \in \soctaa$.
Furthermore,
for any instance $I$ and any $\Sigma\in{\sf SD}$, 
the semi-oblivious chase 
with $\Sigma$ on $I$ terminates in time polynomial in the
number of tuples in $I$.
\end{proposition}

Note that the previous result follows directly also from 
a similar result for the class {\sf sWA} presented by
 \cite{DBLP:conf/pods/Marnette09}.
Still, as shown by the following proposition,
the super-weakly acyclic 
class does not include the class of dependencies that ensures 
termination for the oblivious chase variation, nor does
the inclusion hold in the other direction.

\begin{proposition}\label{PROP:SWAnotOCTAA}
${\sf sWA}$ and $\octaa$ are incomparable wrt inclusion.
\end{proposition}

\noindent
{\em Proof}: (Sketch)
We exhibit the super-weakly acyclic set
$\Sigma_3=\{ {R}(x,y) \rightarrow \exists z\; {R}(x,z) \}$. 
It is clear $\Sigma_3 \notin \octaa$.
For the converse, let
\begin{eqnarray*}
\Sigma_6 & = & \{{S}(x),{R}(x,y) \rightarrow \exists z\;{R}(y,z) \}.
\end{eqnarray*} 
Then $\Sigma_6 \notin {\sf sWA}$,
on the other hand it can be observed that the oblivious chase
with $\Sigma_6$
terminates on all instances. This is because
tuples with new nulls cannot cause the dependency to fire,
as these new nulls will never be present in relation $S$.
$_{\blacksquare}$

%We summarize the acyclicity based classes with the
%following Venn diagram.

%\bigskip
%\begin{figure}[!htbp]
%  \begin{center} 
%\includegraphics[scale=0.2]{WATerm.png}
% \caption{Acyclicity based classes.}
%  \end{center}
%\end{figure} 
%\bigskip

\bigskip 
\bigskip 
\subsection*{Stratification based classes}

\bigskip
Consider $\Sigma_7 = \{\xi_{1},\xi_{2}\}$, where
\begin{eqnarray*}
\xi_{1} & = & R(x,x)  \rightarrow  \exists{z}\;S(x,z), \mbox{ and}  \\ 
\xi_{2} & = & R(x,y),S(x,z)  \rightarrow  R(z,x).
\end{eqnarray*}

In the dependency graph of $\Sigma_7$ we will
have the cycle $(R,1) \rightsquigarrow (S,2) \rightsquigarrow (R,1)$,
and since $(S,2)$ is an existential position,
the set $\Sigma_7$ is not weakly acyclic.
However, it is easy to see that
$\Sigma_7\in\mathsf{CT}^{\sf std}_{\forall\forall}$.
It is also easily seen that if $S$ is empty and $R$ non-empty,
then $\xi_1$ will ``cause'' $\xi_2$ to fire 
for every tuple in $R$. Let us denote this ``causal''
relationship by $\xi_1\prec \, \xi_2$.
On the other hand,
there in no chase sequence in which
a new null in $(S,2)$ can be propagated
back to a tuple in $R$ and fire a trigger based on $\xi_1$,
thereby creating an infinite loop. 
We denote this with
$\xi_2\not\prec\xi_1$.
In comparison,
when chasing with 
\begin{eqnarray*}
\Sigma_8 & = & \{R(x,y)\rightarrow\exists{z}\;R(z,x)\}, 
\end{eqnarray*}
the new
null $z_n$ in $(z_{n},z_{n-1})$ will propagate into
tuple $(z_{n+1},z_{n})$, in an infinite regress.
If we denote the tgd in $\Sigma_8$ with $\xi$,
we conclude that $\xi\prec \, \xi$.
A formal definition of the 
$\prec$ relation is given in the
Appendix.

The preceding observations led 
\cite{DBLP:conf/pods/DeutschNR08} to 
define the class of
{\em stratified dependencies}
by considering the {\em chase graph} 
of a set $\Sigma$,
where the individual tgds in $\Sigma$ are the
vertices and there is an edge from $\xi_1$ to $\xi_2$
when $\xi_1\prec \, \xi_2$. A set $\Sigma$ is then
said to be stratified if the vertex-set of
every cycle in the chase graph forms a weakly
acyclic set. The class of all sets of stratified tgds
is denoted {\sf Str}.
In the previous example,
$\Sigma_7\in{\sf Str}$, and
$\Sigma_8\notin{\sf Str}$.

\cite{DBLP:journals/pvldb/MeierSL09}
observed that 
${\sf Str} \not \subseteq \sctaa$ and that
actually only 
${\sf Str}\subset\mathsf{CT}^{\sf std}_{\forall\exists}$,
and came up with a corrected definition of
$\prec$, which yielded the
{\em corrected stratified} 
{\sf CStr} of tgds\footnote{For a definition 
of {\sf CStr}, see the Appendix.},
for which they showed

\begin{theorem}
{\em \cite{DBLP:journals/pvldb/MeierSL09}}
$$ 
{\sf CStr}\subset \sctaa,\;
{\sf Str} \subset \sctae,\; and \;
{\sf CStr}\subset~{\sf Str}.
$$
\end{theorem}

\bigskip
From the observation that the 
 {\sf CStr} class is closed under semi-enrichment
and from Proposition \ref{PROP:suffsobl}
we have:

\begin{proposition}\label{PROP:cstrinSemiOblivious}
$ $
\begin{enumerate}
\item
${\sf CStr}\subset \soctaa$.
\item
${\sf CStr}$ and $\octaa$ are incomparable wrt
inclusion.
\end{enumerate}
\end{proposition}

\noindent
{\em Proof}: (Sketch)
For the second part we have $\Sigma_3 \in {\sf CStr}$
and $\Sigma_3 \notin \octaa$. For the converse
consider the dependency set $\Sigma_6$ from the proof of
Proposition \ref{PROP:SWAnotOCTAA}.
$_{\blacksquare}$

\medskip
\cite{Meier2009} further observed that the
basic stratification definition
also catches some false negatives.
For this they considered
the dependency set $\Sigma_9 = \{\xi_3,\xi_4\}$, where
%\vspace{-0.6cm}
\begin{eqnarray*}\label{EQ:oblivNonStandardTerminationSuf}
\xi_3 & = & S(x),E(x,y) \rightarrow  {E}(y,x), \mbox{ and}\\
\xi_4 & = & S(x),E(x,y) \rightarrow \exists z\; E(y,z),E(z,x). 
\end{eqnarray*}

Here $\xi_3$ and $\xi_4$ belong to the same
stratum according to the definition of {\sf CStr}.
Since new nulls in both $(E,1)$ and $(E,2)$
can be caused by $(E,1)$ and $(E,2)$,
there will be generating self-loops on
these vertices in the dependency graph.
Hence $\Sigma_9\notin{\sf CStr}$. On the other
hand, it is easy to see that the number of
new nulls that can be generated in the chase 
is bounded by the number of tuples in
relation $S$ in the initial instance.
Consequently $\Sigma_9\in\mathsf{CT}^{\sf std}_{\forall\forall}$.

In order to avoid such false negatives,
\cite{Meier2009} gave an alternative
definition of the $\prec$ relation
and of the chase graph.
Both of theses definitions are rather involved
technically, see the Appendix.
The new {\em inductively restricted} class,
abbreviated  {\sf IR},
restricts each connected component 
in the modified chase graph
to be in {\sf SD}. 
In example above, 
$\Sigma_9 \in {\sf IR}$.

\cite{Meier2009} also observed that {\sf IR}
only catches {\em binary}
relationships $\xi_1\prec \, \xi_2$.
%when $\xi_1$ can ``cause'' a unbounded generation
%of new nulls through $\xi_2$. 
This could be 
generalized to a ternary relation $\prec(\xi_1,\xi_2,\xi_3)$,
meaning that there exists a chase sequence
such that firing $\xi_1$ will cause $\xi_2$ to fire, 
and this in turn causes $\xi_3$ to fire. 
This will eliminate those cases
where $\xi_1$, $\xi_2$ and $\xi_3$ form a connected
component in the (modified) chase graph,
and yet there is no chase sequence that will fire
$\xi_1$, $\xi_2$ and $\xi_3$ in this order.
Thus the tree dependencies should not be in the same stratum.

Similarly to the ternary extension, 
the $\prec$ relation can be generalized to be $k$-ary.
The resulting termination classes are denoted 
${\sf T}[k]$. Thus ${\sf T}[2] = {\sf IR}$,
and in general ${\sf T}[k]\subset{\sf T}[k+1]$ 
as introduced by \cite{Meier2009}. The main property is

\begin{theorem}
{\em \cite{DBLP:journals/pvldb/MeierSL09}}
$$
 {\sf CStr}
\subset {IR} = {\sf T}[2]
\subset {\sf T}[3]
\subset \cdots
\subset {\sf T}[k]
\subset \cdots
\subset\mathsf{CT}^{\sf std}_{\forall\forall}.
$$
\end{theorem}

To complete the picture, we have the following
proposition based on the semi-oblivious closure for the
{\sf T}[k] hierarchy and Proposition \ref{PROP:suffsobl}.

\begin{proposition}\label{PROP:cstrinSemiObliviousH}
$ $
\begin{enumerate}

\item
${\sf T}[k] \subset\soctaa$.
\item
${\sf T}[k]$ and  $\octaa$ are incomparable
wrt inclusion. 
\end{enumerate}
\end{proposition}

Before concluding this section need to mention
that all the classes discussed here are closed under semi-enrichment,
thus they ensure the termination for the less expensive 
semi-oblivious chase in a polynomial number of steps, 
in the size of the input instance. Also need to note that 
all the previous classes were extended by 
\cite{DBLP:journals/pvldb/SpezzanoG10} using an innovative rewriting
approach.

The Hasse diagram in below figure 
summarizes
the all here considered classes and
their termination properties.

\bigskip
\begin{figure}[!htbp]
  \begin{center} 
\includegraphics[scale=0.35]{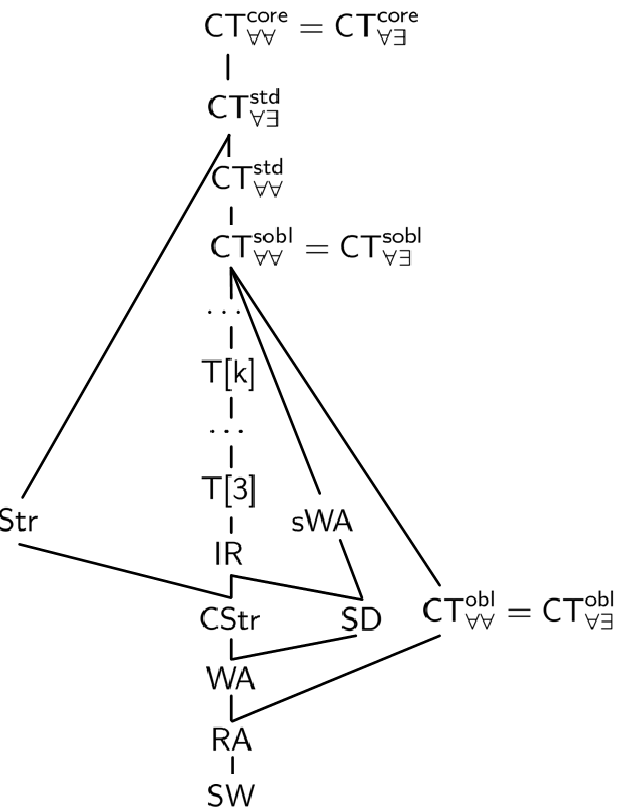}
 \caption{Sufficient classes.}
  \end{center}
\end{figure}

%-- COMPLEXITY 
\bigskip 
\section{Complexity of stratification}\label{complexity}

\bigskip
As we noted in Section \ref{sufficient},
all the acyclicity based classes
have the property that testing whether
a given set $\Sigma$ belongs to it
can be done in {\sf PTIME}. The situation
changes when we move to the stratified classes.
\cite{DBLP:conf/pods/DeutschNR08}
claimed that testing if $\xi_1 \prec \,\xi_2$ is in {\sf NP} for a 
given $\xi_1$ and $\xi_2$,
thus resulting in {\sf Str} having a {\sf coNP} membership problem.
We shall see in Theorem \ref{coNPhard} below
that this cannot be the case,
unless ${\sf NP}={\sf coNP}$.
We shall use the $\prec$ order as it is defined
for the {\sf CStr} class. The results also hold
for the {\sf Str} class. 
First we need a formal
definition.
Given a tgd $\xi$ and a mapping $h$
from the 
universally quantified variables in $\xi$
to $\dom(I)$,
by $h(\xi)$ we denote the formula obtained by replacing all 
universally quantified variables $x$ with $h(x)$.

\begin{definition}\label{meier}{\rm \cite{DBLP:journals/pvldb/MeierSL09}}
Let $\xi_1$ and $\xi_2$ be tgds.
Then $\xi_1$ {\em precedes} $\xi_2$,
denoted
$\xi_1 \prec \, \xi_2$, 
if there exists an instance $I$ and homomorphisms
$h_1$ and $h_2$ from the universal variables in
$\xi_1$ and $\xi_2$
to $\dom(I)$,
such that:
\begin{list}{(\roman{foo})~}{\usecounter{foo}}
\item 
$I\models h_2(\xi_2)$, and
\item
$I\xrightarrow{(\xi_1,h_1)} J$ using an oblivious chase step, and
\item 
$J \not \models 
h_2(\xi_2)$.
\end{list}
\end{definition}
%\begin{list}{\roman{foo}}{\usecounter{foo}}
%\item 
%$I\models
%\alpha_2(h_2(\bar{x}_2),\bar{y}_2)
%\rightarrow\exists\bar{z}_2
%\beta_2(h_2(\bar{x}_2),\bar{z}_2)$,
%\item
%$I\xrightarrow{(\xi_1,h_1)} J$, and
%\item 
%$J \not \models 
%\alpha_2(h_2(\bar{x}_2),\bar{y}_2)
%\rightarrow\exists\bar{z}_2
%\beta_2(h_2(\bar{x}_2),\bar{z}_2)$.
%\end{list}
%\end{definition}
Note that the pair $(\xi_1,h_1)$ in the previous 
definition denotes a trigger, not necessarily an active trigger,
because the chase step considered is the oblivious one.
Intuitively, the instance $I$ in the definition
is a witness to the ``causal'' relationship
between $\xi_1$ and $\xi_2$
(via $h_2$),
as $h_2(\xi_2)$ won't fire at $I$,
but will fire once $\xi_1$
has been applied to $I$.
The notion of stratum of $\Sigma$ is as before, 
i.e.\ we build a chase graph consisting of
a vertex for each tgd in $\Sigma$,
and an edge from $\xi_1$ to $\xi_2$
if $\xi_1\prec \, \xi_2$.
Then  
$\xi_1$ and $\xi_2$ are
in the same {\em stratum} when they both
belong to the same cycle
in the chase graph of $\Sigma$. 
A set $\Sigma$ of tgds is said to be 
{\em C-stratified} ({\sf CStr}) if all
its strata are weakly acyclic
\cite{DBLP:journals/pvldb/MeierSL09}.

\begin{theorem}\label{coNPhard}
$ $ 
\begin{enumerate}
\item
Given two tgds $\xi_1$ and $\xi_2$, the problem of
testing if $\xi_1 \prec \, \xi_2$ is {\sf coNP}-hard.
\item
Given a set of dependencies $\Sigma$,
the problem of  testing if $\Sigma \in {\sf CStr}$ 
is {\sf NP}-hard.
\end{enumerate}
\end{theorem}

\noindent
{\em Proof}:
For part 1 of the theorem
we will use a reduction from the graph 3-colorability 
problem that is known to be {\sf NP}-complete.
It is also well known that a graph $G$ is 3-colorable iff there is 
a homomorphism from $G$ to $K_3$,
where $K_3$ is the complete graph with 3 vertices.
We provide a reduction $G\mapsto\{\xi_1,\xi_2\}$,
such that $G$ is not 3-colorable if and only if
$\xi_1\prec \, \xi_2$.

We identify a graph 
$G=(V,E)$, where $|V|=n$ and $|E|=m$ 
with the sequence
$$
G(x_1,\ldots,x_n) = E(x_{i_{1}},y_{i_{1}}),\ldots,E(x_{i_{m}},y_{i_{m}}),
$$ 
and treat the elements
in $V$ as variables.  
Similarly, we identify the graph
$K_3$ with the sequence 
\begin{eqnarray*}
K_3(z_1,z_2,z_3) =& &
E(z_1,z_2),E(z_2,z_1),E(z_1,z_3),\\
& &E(z_3,z_1),
E(z_2,z_3),E(z_3,z_2)
\end{eqnarray*}
where $z_1,z_2$, and $z_3$ are variables.
With these notations, 
given a graph $G=(V,E)$, 
we construct tgds $\xi_1$
and $\xi_2$ as follows:

\vspace{-0.6cm}
\begin{alignat*}{2}
\xi_1 &= \;\;\;\; R(z)  & \rightarrow\;& \exists z_1,z_2,z_3\; K_3(z_1,z_2,z_3), \mbox{ and} \\
\xi_2 &=  E(x,y)\;& \rightarrow\;& \exists x_1,\ldots,x_n\;G(x_1,\ldots,x_n).
\end{alignat*}

Clearly the reduction is polynomial in the size of $G$.
We will now show that $\xi_1 \prec \,\xi_2$ iff
$G$ is not 3-colorable.

First, suppose that 
$\xi_1 \prec \,\xi_2$.
Then there exists 
an instance $I$ and homomorphisms $h_1$ and $h_2$,
such that $I \models h_2(\xi_2)$.
Consider $J$, where
$I \xrightarrow{(\xi_1,h_1)} J$. 
Thus $R^I$ had to contain at least 
one tuple, and $E^I$ had to be empty,
because otherwise the monotonicity 
property of the chase would imply that
that $J \models h_2(\xi_2)$.

On the other hand, 
we have 
$I \xrightarrow{(\xi_1,h_1')} J$,
where instance $J = I \cup \{K_3(h_1'(z_1),h_1'(z_2),h_1'(z_3))\}$,
and $h_1'$ is a distinct extension of $h_1$.
Since $E^I=\emptyset$, and we assumed that
$J \not \models h_2(\xi_2)$,
it follows that 
there is no homomorphism from 
$G$ into $J$,
i.e.\ 
there is no homomorphism from 
$G(h'_2(x_1),\ldots,h'_2(x_n))$ to 
$K_3(h_1'(z_1),h_1'(z_2),h_1'(z_3))$,
where $h'_2$ is a distinct extension of $h_2$.
Therefore the graph $G$
is not 3-colorable.

For the other direction, let us suppose that graph $G$ is not 3-colorable.
This means that there is no homomorphism from $G$ into $K_3$. 
Consider now $I=\{ R(a) \}$,
homomorphism $h_1=\{ z/a \}$, 
and homomorphism 
$h_2=\{ x/h'_1(z_1), y/h'_1(z_2) \}$.
It is easy to verify that $I$, $h_1$ and $h_2$
satisfy the three
conditions for $\xi_1 \prec \, \xi_2$.

\medskip
For part 2 of the theorem,
consider the set $\Sigma=\{ \xi_1, \xi_2 \}$ 
defined as follows:
{\small 
\begin{eqnarray*}
&\xi_1 \; =& \; R(z_1,v)\; \\
& &  \rightarrow \exists z_2,z_3,w\; K_3(z_1,z_2,z_3),R(z_2,w), R(z_3,w), S(w), \mbox{ and} \\
& \xi_2 \; =& \; E(x,y)\; \rightarrow  \exists x_1,\ldots,x_n,v\; G(x_1,\ldots,x_n),R(x,v).
\end{eqnarray*}
}

It is straightforward to verify
that $\Sigma \notin {\sf WA}$ and 
that $\xi_2 \prec \, \xi_1$. 
Similarly to the proof of part~1,
it can be shown that $\xi_1 \prec \, \xi_2$ iff the graph $G$ is not 
3-colorable. From this follows that $\Sigma \in {\sf CStr}$
iff there is no cycle in the chase graph
iff $\xi_1 \not \prec \,\xi_2$ iff 
$G$ is 3-colorable. 
$_{\blacksquare}$

\bigskip
Note that the 
reduction in the previous proof
can be used to show that  the 
problem ``$\Sigma \in {\sf Str}$?'' is {\sf NP}-hard.
Similar result can be also obtained
for the {\sf IR} class and also for the 
{\em local stratification} based classes introduced by  
\cite{DBLP:journals/pvldb/GrecoST11}.
The obvious upper bound for the problem $\xi_1 \prec \,\xi_2$ 
is given by:

\bigskip
\begin{proposition}\label{easy}
Given dependencies $\xi_1$ and $\xi_2$, 
the problem of determining whether
$\xi_1 \prec \, \xi_2$ is 
in $\mathsf{\Sigma^p_2}$.
\end{proposition}

\noindent
{\em Proof}:
From \cite{DBLP:conf/pods/DeutschNR08} 
we know that if $\xi_1 \prec \, \xi_2$
there is an instance $I$ satisfying 
Definition \ref{meier}, such that 
size of $I$ is bounded by a polynomial in the size of 
$\{\xi_1,\xi_2\}$. 
Thus, we can guess instance~$I$, homomorphisms $h_1$ and 
$h_2$ in {\sf NP} time. 
Next, with a {\sf NP} oracle 
we can check if 
$I \models h_2(\xi_2)$ and 
$J \not \models h_2(\xi_2)$, where
$I \xrightarrow{(\xi_1,h_1)} J$.$_{\blacksquare}$ 

\bigskip
We shall see that the upper bound of the proposition
actually can be lowered to $\mathsf{\Delta^p_2}$.
For this we need the following characterization theorem. 

\begin{theorem}\label{THEO:charac}
Let $\xi_1=\alpha_1 \rightarrow \beta_1$ and 
$\xi_2=\alpha_2 \rightarrow \beta_2$ be tgds. Then,
$\xi_1 \prec \, \xi_2$ if and only if
there is an atom $t$, and (partial) mappings
$h_1$ and $h_2$ on {\sf Vars},
such that the following hold.
\begin{list}{(\alph{qcounter})~}{\usecounter{qcounter}}
\item 
$t \in \beta_1$,
\item 
$h_1(t) \in h_2(\alpha_2)$, 
\item 
$h_1(t) \notin h_1(\alpha_1)$, and
\item There is no idempotent homomorphism from
$h_2(\beta_2)$ to
$h_2(\alpha_2) \cup h_1(\alpha_1) \cup h_1(\beta_1)$.
\end{list}
\end{theorem}

\noindent 
{\em Proof}:
We first prove the ``only if`` direction. 
For this, suppose that
$\xi_1 \prec \, \xi_2$, that is, 
there exists an instance
$I$ and mappings
$g_1$ and $g_2$,
such that conditions
$(i) - (iii)$ of Definition \ref{meier}
are fulfilled.

From conditions $(ii)$ and $(iii)$ 
we have that
$g_1(\alpha_1)  \subseteq I$ and 
$g_1'(\beta_1) \not \subseteq I$,
for any distinct extension $g_1'$ of $g_1$. 

Now, 
consider $h_1=g_1$ and $h_2=g_2$.
Let $t$ be an atom from $\beta_1$ such that 
$h'_1(t) \in h'_1(\beta_1) \cap h_2(\alpha_2)$  and 
$h'_1(t) \notin h_1(\alpha_1)$, for an extension $h'_1$ of $h_1$. 
Such an atom $t$ must exists, since
otherwise it will be that  
$h'_1(\beta_1) \cap h_2(\alpha_2) \subseteq h_1(\alpha_1)$,
which is not possible
because of conditions $(i)$ and $(iii)$ (note that $h_1=g_1$).
It is now easy to see that $t$, $h'_1$ and $h_2$ satisfy conditions
$(a), (b)$, and $(c)$ of the theorem.
It remains to show that condition $(d)$ also is satisfied.
By construction
we have $J= I \cup h'_1(\beta_1)$.
It now follows
that $I \cup h'_1(\beta_1) \not \models h_2(\xi_2)$.
Because $h_1(\alpha_1) \subseteq I$,
condition $(d)$  is indeed satisfied.

\medskip 
For the ``if'' direction of the theorem,
suppose that there exists an atom $t$
and homomorphisms $h_1$ and $h_2$,
such that 
conditions $(a),(b),(c)$ and $(d)$ holds.
Let $g_1=h_1$, $g_2=h_2$
and let $I=(h_1(\alpha_1) \cup h_2(\alpha_2)) \setminus h'_1(t)$,
for a distinct extension $h'_1$ of $h_1$.
Because $h'_1(t) \notin I$ and $h'_1(t) \in h_2(\alpha_2)$,
it follows that $h_2(\alpha_2) \not \subseteq I$.
Thus we have $I \models h_2(\xi_2)$, proving 
point $(i)$ of Definition
\ref{meier}. 
On the other hand, 
because point $(c)$ of the theorem is assumed, 
it follows that
$h_1(\alpha_1) \subseteq I$, 
from which we get
$I \xrightarrow{(\xi_1,h_1)} J$, 
where $J = I \cup h'_1(\beta_1)$,
proving  points $(i)$ and $(ii)$ from Definition \ref{meier}.  
Since
$I \cup h'_1(\beta_1)=h_1(\alpha_1) \cup h_2(\alpha_2) \cup h'_1(\beta_1)$,
and point $(d)$ holds, we get
$J \not \models h_2(\xi_2)$,
thus showing that condition $(iii)$ of Definition \ref{meier}
is also satisfied.\footnote{
It is easy to note that by adding the extra 
condition
``{\em $(e)$  there is no idempotent homomorphism from $\beta_1$ to $\alpha_1$}''
in the previous theorem we obtain a characterization of the 
stratification order associated with the {\sf Str} class.}
$_{\blacksquare}$

\bigskip
With this characterization result we can now 
tighten the
$\Sigma_2^p$ upper bound of Proposition \ref{easy}
as follows:

\begin{theorem}
Given dependencies $\xi_1$ and $\xi_2$,
the problem of determining whether
$\xi_1 \prec \, \xi_2$ is 
in $\mathsf{\Delta^p_2}$.
\end{theorem}

\noindent
{\em Proof}:
For this proof we will use the 
characterization Theorem \ref{THEO:charac},
and the observation  that 
$\mathsf{\Delta^p_2} = {\sf P}^{\sf NP}={\sf P}^{\sf coNP}$.
Consider the following {\sf PTIME} algorithm that 
enumerates all possible $h_1, h_2$ and $t$:

%\begin{codebox}
%\li \For $t \in \beta_1$ 
%\li  \Do \For all $(h_1,h_2)\in\mbox{mgu}(t,\alpha_2)$
%\li     \Do \If $h_1(t)\notin h_1(\alpha_1)$ \Return $t, h_1, h_2$ \End \End
%\end{codebox}

\begin{quote}
{\bf for all} $t \in \beta_1$\\ 
\hspace*{3ex}{\bf for all} $(h_1,h_2)\in\mbox{\sl mgu}(t,\alpha_2)$\\
\hspace*{6ex}{\bf if} $h_1(t)\notin h_1(\alpha_1)$ {\bf return} $t, h_1, h_2$

\end{quote}

\noindent 
In the algorithm, 
$\mbox{\sl mgu}(t,\alpha_2)$ denotes all 
pairs $(h_1,h_2)$
such that there exists an atom $t' \in \alpha_2$,
with $h_1(t)=h_2(t')$,
and there is no $(g_1,g_2)$ 
and $f$ different from the identity mappings, 
such that
$h_1=g_1 \circ f$
and 
$h_2=g_2 \circ f$.

\noindent
Using the values returned by previous algorithm and
with a {\sf coNP} oracle 
we can test if point $(d)$
holds.
Thus, the problem is in $\mathsf{\Delta^p_2}$. 
$_{\blacksquare}$
\bigskip

Armed with these results we can now state the 
upper-bound for the complexity 
of the {\sf CStr} 
membership problem.

\begin{theorem}\label{THEO:cstrUpperBound}
Let $\Sigma$ be a set of tgds. Then
the problem of testing if $\Sigma\in {\sf CStr}$ 
is in $\mathsf{\Pi^p_2}$.
\end{theorem}

\noindent
{\em Proof}: 
(Sketch)
To prove that $\Sigma$ is not in {\sf CStr}
guess a set of tuples $(\xi_1,t^1,h^1_1,h^1_2)$,
$\ldots$,$(\xi_k,t^k,h^k_1,h^k_2)$,
where $\xi_1, \ldots, \xi_k $ are tgds in $\Sigma$,
$t_1,\ldots,t_k$ are atoms, and $h^i_1$,$h^i_2$
are homomorphisms, for $i \in \{ 1,\ldots,k \}$.
Then, using an ${\sf NP}$ oracle 
check that 
$\xi_i < \xi_{i+1}$, for $i \in \{1,\ldots, k-1 \}$, 
and that $\xi_k < \xi_1$, using
the characterization Theorem \ref{THEO:charac}
with $t^i$, $h^i_1$, $h^i_2$ and
$t^k$, $h^k_1$, $h^k_2$ respectively.
And then check in  {\sf PTIME}
if the set of dependencies $\{ \xi_1, \ldots, \xi_k \}$
is not weakly acyclic.
Thus, the complexity is in $\mathsf{\Pi^p_2}$. 
$_{\blacksquare}$

\bigskip
We note that using the obvious upper-bound $\mathsf{\Sigma^p_2}$ for 
testing if $\xi_1 \prec \xi_2$, the membership problem for 
the class {\sf CStr} would be in $\mathsf{\Pi^p_3}$.
As mentioned the same results apply also for the class {\sf Str}.
Even if the complexity bounds for testing if $\xi_1 \prec \; \xi_2$
are not tight, it can be noted that a {\sf coNP} upper bound 
would not lower the $\mathsf{\Pi^p_2}$ upper bound of the membership problem
for {\sf CStr}.

The complexities of the various restricted classes is shown in
the diagram below. The diagram also shows the termination classes
$\octaa$, $\soctaa$, $\sctaa$ and $\cctaa$.

\bigskip
\begin{figure}[!htbp]
  \begin{center} 
\includegraphics[scale=0.16]{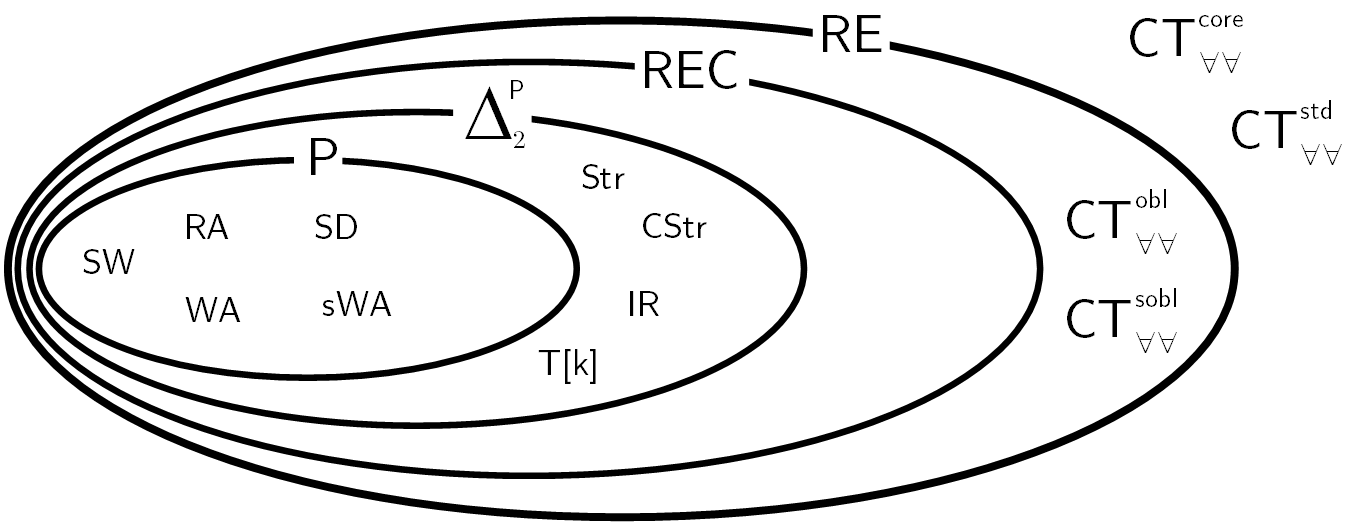}
 \caption{Membership complexity for restricted dependency classes.}
  \end{center}
\end{figure}

%-- CONCLUSION 
 \section{Conclusions}

\bigskip
We have undertaken a systematization of the
somewhat heterogeneous area of the chase.
Our analysis produced a taxonomy of the 
various chase versions and their termination
properties,
showing that the main sufficient classes 
that guarantee termination for the standard chase
also ensures termination for the complexity-wise less 
expensive semi-oblivious chase.
Even if the standard chase procedure 
in general captures more sets of dependencies 
that ensure the chase termination than the 
semi-oblivious chase, 
we argue that for most
practical constraints 
the semi-oblivious chase is a better choice.
We have also proved that the membership
problem for the classes $\cctaa,\cctae$ and $\sctae$
are {\sf coRE}-complete,
and in case we also at least one 
denial constraint the 
same holds for  $\sctaa$, $\soctaa$ and $\octaa$.
Still, 
it is yet unknown if the membership problem 
for  $\sctaa$ remains  {\sf coRE}-complete
without denial constraints.
The same also holds
for the classes $\soctaa$ and $\octaa$.
Finally we have analyzed the
complexity of the membership problem
for the class of stratified sets of dependencies.
Our bounds for this class are not tight, and 
it remains an open problem to determine
the complexity exactly.

\section*{Acknowledgements}
Many thanks to Friedrich Otto for pinpointing
  the complexity of the normal form existence
  problem.

%-- BIBLIOGRAPHY 
\bigskip
\bibliographystyle{abbrv}

\bibliography{Bibliography}

%-- APPENDIX 
\newpage
\appendix
\bigskip
\section{Termination Classes}

This appendix includes the full definitions 
of the dependency classes considered in 
Section \ref{sufficient}.

\subsection{Dependencies with Stratified-Witness ({\sf SW})}
\label{SUBSEC:stratifiedWitness}

\begin{definition}\label{DEF:position}
{\hspace*{-0.1ex}
\cite{DBLP:conf/icdt/FaginKMP03,DBLP:conf/kr/CaliGK08}}
For a database schema $\mathbf{R}$, 
define a {\em position}
in $\mathbf{R}$ to be a pair $(R,k)$, 
where $R$ is a relation symbol from 
$\mathbf{R}$ and $1 \leq  k \leq arity(R)$.
\end{definition}

\begin{definition}\label{DEF:flowGraph}\index{Chase Termination!Graphs!chase flow graph}
{\hspace*{-0.1ex}
\cite{DBLP:conf/icdt/DeutschT03}}
Let $\Sigma$ be a set of tgds over schema $\mathbf{R}$. 
The {\em chase flow graph} associated with 
$\Sigma$ is a directed edge-labeled graph $G^{F}_{\Sigma}=(V,E)$, 
such that each vertex in $V$ represents a position in $\mathbf{R}$ and 
 $((R,i),(S,j)) \in E$ if
there exists a tgd
$\xi \in \Sigma$ of the form
%$\forall\bar{x}\forall\bar{y}\; 
$\alpha(\bar{x},\bar{y}) \rightarrow \exists \bar{z}\; 
\beta(\bar{x},\bar{z})$,
a variable $u \in \bar{x}\cup\bar{y}$ occurring in position 
$(R,i)$ in $\alpha$ and a variable $v \in \bar{x}\cup\bar{z}$ 
that occurs in position $(S,j)$  in  $\beta$.
In case $v \in \bar{z}$, the edge is labeled as {\em existential},
otherwise the label is considered {\em universal}.
\end{definition}

\begin{definition}\label{DEF:stratifiedWitnesses}
{\hspace*{-0.1ex} \cite{DBLP:conf/icdt/DeutschT03}}
A set $\Sigma$ of tgds has {\em stratified-witness} if 
%its 
$G^{F}_{\Sigma}$ 
%corresponding chase flow graph 
has no cycles through an existential edge.
The class of all sets of dependencies with stratified-witness 
is denoted {\sf SW}. 
\end{definition}

As an example consider
a schema with two binary relations $R$
and $S$.
The positions in this schema 
are
$\{ (R,1), (R,2), (S,1), (S,2) \}$.
Let $\Sigma_1$ contain the following tgd:

\begin{eqnarray*}
\xi_{11}\;&:&\; S(x,y) \rightarrow \exists z\; R(x,z) \hspace{0cm}
\end{eqnarray*}

\noindent 
and let $\Sigma_2$ be:

\vspace{-0.7cm}
\begin{eqnarray*}
\xi_{21}\;&:&\; S(x,y) \rightarrow \exists z\; R(x,z) \hspace{0cm}\\
\xi_{22}\;&:&\; R(x,y) \rightarrow S(x,x) \hspace{0cm}
\end{eqnarray*}

\bigskip
It is easy to observe that $\Sigma_1 \in$ {\sf SW}.
On the other hand,  $\Sigma_2 \notin$ {\sf SW}, since the flow graph
of $\Sigma_2$
has the cycle given by the existential edges $((S,2),(R,2))$ and 
universal edge $((R,2),(S,2))$.
The following theorem
by \cite{DBLP:conf/icdt/DeutschT03}
guarantees standard chase termination for all sets
of dependencies in {\sf SW}.

\begin{theorem}
{\em \hspace*{-0.1ex} \cite{DBLP:conf/icdt/DeutschT03}}
{\sf SW} $\subset \sctaa$. 
\end{theorem}

\subsection{Rich Acyclicity ({\sf RA})}\label{SUBSEC:reachLyAcyclic}

\begin{definition}\label{DEF:extendedDependencyGraph}
{\hspace*{-0.1ex}
\cite{DBLP:conf/pods/HernichS07}}
Let $\Sigma$ be a set of tgds over schema $\mathbf{R}$. 
The {\em extended-dependency graph} associated with 
$\Sigma$ is a directed edge-labeled graph $G^{E}_{\Sigma}=(V,E)$, 
such that each vertex represents a position in $\mathbf{R}$ and 
 $(({R},i),({S},j)) \in E$, if
there exists a tgd
$\xi \in \Sigma$ of the form
$\alpha(\bar{x},\bar{y}) \rightarrow \exists \bar{z}\; 
\beta(\bar{x},\bar{z})$,
and if one of the following holds: 
\begin{enumerate}
\item 
$x \in \bar{x}$ and $x$ occurs in $\alpha$ on position $({R},i)$
and in  $\beta$ on position $({S},j)$. In this case the edge is labeled
as universal;
\item 
$x \in \bar{x}\cup\bar{y}$ 
and $x$ occurs in $\alpha$ 
on position $({R},i)$, 
and a variable $z \in \bar{z}$ 
occurs in  $\beta$ on position $({S},j)$.
In this case the edge is labeled as existential.
\end{enumerate}
\end{definition}

\begin{definition}\label{richlyAcyclic}
{\hspace*{-0.1ex} \cite{DBLP:conf/pods/HernichS07}}
A set of tgds $\Sigma$ is said to be {\em richly acyclic}
%if its extended dependency graph 
$G^{E}_{\Sigma}$ 
does not contain a cycle going 
through an existential edge. 
The class of all richly acyclic tgd sets
is denoted ${\sf RA}$. 
\end{definition}

As an example consider database schema 
$\mathbf{R}=\{S,R\}$, with $arity(S)=1$ and $arity(R)=2$. 
The following set
$\{ (S,1), (R,1), (R,2) \}$ represents all positions in $\mathbf{R}$.
Let $\Sigma_1$ contain the following dependency:
\begin{eqnarray*}
\xi_{1}\;&=&\; S(x) \rightarrow \exists y\; R(x,y)
\end{eqnarray*}

\noindent
and let $\Sigma_2$ contain:
\begin{eqnarray*}
\xi_{2}\;&=&\; R(x,y) \rightarrow \exists z\; R(x,z).
\end{eqnarray*}

\noindent
It is easy to see that $\Sigma_1 \in$ {\sf RA} as there are only 
outgoing edges from $(S,1)$ in the extended dependency graph.
On the other hand, $\Sigma_2 \notin$ {\sf RA} because 
there is a an existential self-loop on node $(R,2)$ 
in the extended dependency graph.
Note that the problem of testing if $\Sigma \in {\sf RA}$
is polynomial in size of $\Sigma$.

\begin{theorem}\label{obvChaseTermination}{\em \hspace*{-0.1ex}
\cite{DBLP:conf/amw/GrahneO11}}
${\sf RA} \subset \octaa$ 
\end{theorem}

\subsection{Weak Acyclicity ({\sf WA})} 

\begin{definition}\label{depGraph}
{\hspace*{-0.1ex} \cite{DBLP:conf/icdt/FaginKMP03}}
Let $\Sigma$ be a set of tgds over schema $\mathbf{R}$. 
The {\em dependency graph} associated with 
$\Sigma$ is a directed edge-labeled graph $G_{\Sigma}=(V,E)$, such that
the set of vertices $V$ represents the positions in $\mathbf{R}$. 
There is an edge $((R,i),(S,j)) \in E$, 
if 
there exists a dependency 
$\xi \in \Sigma$ of the form
$\alpha(\bar{x},\bar{y}) \rightarrow \exists \bar{z}\; 
\beta(\bar{x},\bar{z})$,
and a variable
$x\in \bar{x}$ such that
$x$ occurs in position 
$(R,i)$ in $\alpha$
and one of the following holds:
\begin{enumerate}
\item 
 $x$ occurs in $\beta$ in position $(S,j)$. In this case the 
edge is labeled as universal;
\item there exists variable $z \in \bar{z}$ which occurs 
in position $(S,j)$ in $\beta$. In this case the 
edge is labeled as existential.
\end{enumerate}
\end{definition}

\begin{definition}\label{weaklyAcyclic}
{\hspace*{-0.1ex} \cite{DBLP:conf/icdt/FaginKMP03}}
A set of tgds $\Sigma$ is said to be {\em weakly acyclic}
if 
%the corresponding dependency graph 
$G_{\Sigma}$
does not have any cycle going
through an existential edge. 
The class  of all weakly acyclic sets of tgds
is denoted ${\sf WA}$. 
\end{definition}

Note that the problem of testing if $\Sigma \in {\sf WA}$
is polynomial in size of $\Sigma$.

As an example consider $\Sigma_1$ containing the same dependencies
as in the example used for the {\sf RA}, $\Sigma_1$:
\begin{eqnarray*}
\xi_{1}\;&=&\; R(x,y) \rightarrow \exists z\; R(x,z)
\end{eqnarray*}

\noindent 
and let $\Sigma_2$ containing a slight variation of the dependency
from $\Sigma_1$:
\begin{eqnarray*}
\xi_{2}\;&=&\; R(x,y) \rightarrow \exists z\; R(y,z).
\end{eqnarray*}

\noindent
In this case $\Sigma_1 \in$ {\sf WA} 
as the dependency graph does not contain any cycles (note that
compared with the extended dependency graph the existential
self loop on node $(R,2)$ is not part of the dependency graph).
On the other hand, 
$\Sigma_2$ is not weakly acyclic as it has an existential self loop
on node $(R,2)$.

\begin{theorem}\label{WATermination}{\em \hspace*{-0.1ex}
\cite{DBLP:conf/icdt/FaginKMP03}}
{\sf WA} $\subset \sctaa$.
\end{theorem}

\subsection{Safe dependencies ({\sf SD})}\label{safeConstraints}

\begin{definition}\label{affectedPosition}
{\hspace*{-0.1ex}\cite{DBLP:conf/kr/CaliGK08}}
The set of {\em affected positions} {\sl aff}$\,(\Sigma)$ 
for a set of tgds $\Sigma$
is defined as follows. For all positions 
$(R,i)$ that occur in the head of some tgd $\xi \in \Sigma$, then
\begin{enumerate}
\item if an existential variable appears in position $(R,i)$ in $\xi$, then
$(R,i) \in$ {\sl aff}$\,(\Sigma)$; 
\item if universally quantified variable $x$ appears in position 
$(R,i)$ in the head and $x$ appears only in affected positions in the 
body, then $(R,i) \in$ {\sl aff}$\,(\Sigma)$.
\end{enumerate}
\end{definition}

Intuitively, the affected positions are those where new null values
can occur during the chase process.
For example, the set of affected positions associated with the 
set of dependencies    
$\Sigma = \{  R(x,y,z),S(y) \rightarrow \exists w\; R(y,w,x) \}$
is  {\sl aff}$\,(\Sigma)=\{ (R,2) \}$.

\begin{definition}\label{propagationGraph}
{\hspace*{-0.1ex}\cite{DBLP:journals/pvldb/MeierSL09}}
The {\em propagation graph} for a
set of tgds $\Sigma$ is a directed edge labeled graph $G^P_{\Sigma}=(\mbox{\sl aff}\,(\Sigma),E)$.
An edge $((R,i),(S,j))$ belongs to $E$ if 
there exists a dependency 
$\xi \in \Sigma$ of the form
$\alpha(\bar{x},\bar{y}) \rightarrow \exists \bar{z}\; 
\beta(\bar{x},\bar{z})$,
a variable $x$ that occurs in $\alpha$ at position 
$(R,i)$, such that 
$x$ occurs only in affected positions in $\alpha$
and one of the following holds:
\begin{enumerate}
\item $x$ appears in $\beta$ at affected position $(S,j)$.
In this case the edge is labeled as universal;
\item there exists variable $z \in \bar{z}$ which occurs 
at position $(S,j)$ in $\beta$.
In this case the edge is labeled existential.
\end{enumerate}
\end{definition}

\begin{definition}\label{safeDependencies}
{\hspace*{-0.1ex}\cite{DBLP:journals/pvldb/MeierSL09}}
A set of tgds $\Sigma$  is called {\em safe} if its propagation  
graph $G^P_{\Sigma}$ does not have a cycle going through an existential edge.
The class of all 
safe sets of tgds
is denoted {\sf SD}.  
\end{definition}

Consider for example 
$\Sigma =
\{R(x,y,z),S(y) \rightarrow \exists w\; R(y,w,x)\}$. 
It is easy to see that $\Sigma \notin$ {\sf WA}, but 
because the only affected position in $\Sigma$ is $(R,2)$ 
and there are no edges going in or out from this position 
in $G^P_{\Sigma}$, 
it follows that $\Sigma \in$ {\sf SD}.

We note that the problem of testing 
if $\Sigma \in$ {\sf SD},
for a given $\Sigma$,
is polynomial in size of~$\Sigma$.

\begin{theorem}\label{teo:safePolynomial}
{\em \hspace*{-0.1ex}\cite{DBLP:journals/pvldb/MeierSL09}}
${\sf SD} \subset \sctaa$. 
\end{theorem}

\subsection{Super-weak acyclicity ({\sf sWA})}

Let $\Sigma$ be a set of tgds such that no two tgds share 
common variable names.
First Skolemize $\Sigma$ by replacing each 
existential variable $y$ in $\xi \in \Sigma$ 
by a Skolem function $f_y(x_1,\ldots,x_n)$,
where $x_1,\ldots,x_n$ are the universal variables
that occur both in the body and head of $\xi$.
The Skolemized $\Sigma$ can then be viewed as a
logic program $P_{\,\Sigma}$.

\begin{definition}
{\hspace*{-0.1ex}\cite{DBLP:conf/pods/Marnette09}}
A {\em place} for a logic program $P_{\,\Sigma}$ is a
pair $(A,i)$, where $A$ is an atom in $P_{\,\Sigma}$
and $1 \leq i \leq arity(A)$.
\end{definition}

\begin{definition}
{\hspace*{-0.1ex}\cite{DBLP:conf/pods/Marnette09}}
Let $\xi \in \Sigma$ be a tgd, and $y$ an existential 
variable in $\xi$. The set of {\em output places} for
$y$ in $\xi$, denoted {\sl Out}$(\xi,y)$,
is the set of places in the head of $P_{\{\xi\}}$
that contains the Skolem term $f_y(\ldots)$.
\end{definition}

\begin{definition}
{\hspace*{-0.1ex}\cite{DBLP:conf/pods/Marnette09}}
Given a tgd $\xi \in \Sigma$, and $x$ a universal 
variable in $\xi$. The set of {\em input places} for
$x$ in $\xi$, denoted  {\sl In}$\,(\xi,x)$,
is the set of places in the body of $P_{\{\xi\}}$
where $x$ occurs.
\end{definition}

In the following definition a {\em substitution}
is a function from variables to variables and constants.
Substitutions are extended to atoms containing function terms
in the natural way.

\begin{definition}
{\hspace*{-0.1ex}\cite{DBLP:conf/pods/Marnette09}}
Places $(A,i)$ and $(B,j)$ are unifiable, denoted 
$(A,i) \sim (B,j)$, if $i=j$ and there exists 
substitutions $\theta$ and $\theta'$ such that 
$\theta(A)=\theta'(B)$.
\end{definition}

Given two sets of places $Q$ and $Q'$, 
the relationship $Q \sqsubseteq Q'$
means that 
for all $q \in Q$ there exists $q' \in Q'$,
such that $q \sim q'$. 

Let $S$ be a set 
of atoms and $x$
a variable.
Then $\Gamma_x(S)$ denotes the set of all places where $x$ occurs
in some atom in $S$.

Let $\Sigma$ be a set of tgds and $Q$ a set of places.
Then {\sl Move}$\,(\Sigma,Q)$ denotes the smallest set of places
such that $Q \sqsubseteq \mbox{\sl Move}\,(\Sigma,Q)$, 
and for every rule
$\xi \in P_{\Sigma}$ of the form $\alpha \rightarrow \beta$,
and for every variable $x$,
if $\Gamma_x(\alpha) \sqsubseteq \mbox{\sl Move}\,(\Sigma,Q)$,
then $\Gamma_x(\beta) \sqsubseteq \mbox{\sl Move}\,(\Sigma,Q)$.

\begin{definition}
{\hspace*{-0.1ex}\cite{DBLP:conf/pods/Marnette09}}
Let $\Sigma$ be a set of tgds and let $\xi, \xi' \in \Sigma$. 
Then $\xi$ is said to trigger $\xi'$, denoted $\xi \sim_{\Sigma} \xi'$,
if there exists an existential variable $y$ in $\xi$, 
and a universal variable $x$ 
that appears both in the head and body of $\xi'$,
such that 
{\sl In}$\,(\xi',x) \sqsubseteq 
\mbox{\sl Move}\,(\Sigma, \mbox{\sl Out}\,(\xi,y))$.
A set of constraints $\Sigma$ is 
said to be {\em super-weakly acyclic} iff
the trigger relation $\sim_{\Sigma}$ is acyclic.
The set of all 
super-weakly acyclic sets of tgds
is denoted {\sf sWA}. 
\end{definition}

\begin{theorem}
{\hspace*{-0.1ex}\cite{DBLP:conf/pods/Marnette09}}
{\sf sWA} $\subset \soctaa$.
\end{theorem}

\cite{DBLP:conf/pods/Marnette09} also shows that
testing if a set of tgd $\Sigma$ is super-weakly acyclic 
is polynomial.

\subsection{Stratification ({\sf Str}, {\sf CStr})}\label{cStratifiction}

Let $\xi$ be a tgd
$\alpha(\bar{x},\bar{y})\rightarrow \exists \bar{z}\; \beta(\bar{x},\bar{z})$,
and $\bar{a} = \bar{a}_1\bar{a}_2$ a sequence cf constants of same length
as $\bar{x}\bar{y}$. Then $\xi(\bar{a})$ denotes the tgd
$\alpha(\bar{a}_1,\bar{a}_2)\rightarrow \exists \bar{z}\; \beta(\bar{a}_1,\bar{z})$.

\begin{definition}\label{stratTigger}
{\hspace*{-0.1ex}\cite{DBLP:conf/pods/DeutschNR08}}
Let $\xi_1$ and $\xi_2$ be tgds. 
We write 
$\xi_1 \prec \xi_2$, 
if there are instances $I$ and $J$, 
and a sequence $\bar{a}$ with values from $\dom(J)$,
such that:
\begin{enumerate}
\item $I \models \xi_2(\bar{a})$, and
\item there exists an active trigger $(\xi_1,h)$, such that 
$I \xrightarrow{(\xi_1,h)} J$, and
\item $J \not \models \xi_2(\bar{a})$.
\end{enumerate}
\end{definition}

As an example consider $\Sigma=\{ \xi_1, \xi_2 \}$, where:
\begin{eqnarray*}
\xi_1&=&\; R(x,y) \rightarrow S(x),\mbox{ and} \\
\xi_2&=&\; S(x) \rightarrow R(x,x). 
\end{eqnarray*}
With the instance $I=\{ R(a,b) \}$ and 
the sequence $\bar{a}=(a)$ we have that
$I \models \xi_2(a)$; and for the homomorphism 
$h=\{ x/a, y/b \}$ we have
$I \xrightarrow{(\xi_1,h)} J$, where
$J= \{ R(a,b), S(a) \}$.
Because $J \not \models \xi_2(a)$, it follows that
$\xi_1 \prec \xi_2$.
On the other hand, $\xi_2 \not \prec \xi_1$
because for any instance $I$ and sequence
of constants $\bar{b}$ such that
$I \models \xi_1(\bar{b})$
and 
$I \xrightarrow{(\xi_2,h)} J$, for some active trigger $(\xi_2,h)$,
it follows that 
$J \models \xi_2(\bar{b})$.

\bigskip

Given a set of tgds $\Sigma$, the {\em chase graph} associated with
$\Sigma$ is a directed graph 
$G^C_{\Sigma}=(V,E)$, 
where $V=\Sigma$, and 
$(\xi_1,\xi_2) \in E$ iff $\xi_1 \prec \xi_2$.

\begin{definition}\label{chaseGraph}
{\hspace*{-0.1ex}\cite{DBLP:conf/pods/DeutschNR08}}
A set of tgds $\Sigma$ is said to be {\em stratified}
if the set of dependencies in every simple cycle
in $G^C_{\Sigma}$ 
 is weakly acyclic. The set of all stratified tgd sets is denoted
{\sf Str}.
\end{definition}

\begin{theorem}\label{teo:stratified}
\hspace*{-0.1ex}{\em \cite{DBLP:journals/pvldb/MeierSL09}}
{\sf Str} $\subset \sctae$. 
\end{theorem}

Next is the definition of the class {\sf CStr}.

\begin{definition}\label{cprec}
{\hspace*{-0.1ex}\cite{DBLP:journals/pvldb/MeierSL09}}
Let $\xi_1$ and $\xi_2$ be tgds. 
We write 
$\xi_1 \prec_{\mbox{c}} \xi_2$, 
if there are instances $I$ and $J$, 
and sequence
$\bar{a}$ with values from $\dom(J)$,
such that:
\begin{enumerate}
\item 
$I \models \xi_2(\bar{a})$, and
\item 
there exists (not necessarily active) trigger $(\xi_1,h)$, 
such that 
$I \xrightarrow{(\xi_1,h)} J$ in an oblivious 
chase step, and
\item 
$J \not \models \xi_2(\bar{a})$.
\end{enumerate}
\end{definition}

Given a set of tgds $\Sigma$, the {\em c-chase graph} associated with
$\Sigma$ is a directed graph $G^{CC}_{\Sigma}=(V,E)$, 
where $V=\Sigma$ and 
$(\xi_1,\xi_2) \in E$ iff $\xi_1 \prec_{\mbox{c}} \xi_2$. 
A set of tgds $\Sigma$ is said to be {\em c-stratified}
if the set of dependencies in every simple cycle in $G^{CC}_{\Sigma}$ 
is weakly acyclic. 
The set of all c-stratified tgd sets is denoted
{\sf CStr}.

\begin{theorem}\label{Cstratified}
{\em \hspace*{-0.1ex}\cite{DBLP:journals/pvldb/MeierSL09}}
${\sf CStr} \subset \sctaa$. 
\end{theorem}

\subsection{Inductively Restricted Dependencies ({\sf IR}) and the {\sf T}-hierarchy}\label{terminationHierarchy}

The definitions below are taken from the erratum 
(http://arxiv.org/abs/0906.4228) and not from 
\cite{DBLP:journals/pvldb/MeierSL09}, where the presented condition,
as mentioned in the erratum, does not guarantee the standard-chase termination 
for all sequences on all instances.

Let $\Sigma$ be a set of tgds, $I$ an instance and $N$ a set of nulls.
The set of all positions $(R,i)$, 
such that there is a tuple in $I$ that contains a null from $N$
in position $(R,i)$, 
is denoted {\sl nullpos}$\,(N,I)$.

\begin{definition}
{\hspace*{-0.1ex}\cite{DBLP:journals/pvldb/MeierSL09}}
Let $\Sigma$ be a set of tgds and $P$ a set of positions.
Let $\xi_1,\xi_2\in\Sigma$. 
Then
$\xi_1 \prec_{P} \xi_2$ if there are instances $I$, $J$ and sequence
$\bar{a}$ of values  
from $\dom(J)$, such that:
\begin{enumerate}
\item 
$I\models \xi_2(\bar{a})$, and
\item 
there exists (not necessarily active) trigger $(\xi_1,h)$, 
such that 
$I \xrightarrow{(\xi_1,h)} J$ in an oblivious 
chase step, and
\item 
$J \not \models \xi_2(\bar{a})$, and
\item 
there is an null $x$ in the head of $\xi_2(\bar{a})$, 
such that {\sl nullpos}$\,(\{x\},I) \subseteq P$.
\end{enumerate}
\end{definition}

As an example, consider $\Sigma$ containing a single tgd
$\xi=\; R(x,y) \rightarrow \exists z\; R(y,z)$.
Note that $\Sigma \notin \sctaa$.
It is easy to see 
that with instances $I=\{ R(a,b) \}$, 
$J=\{ R(a,b), R(b,X) \}$, 
and sequence $\bar{a}=(b,x)$, 
conditions 1,2 and 3 from the previous definition are fulfilled.
For the 4$^{\mbox{{\small th}}}$ condition, note that
$\xi(\bar{a})$ represents the formula 
$R(a,x) \rightarrow \exists z R(x,z)$.
Thus, $x$ occurs in the head of  $\xi(\bar{a})$. 
On the other hand, 
{\sl nullpos}$\,(\{ X\}, I)=\emptyset$
and
instance $I$ does not contain
any nulls, hence for any set $P$,  
{\sl nullpos}$\,(\{ X\}, I) \subseteq P$. 
Consequently $\xi \prec_{P} \xi$, for any set 
of positions $P$.

\bigskip
For a tgd $\xi$,
with {\sl vars}$_{\forall}(\xi)$ 
we denote the set of all universally quantified variables in $\xi$ and
with {\sl vars}$_{\exists}(\xi)$  
the set of all existentially quantified
 variables in $\xi$.

\begin{definition}\label{affcl}
{\hspace*{-0.1ex}\cite{DBLP:journals/pvldb/MeierSL09}}
Let $P$ be a set of positions and $\xi$ a tgd. 
Then {\sl affcl}$\,(\xi,P)$ denotes the
set of positions $(R,i)$ from the head of $\xi$, 
such that one of the following holds:
\begin{enumerate}
\item 
for all $x \in {\sl vara}_{\forall}(\xi)$, where $x$ occurs in $(R,i)$,
the variable $x$ occurs in the body of $\xi$ only in positions from $P$, or
\item 
position $(R,i)$ contains a variable $x \in {\sl vars}_{\exists}(\xi)$.
\end{enumerate}
\end{definition}

For the previous example 
{\sl affcl}$(\xi,P) =\{ (R,1), (R,2) \}$, 
where $P=\{ (R,2) \}$.
Given a set of dependencies $\Sigma$,  the set of all 
positions in $\Sigma$ is written as {\sl positions}$\,(\Sigma)$. 

\begin{definition}\label{2restrictionSystem}
{\hspace*{-0.1ex}\cite{DBLP:journals/pvldb/MeierSL09}}
A {\em 2-restriction system} is a pair $(G_{\Sigma},P)$,
where $G_{\Sigma}$ is a directed graph $(\Sigma,E)$ and 
$P \,\subseteq$ {\sl  positions}$(\Sigma)$ such that:
\begin{enumerate}
\item for all $(\xi_1, \xi_2) \in E$, 
{\sl affcl}$\,(\xi_1,P) \;\cap$ {\sl position}$\,(\Sigma) \subseteq P$ and 
{\sl affcl}$\,(\xi_2,P) \;\cap$ {\sl positions}$\,(\Sigma) \subseteq P$, and
\item for all $\xi_1 \prec_{P} \xi_2$, $(\xi_1,\xi_2) \in P$.
\end{enumerate}
\end{definition}

A 2-restriction system is {\em minimal} if it is obtained from 
$((\Sigma,\emptyset),\emptyset)$, 
that is the graph over $\Sigma$ without any edge
and the set of position empty,
by a repeated application of constraints 
1 and 2, from the previous definition, such that $P$ is extended only by those 
positions that are required to satisfy condition 1.
Let us denote by $part(\Sigma,2)$ the set that contains 
the sets of all 
strongly connected components in a minimal 2-restriction system.

Returning to our example 
the minimal 2-restriction system is computed as follows. 
Consider pair 
$((\{ \xi\},\emptyset),\emptyset)$.
Previously we showed that 
$\xi \prec_{P} \xi$, for any set of positions $P$,
by particularization we have $\xi \prec_{\emptyset} \xi$.
Thus, we add edge $(\xi,\xi)$ to $E$. 
Using condition 1 from
Definition \ref{2restrictionSystem} we have $affcl(\xi,\emptyset)=\{ (R,2)\}$.
That is we add position $(R,2)$ to $P$. 
By repeating this process once again
with $P=\{ (R,2) \}$, we add to $P$ the position $(R,1)$ too. 
Hence, the minimal 2-restriction system is 
$((\Sigma,\{ (\xi,\xi)),\{ (R,1), (R,2) \} \})$. 
The only connected 
component in this restriction system is $\{ \xi \}$.
\cite{DBLP:journals/pvldb/MeierSL09} provide a simple 
 algorithm that computes the set $part(\Sigma,2)$.

\begin{definition}\label{indRestDep}
{\hspace*{-0.1ex}\cite{DBLP:journals/pvldb/MeierSL09}}
A set $\Sigma$ of tgds is called {\em inductively restricted} iff
every $\Sigma' \in part(\Sigma,2)$ is in {\sf SD}.
The set of all inductively restricted tgd sets is denoted by
${\sf IR}$.
\end{definition}

\begin{theorem}\label{teo:indRestPolynomial}
{\em \hspace*{-0.1ex}\cite{DBLP:journals/pvldb/MeierSL09}}
{\sf IR} $\subset \sctaa$.
\end{theorem}

\cite{DBLP:journals/pvldb/MeierSL09} observed
that the inductive restriction criterion can be extended to form a hierarchy
of classes that ensure the standard-chase termination on all branches for all
instances. 
Intuitively, the lowest level of this hierarchy, noted $T[2]$, 
is the class of 
inductively restricted dependencies. 
Level $T[k]$, $k>2$ is obtained by extending the binary relation $prec_{P}$
to a $k$-ary relation $prec_{k,P}$. 
Intuitively, $prec_{k,P} (\xi_1,\ldots,\xi_k)$
means that there exists a standard-chase sequence 
such that firing $\xi_1$ will also cause $\xi_2$ to fire.
This in turn will
cause $\xi_3$ to fire and so on until $\xi_k$.
Based on this new relation, the set
$part(\Sigma,k)$ is computed similarly to  $part(\Sigma,2)$.
The algorithm that computes $part(\Sigma,k)$ was introduced by  \cite{DBLP:journals/pvldb/MeierSL09}.
For all $k\leq 2$, it is shown that $T[k] \subset T[k+1]$.

\end{document}